\documentclass{aa}
\usepackage{natbib}
\bibpunct{(}{)}{;}{a}{}{,}
\usepackage{graphicx}
\usepackage{txfonts}
\usepackage{lscape}
\usepackage{longtable}
\newcommand{\spi}{{\it Spitzer}}
\newcommand{\hypz}{{\it Hyperz}}
\newcommand{\hst}{{\it Hubble Space Telescope}}
\newcommand{\sex}{{\it SExtractor}}
\begin{document}

   \title{Searching for massive galaxies at $z\geq3.5$ in GOODS-North}

   \author{C. Mancini\inst{1,2,4}, 
     I. Matute\inst{2}, A. Cimatti\inst{3}, E. Daddi\inst{4}, M. Dickinson\inst{5}, G. Rodighiero\inst{6},
	 M. Bolzonella\inst{7},  L. Pozzetti\inst{7}}
   \offprints{aaa}

   \institute{
     Dipartimento di Astronomia e Scienza dello Spazio, Universit\`a
     degli Studi di Firenze, Largo E. Fermi 3 50125, Firenze, Italy
     \email{mancini@arcetri.astro.it}
     \and           
     Osservatorio Astrofisico di Arcetri (OAF), INAF-Firenze, Largo
     E. Fermi 5, 50125 Firenze
     \and
      Dipartimento di Astronomia, Universit\`a 
     di Bologna, via Ranzani 1, I-40127 Bologna, Italy
     \and
     CEA-Saclay,DSM/DAPNIA/Service d'Astrophysique, 91191 Gif-Sur Yvette
     Cedex, France
     \and
     NOAO, 950 N. Cherry Ave. P.O. 26732, Tucson, AZ 85726-6732, USA
     \and
     Dipartimento di Astronomia, Universit\`a di Padova, 
     Vicolo Osservatorio 2, I-35122, Padova, Italy
     \and
     INAF-Bologna, Via Ranzani, I-40127 Bologna, Italy
}

 \abstract
{}
{We constrain the space density and properties of massive galaxy candidates
at redshifts of $z\geq3.5$ in the Great Observatories
Origin Deep Survey North (GOODS-N) field. 
By selecting sources in the \spi\ + IRAC bands, 
a highly stellar mass-complete sample is assembled, including massive galaxies 
which are very faint in the optical/near-IR bands that would
be missed by samples selected at shorter wavelengths.}
{The $z\geq3.5$ sample was selected down to $m_{AB}=23$ mag at $4.5~\mu$m using photometric redshifts that have been
obtained by fitting the galaxies spectral energy distribution at optical, near-IR bands and IRAC bands. 
We also require that the brightest band (in AB scale) in which candidates are detected is the IRAC  $8.0~\mu$m band in order to ensure that
the near-IR $1.6~\mu$m (rest-frame) peak is
falling in or beyond this band.} 
{ We found 53 $z~\geq3.5$ candidates, with masses in the range of
  M$_{\star}\sim10^{10}-10^{11}$~M$_{\odot}$. At least $\sim $81\% of these
  galaxies are missed by traditional 
Lyman Break selection methods based on ultraviolet light. 
\spi\ + MIPS emission is detected for 60\% of the sample of $z\geq3.5$ 
galaxy candidates. Although in some cases 
this might suggest a residual contamination from lower redshift star-forming galaxies or Active Galactic Nuclei, 37\%
of these objects are also detected in the sub-mm/mm bands in recent SCUBA, AzTEC and MAMBO
surveys, and have properties fully consistent with vigorous starburst galaxies at $z\geq3.5$. 
The comoving number density of galaxies with stellar masses $\geq 5\times 10^{10}$~M$_{\odot}$ (a reasonable stellar mass completeness limit for our sample)
is $2.6\times10^{-5}$~Mpc$^{-3}$ (using the volume within $3.5<z<5$), and the corresponding
stellar mass density is $\sim(2.9\pm 1.5)\times10^6$~M$_{\odot}$Mpc$^{-3}$, or
about 3\% of
the local density above the same stellar mass limit. For the sub-sample of MIPS-undetected
galaxies, we find a number density of $\sim 0.97\times10^{-5}$~Mpc$^{-3}$ and a stellar mass
density of $\sim(1.15\pm 0.7)\times10^6$~M$_{\odot}$Mpc$^{-3}$. Even in the unlikely case that
these are all truly quiescent galaxies, this would imply an increase in the space density 
of passive galaxies by a factor of $\sim15$ from $z\sim 4$ to $z=2$ , and by $\sim100$ to $z=0$.} 
{}

\titlerunning{Searching for massive galaxies at $z\geq3.5$ in GOODS-North}
\authorrunning{Mancini C. et al.}
\maketitle
%

\section{Introduction}
The question of the galaxy mass assembly at early cosmological epochs is still open. In a $\Lambda$CDM Universe, the standard formation scenario predicts hierarchical growth of structures. Local galaxies would have been formed through repeated merging events. The first semi-analytic galaxy-formation models \citep{1998astro.ph.10031K} predicted that the assembly of the most massive systems would only have been completed in the most recent epochs ($z<1$). Nevertheless, in recent years a substantial population of massive galaxies has been found beyond $z\sim1$. 
Up to $z\simeq2.5$ the most massive objects spectroscopically confirmed are
similar to the local Early Type Galaxies (ETGs): old (age of $\sim$1~Gyr),
passively evolving, and with M$_{\star}>10^{11}$~M$_{\odot}$
\citep{2004Natur.430..184C,2004ApJ...614L...9M, 2005ApJ...626..680D,
  2005MNRAS.357L..40S, 2006ApJ...649L..71K}. 

At higher redshifts the availability of multi-wavelengths data from the new generations of deep surveys has allowed searches for massive galaxy candidates up to $z~\simeq5-6.5$. 
The dropout-technique  pioneered by \citet{1996ApJ...462L..17S} to find Lyman-break galaxies
(LBGs) at $z\simeq3$,
has proven successful also for identifying blue star-forming
galaxies at higher redshifts \citep[$z\sim4-6$][]{1999ApJ...519....1S, 1998hdf..symp..219D,
  2003ApJ...592..728S,2004ApJ...600L..93G, 2006MNRAS.372..357M, 2006ApJ...651...24Y}.

A substantial amount of  spectroscopic confirmations is available for LBGs at $z\sim 5-6$
\citep{2003MNRAS.342L..47B,2004ApJ...607..704S,
2004ApJ...600L..99D, 2007MNRAS.374..910E, 2007ApJ...659...84S,
2008A&A...478...83V}.  This allowed early estimates of the stellar mass density at those redshifts. For instance 
  \citet{2006ApJ...651...24Y} measured the comoving stellar mass density at
  $z\sim 6$ in the total GOODS field and found it in good agreement with the simulations of
  \citet{2006MNRAS.366..705N}. The stellar mass density found in the GOODS-S
  field by
  \citet{2007MNRAS.374..910E} for a sample of
  i-dropout LBGs in the same
  redshift range is also consistent with the
  \citet{2006ApJ...651...24Y} results. However it is larger by a factor of 4
  ($\sim2$, if the different IMFs are taken into account) with respect to the
 \citet{2006MNRAS.370..645B} semi-analytical model predictions. 
The general picture is, however, still controversial and further studies
are necessary to better constrain galaxy formation models. In fact, the Lyman-break technique can only be used 
to select actively star-forming
galaxies having sufficiently luminous UV  emission to allow their detection and color selection. 
Galaxies, and especially massive ones, could be faint or undetected in the
optical and near-infrared bands. Such sources could be either old systems with small quantities of dust and low levels of star-formation rate, 
or dust reddened starburst galaxies.  
If such $z>3.5$ objects really exist, they would be 
too faint to be studied spectroscopically with the instrumentation that is currently available, and the only possible approach that
can be used to constrain their number density and their properties
remains a photometric one, based on the Spectral Energy Distribution (SED) fitting analysis.

The most popular criterion used so far to select high-$z$ galaxies with red rest-frame colours is based on the redshifted 4000~\AA/Balmer break, i.e. the typical features of galaxies with evolved stellar populations. It was pioneered by \citet{2003ApJ...587L..79F} to pick out `distant red galaxies' (DRGs) at $z>2$ ($J_s-K_s>2.3$, AB system).
Recently it was extended by \citet{2007ApJ...654L.107B} to identify DRGs up to $z\sim3.7$ ($H-K>0.9$, AB system). By comparing the properties and the space density of DRGs at $z\sim3.7$ and $z\sim2.4$ in the same field, they found a stellar mass density of about a factor of 5 lower in the higher redshift bin.

In the recent years the availability of \spi\ data has also enabled  us to
search for galaxies with red rest-frame optical colors at high
redshifts, and to better constrain their stellar masses \citep{
  2006A&A...459..745F,2007MNRAS.376.1054D, 2007A&A...470...21R,
  2008ApJ...676..781W, 2005MNRAS.364..443E}. Some of the objects identified may be among the most massive stellar systems found at $z\sim4-5$.
In particular, \cite{2007A&A...470...21R} extracted a $\spi$+IRAC selected sample in the GOODS-South field, limited to galaxies undetected in the optical and close
to the detection limit in the $K$-band. Their criterion is complementary with respect to those adopted by previous studies \citep[i.e. the $K<23.5$ selection used
by][]{2007MNRAS.376.1054D} and is aimed at identifying high-$z$ massive galaxies that are missed by conventional selection techniques based on optical and near-IR observations. They found a potential population of optically obscured massive galaxies at $z~\geq~4$. 
In the same field \cite{2008ApJ...676..781W} found 11 $K$-selected $z\geq 5$ massive and evolved galaxy candidates. They selected these objects by applying color
criteria aimed at the identification of the 4000~\AA/Balmer break between the $K$ and 3.6~$\mu$m passbands, in combination with SED fitting. 

It should be noted that in both of these works roughly half of the high-$z$
massive candidates are detected at 24~$\mu$m. The 24~$\mu$m emission might be due to obscured AGN
activity at high redshift, but also from Polycyclic Aromatic Hydrocarbon (PAH) emission in dusty star forming
galaxies at lower redshifts. 
In fact, one has to take into account the  degeneracy between reddening and redshift while selecting high redshift sources based on photometry. 
Recent studies of sub-mm galaxies (SMGs) have shown that moderate 24~$\mu$m emission can arise from massive starburst galaxies at high redshift \citep[$z\geq3.5$, ][]{Daddietal2008,2008arXiv0806.3106G,2008arXiv0806.3791P}.
Hence one of the main difficulties in selecting bona fide high-$z$ galaxies through photometric redshifts
is to efficiently remove the contamination from dust-reddened galaxies at lower redshift.
As an example, one of the most debated cases is the galaxy HUDF-JD2 in the
Hubble Ultra Deep Field. \citet{2005ApJ...635..832M} suggested that such a galaxy is a massive ($M \simeq 6 \times 10^{11}$~M$_{\odot} $), low-reddened, old galaxy (with an age of $\sim 10^8$ yrs and a $z_{form} > 9$) at $z\simeq 6.5$.  However, other studies \citep[e.g.][]{2007MNRAS.376.1054D,2007A&A...470...21R,2007ApJ...665..257C} suggest that HUDF-JD2 is instead a lower redshift galaxy with very strong dust reddening \citep[$z\simeq 2.2$ and $A_V\simeq 3.8$,][]{2007MNRAS.376.1054D}.

The principal aim of this paper is to study the properties and space density
of massive galaxies at $z\geq3.5$ in the Great Observatory Origin Deep
Survey-North (GOODS-N). This survey is especially suitable for our purpose
because of the quality and depth of the available multi-wavelength data. 
In particular it has the advantage of a better coverage at sub-millimetric wavelengths with respect to GOODS-S. 
In addition, while similar searches have been carried out in the GOODS-South field, none has been performed so far in the northern hemisphere field of the GOODS Survey. 
We chose to select the sample in the IRAC channel 2 ($4.5~\mu$m) for two reasons. 
First, to pick out the most massive systems. At $z\geq 3.5$ the $4.5~\mu$m band samples the rest-frame near-IR emission, which is well correlated to the galaxy
stellar mass \citep{1998astro.ph.10031K}.
Second, the $4.5~\mu$m band selection allows us to have a more complete sample with respect to an optical or near-IR selection.
This is one of the main differences with the previous works: we attempted to construct a sample as complete as possible, including objects undetected in optical and near-IR bands, and hence missed by the optical or $K$ selection criteria used so far in the literature \citep[][]{2004ApJ...608..742D,2006A&A...459..745F,2007MNRAS.376.1054D,2008ApJ...676..781W,2007ApJ...654L.107B}.  As already mentioned, \cite{2007A&A...470...21R} also used an IRAC band (3.6~$\mu$m) selection, but their additional conditions about the lack of optical emission and the faint near-IR detection ($K>23$, AB system) make their sample incomplete by construction.

The structure of the paper is as follows. In Sect. \ref{data} we present the optical, near-IR and IR data available for the GOODS-N field. In Sect. \ref{sample
extraction} we describe the IRAC selection and the photometry of the sample. Sect. \ref{SED} describes the results from the SED fitting analysis with 
the measurement of photometric redshift and the stellar mass estimates. In Sect. \ref{zge3.5} we discuss the selection of $z\geq3.5$ galaxy candidates. In the same
section we discuss the detections of the sample in the MIPS-$24~\mu$m and in the sub-mm/mm. In Sect. \ref{complet} we study the effects of incompleteness and the
bias against low mass galaxies in our magnitude-limited sample. We give estimates for the comoving number density and stellar mass density for the full sample, and
for the sub-sample of MIPS-undetected galaxies. Sect. \ref{concl} presents the summary and conclusions.

Throughout this paper, we adopt the following cosmological parameters H$_0=70$km s$^{-1}$~Mpc$^{-1}$, $\Omega_{\Lambda}=0.7$ and $\Omega_{m}=0.3$, and all magnitudes are given in the AB system.

\section{The GOODS-N data.}\label{data}
This work is based on the analysis of the Great Observatory Origins Deep Survey - North (GOODS-N) data.
The GOODS-N is a multi-wavelength deep survey centered in the Hubble Deep Field
North \citep[HDFN  $ 12^{h}36^{m}55^{s}$,$ +62^{\circ}14'15''$][]{2004ApJ...600L..93G} with size of $ \sim 10'\times
16'$. 
 The field is part of the GOODS \spi\ Legacy Program (PI: M. Dickinson) and the
\hst\ (HST) Treasury Program (PI:  M. Giavalisco), divided in two fields (GOODS-N and
GOODS-S), from the northern and southern hemispheres \citep[see][]{2003mglh.conf..324D}.
The GOODS-N field observations span from the X-rays to the sub-mm wavelengths. In particular, for the analysis presented in this paper, we make use of the following data sets:
\begin{itemize}
\item $U$ band imaging, taken at the Kitt Peak National Observatory (KPNO) 4m telescope, from the  Hawaii Hubble Deep Field North (H-HDFN) survey datasets \citep{2004AJ....127..180C}.

\item Optical imaging from the \hst\ (HST) with the Advanced Camera for Survey (ACS) in the F435W ($B$), F606W ($V$), F775W ($i$), F850LP ($z$) pass-bands \citep{2004ApJ...600L..93G}.

\item Near-Infrared imaging, in the $J$, $H$ and $Ks$ bands, from the KPNO-4m telescope (M. Dickinson private communication).

\item Infrared observations from the \spi\ Space Telescope \citep{2004ApJS..154....1W} + Infra-Red Array Camera \citep[IRAC;][]{2004ApJS..154...10F}

\item $24~\mu$m observations from the \spi\ Space Telescope and Multi-band Imaging Photometer for \spi\ \citep[MIPS;][]{2004ApJS..154...25R}.
\end{itemize}
Specifics of the data are given below.

\subsection{KPNO 4m U-band imaging} 

The H-HDFN Survey is a deep multi-colour imaging survey of 0.2 $degrees^2$, centered in the Hubble Deep Field North. The data consist of deep images in the U, B, V,
R, I, and $z'$ bands. Out of these we used only the U-band images, due to the availability of HST/ACS imaging from the $B$ through $z$ bands.
The U-band observations were taken in March 2002, using the KPNO-4m telescope with the MOSAIC prime focus camera~\citep{1998SPIE.3355..577M}, covering a field of view of $36' \times 36'$ with a seeing of $1\farcs 26$. We analyzed the part overlapping with the GOODS-N field (an area of about 165~ $arcmin^2$ around the HDF-N). The U image reaches a 5$\sigma$ sensitivity limit of 27.1 mag.

\subsection{ACS+HST optical imaging}
The GOODS-N optical data from the Advanced Camera for Surveys (ACS), on-board
HST, are in the F435W (B), F606W (V), F775W (i), F850LP (z) pass-bands (v1.0
data products). The 5$\sigma$ point source sensitivity limits are 28.5, 28.8,
  28.1, 27.6 respectively, scaled from values reported in
  \citet{2004ApJ...600L..93G} to the full exposure time of ACS v1.0 images.
The HST/ACS high-resolution data is organized in 17 images, called
``sections''.  Each section is an image $8\,192~\times~8\,192$ pixels in size,
 drizzled from the native ACS pixel scale ($0 \farcs 05$/pixel)
  to an image subsampled at $0\farcs 03$/pixel. 
The huge size ($\sim40\,000\times 40\,000$ pixels) of the combined HST images in the field made source extraction an extremely intensive computing task. Therefore we
re-binned the final image using an $8\times 8$ pixel kernel, requiring flux conservation. The new re-binned mosaic obtained in this way has a size of $5\,120\times 5\,120$ pixels, and a pixel scale of $0\farcs 24$/pixel.

\subsection{KPNO 4m FLAMINGOS near-infrared imaging}\label{flami}
We complemented the public data with near-IR
imaging taken at the KPNO-4m telescope with the Florida Multi-object Imaging
Near-IR Grism Observational Spectrometer (FLAMINGOS; observations by Mark Dickinson et al.).
 The FLAMINGOS images in the J, H and Ks pass-bands, are mosaics of data
  taken at different pointings and orientations to cover the whole GOODS-N. The
  average exposure times per band in the main part of the GOODS-N area are
  20250~s, 15770~s and 35440~s at $J$-, $H$- and $Ks$, respectively.
The three images reach a $5~\sigma$ point source sensitivities of 24.03,
23.77, 23.81 mag. The pixel-scale is $0\farcs 316 $/pixel, and the seeing
$1\farcs 27$, $1\farcs 2$ and $1\farcs 2$, for the J, H, and Ks pass-bands, respectively.   

We performed some photometric checks before measuring source fluxes.
We verified the zero-points reliability of FLAMINGOS images, comparing the
bright and unsaturated star magnitudes with stars from the 2MASS catalogue.  

\subsection{Spitzer imaging at 3.6 to 24~$\mu$m}
The infrared imaging is from \spi\ Space Telescope +
IRAC between 3.6 and $8.0~\mu$m, and from \spi\
+ MIPS at $24~\mu$m.

The four IRAC channels, centered on $3.6~\mu$m, $4.5~\mu$m, $5.8~\mu$m,
$8.0~\mu$m, have formal 5~$\sigma$ limits for isolated point sources of 26.1, 25.5,
23.5, 23.4 respectively. The mean FWHM of a point source for the
  IRAC-bands are $1\farcs 66$, $1\farcs 72$, $1\farcs 88$ and $1\farcs 98$,
  for channels 1, 2, 3 and 4 respectively.  The initial image pixel-scale of
  $1\farcs 22$ is reduced to $0\farcs 6$/pixel, after the dithering and
  drizzling process.

The MIPS public dataset includes calibrated maps and a catalogue of $24~\mu$m sources with flux densities $S_{24}>80~\mu$Jy. 
The PSF was generated from isolated sources in the image, and renormalized
based on the aperture correction published in the MIPS data Handbook (v2.1,
section 3.7.5, table 3.12). Then an independent PSF algorithm was run to
extend the $24~\mu$m sample to fainter sources~\citep[see][]{2006MNRAS.371.1891R}. By  this procedure the detection threshold was extended to S$_{24}>35~\mu$Jy.
Since the $24~\mu$m information can help to better constrain the SFR
\citep{2001ApJ...556..562C} and the nature of high-redshift sources, we also
exploited a deeper $24~\mu$m catalogue, reaching a 5$\sigma$ limit of
$20~\mu$Jy \citep[Chary et al. 2008 in preparation, ][]{2007ApJ...670..156D}.

\section{Sample selection}\label{sample extraction}
In order to have an accurate measurement of the object fluxes in all
bands, we performed aperture photometry in
each band using the \sex\ software \citep{1996A&AS..117..393B}.
The initial catalogue was built by selecting the sources in the $4.5~\mu$m public image obtained with {\itshape Spitzer} Space Telescope + IRAC, down to a limiting
magnitude of $m_{4.5}=23.0$ ($2.3~\mu$Jy). Simulations indicate that a sample selected in this way is complete at the 80\%  level at $m_{4.5}\sim$ 23.0 
(M. Dickinson private communication),
where 20\% of the sources are lost due to blending.

The infrared selection allows us to search for the presence of massive
galaxies up to very high redshifts. In fact for galaxies at  3.5$<z<$7 the
rest-frame light emitted at $\lambda > 0.6~\mu$m is redshifted into the
IRAC 4.5$\mu$m band. Therefore, the  4.5$\mu$m  selection is directly sensitive to stellar mass, rather than to the ongoing star formation activity. 
From a technical point of view, the choice of selecting the sample at
$4.5~\mu$m represents the best compromise amongst the IRAC bands in terms of
image quality, blending problems, and sensitivity. 

In order to detect galaxies in the 4.5$\mu$m IRAC band and perform the photometry
we used \sex\ with the following set of parameters. The detection limit was
set at $\sim 1~\sigma$ above the sky background. A Gaussian filter was used to
improve the detection of faint extended objects. Due to the high crowding of IRAC images, and the large PSF, the minimum contrast parameter was set to a small value
of $5\times 10^{-9}$, to help with source deblending.

We measured fluxes in $4\farcs0$ diameter apertures, as this size allows us to minimize uncertainties in the photometry in the \spi\ +IRAC bands
(as also suggested by the SWIRE team{\footnote{http://data.spitzer.caltech.edu/popular/swire/20050603\_enhanced\_v1/}}).
In the 4.5~$\mu$m IRAC filter we applied an aperture correction of 0.25 mag that we independently computed from point-source objects, by measuring the total flux in $8\farcs 0$ diameter apertures.

The final IRAC selected catalogue contains $4\,142$ objects ($m_{4.5}< 23.0$).

\subsection{Multi-band Photometry}\label{sec:pcheck}
We searched for the counterparts of the $4\,142$ 
IRAC-selected objects in the other datasets.
We detected the sources and measured the photometry independently in each
dataset and associated counterparts to the 4.5~$\mu$m IRAC detected galaxies using 
a search radius of $1\farcs~0$.

For the U-band we measured fluxes $4\farcs 0$ (diameter) apertures. We applied an aperture-correction of 0.1 mag, computed by considering the total flux in $8\farcs 0$ diameter aperture. For undetected sources we used the $3~\sigma$ flux upper limit of 26.7 mag. This value was computed by measuring the standard deviation of the sky signal at random position over the field.

For the HST/ACS bands circular apertures with diameters of $2\farcs 0$ were used. The following  aperture-corrections for point-sources were found by measuring the total flux in $5\farcs 0$  diameter-aperture, for the B, V, i, z bands: 0.050, 0.040, 0.030, 0.045 mag, respectively.
For undetected objects we obtained $3~\sigma$ upper limits in the same way as described above for the U-band. We measured 
upper limits to be 26.3, 26.9, 26.3, 25.5, for the B, V, i and z bands respectively.

The near-IR magnitudes were measured in $4\farcs 0$ diameter apertures, and corrected to total (within $8 \farcs0$ diameter) by 0.1, 0.14, 0.15 mag for the J-, H- and Ks- band, respectively. 
The 3~$\sigma$ detection limits are 23.3, 23.0 and 23.1, for the J-, H- and Ks- band.

In the other IRAC channels we used the same aperture diameters as in
IRAC-$4.5~\mu$m ($4\farcs 0$), and the same approach to compute  aperture corrections.
The final IRAC magnitudes were obtained by
applying aperture corrections of 0.23, 0.25, 0.40, 0.44~mag for 3.6, 4.5,
5.8 and 8.0~$\mu$m bands, respectively.

\begin{figure}[htbp]
\includegraphics[width=\columnwidth]{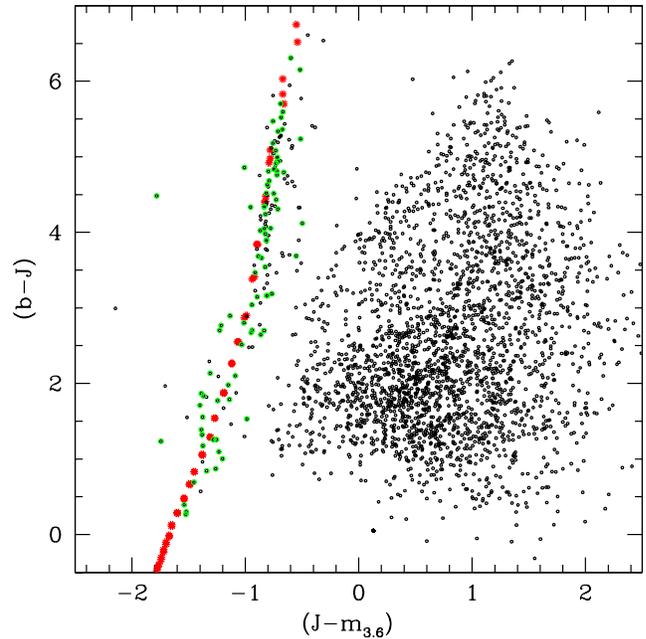} 
\caption{The color-color diagram $(b-J)$ $vs$ $(J-m_{3.6})$. The star
  sequence is clearly visible on the left side of the diagram.  We found a
  very good agreement between the observed colours (green circles) and the expected
  ones \citep[red asterisks from the ``empirically corrected'' templates of][]{1997A&AS..125..229L} for spectroscopically identified stars \citep[public
  database of the TKRS survey][]{2004AJ....127.3121W}.}
\label{fig:fig1}
\end{figure}

To check the reliability of our photometry we compared the colors of stars with those expected
based on the models of \citet{1997A&AS..125..229L} in color-color 
diagrams involving one band from
each one of the three datasets (ACS/HST optical, Near-Infrared/FLAMINGOS and
infrared/IRAC-\spi). The case of $(b-J)$ $vs$ $(J-m_{3.6})$ is shown in
Fig.~\ref{fig:fig1}.  The star sequence is clearly visible on the left
  side of the panel. As it appear from the figure we found a very good agreement
  between the observed colours (green circles) and the expected ones
  \citep[red asterisks from the ``corrected'' templates
    of][]{1997A&AS..125..229L} for spectroscopically identified stars \citep[public
  database of the TKRS survey][]{2004AJ....127.3121W}. 
  Therefore, thanks to the very good concordance between optical, near-IR and IR datasets we did not have to apply further
  photometric corrections to our data magnitudes.

The photometric errors given by \sex\ software are generally underestimated, because the software does not take into 
account the possible correlation between neighboring pixels that can result from image processing.
Hence we applied corrections to the photometric errors in each data-set.
For the four IRAC channels we added 0.1 mag quadratic to \sex\ errors.
 This is a conservative choice to account for uncertainties on the
  photometric zeropoints and on the aperture corrections, as also suggested in
  \citet{2006ApJ...652...85M}. In the other filters we measured the signal in
a number of random sky positions in the field, within the same aperture used
to perform galaxy photometry in each band. Comparing the standard deviation of
these measurements to the actual errors in the photometry of faint galaxies inferred by
SExtractor we derived scaling factors to apply to the formal errors reported in
the {\it SExtractor} catalogues.

\section{Spectral Energy Distribution (SED) fitting analysis}\label{SED}

\subsection{Photometric redshifts estimate}
To estimate the photometric redshifts of our sample of galaxies we used the
\hypz\ code \citep{2000A&A...363..476B}. This code compares the
multi-wavelength photometry of each source to a database of theoretical and/or
empirical templates at different redshifts. The best fit solution, for a given
set of templates, is derived through a standard $\chi^2$ minimization. 

For the determination of the photometric redshifts we excluded the $24~\mu$m band, because the
templates of stellar populations that we used do not include dust emission. For a
similar reason we decided to omit also the $8~\mu$m band, which includes
thermal dust and PAH (i.e. non-stellar) emission for the lower redshift galaxies and can be contaminated by star formation or AGN also at higher redshift \citep{2007ApJ...670..156D}. 
We used the template library of the observed SED constructed by \citet[][hereafter CWW]{1980ApJS...43..393C}, extended in the UV and IR regions by means of \citet{2003MNRAS.344.1000B} synthetic spectra, as explained in the \hypz\ User's manual\footnote{http://webast.ast.obs-mip.fr/hyperz/}. This set of templates includes four types of spectra with characteristics corresponding to the different local types of galaxies: Elliptical/Lenticular (E/S0), Spiral (Sbc and Scd) and Irregular (Im). The CWW library was complemented with a spectrum of a Young Star-Burst (SB) galaxy, with constant star formation rate (SFR) and age of 0.1 Gyr, from the \citet{2003MNRAS.344.1000B} models.

We took into account the dust extinction by applying the \cite{2000ApJ...533..682C} law. The extinction parameter ($A_V$) was allowed to vary from 0 to 0.8 in steps of $d(A_V)=0.1$. Only a small range of $A_V$ was allowed since the CWW templates include already intrinsically reddened SEDs. The permitted redshift range span the large interval of values of $0<z<10$, with a step of $d(z)=0.07$. 

We used the prescription of \citet{1995ApJ...441...18M} to represent the
average Lyman-$\alpha$ forest opacity as a function of wavelength and redshift. 
\subsection{Photometric redshift reliability}
We checked the reliability of our photometric redshifts for 836
objects of the GOODS-N field having  high quality spectroscopic redshifts. Unobscured AGN were excluded from the list by means of the X ray information from the publicly available X-ray catalogue of the Chandra Deep Field-North Survey\citet{2003AJ....126..539A}. The largest number 
of spectroscopic redshifts (721) are from the public database of the Team Keck Treasury Redshift Survey \citep[TKRS;][]{2004AJ....127.3121W}, and are accurate to 100~km$s^{-1}$. The median redshift of this survey is rather low ($z=0.65$). In order to increase the number of available objects with the highest spectroscopic redshifts, we complemented the TKRS database with the galaxies found by \cite{2006ApJ...653.1004R} in the redshift interval of $2.0\lesssim~z~\lesssim3.5$ (108). We also used some objects (5) up to $z\simeq~5$ from the NASA/IPAC  Extragalactic Database\footnote{http://nedwww.ipac.caltech.edu/} (NED).

\begin{figure}[htbp]
\includegraphics[width=\columnwidth]{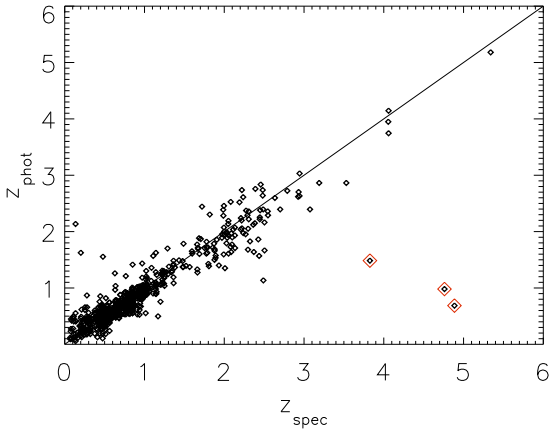} 
\caption{Comparison between photometric $z_{phot}$ and spectroscopic
  $z_{spec}$ redshifts for 836 objects of our sample. The
  $z_{spec}$-$z_{phot}$ relation has a combined mean offset of
  $(z_{spec}-z_{phot})/(1+z_{spec})=0.004$ with an rms scatter of
  $\sigma~[(z_{spec}-z_{phot})/(1+z_{spec})]= 0.09$. For the flagged objects
  (red open diamonds), appearing as catastrophic outliers we suggest the lower
  redshift solution, as detailed in the text.}
\label{fig:fig2}
\end{figure}

The comparison between photometric ($z_{phot}$) and spectroscopic
($z_{spec}$) redshifts is shown in Fig.\ref{fig:fig2}. After
clipping 5~$\sigma$ outliers we found a combined mean offset of
$(z_{spec}-z_{phot})/(1+z_{spec})=0.004$ and a rms scatter of
$\sigma~[(z_{spec}-z_{phot})/(1+z_{spec})]= 0.09$. This result  was obtained by
means of an iterative optimization procedure applied to the \hypz\ results. It consists
in repeating the SED fitting analysis for the same objects, fixing
the redshifts to the known spectroscopic values. For each filter the median
offset between empirical and ``theoretical'' magnitudes (the latter deriving
from the best fit SEDs) was obtained. The largest photometric offsets that we
found were for the $U$-, $z$- and $K$-bands (+0.2, +0.09, +0.14, respectively) and $\mid\Delta$mag$\mid\sim0.05$ for the others.
Adding these corrections to the observed magnitudes, the rms of the $\Delta~z/(1+z_{spec})$ quantity was slightly reduced.
The same offsets were finally applied to the whole sample for estimating the photometric redshifts.

\begin{figure}[htbp]
\includegraphics[width=\columnwidth]{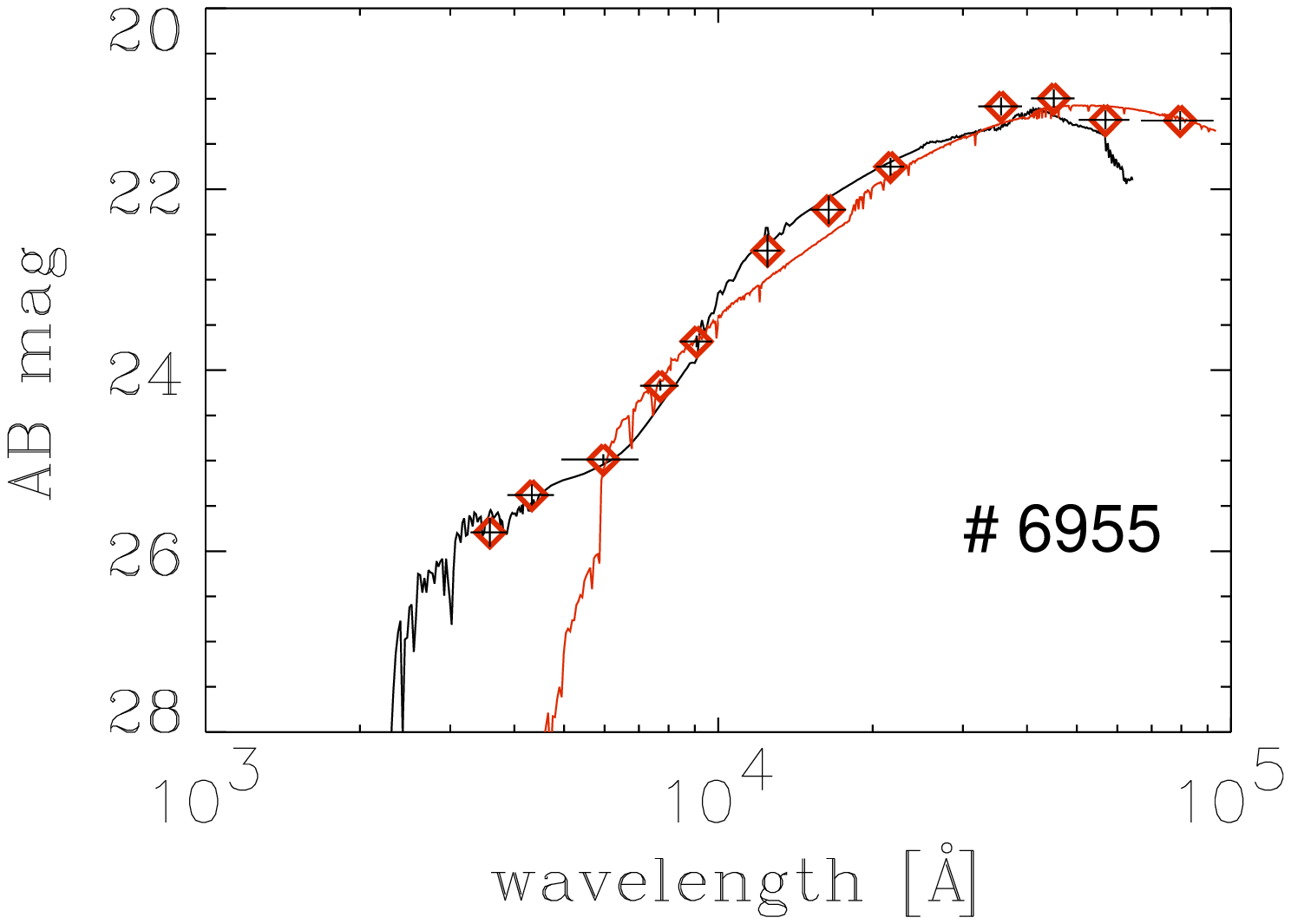} 
\includegraphics[width=\columnwidth]{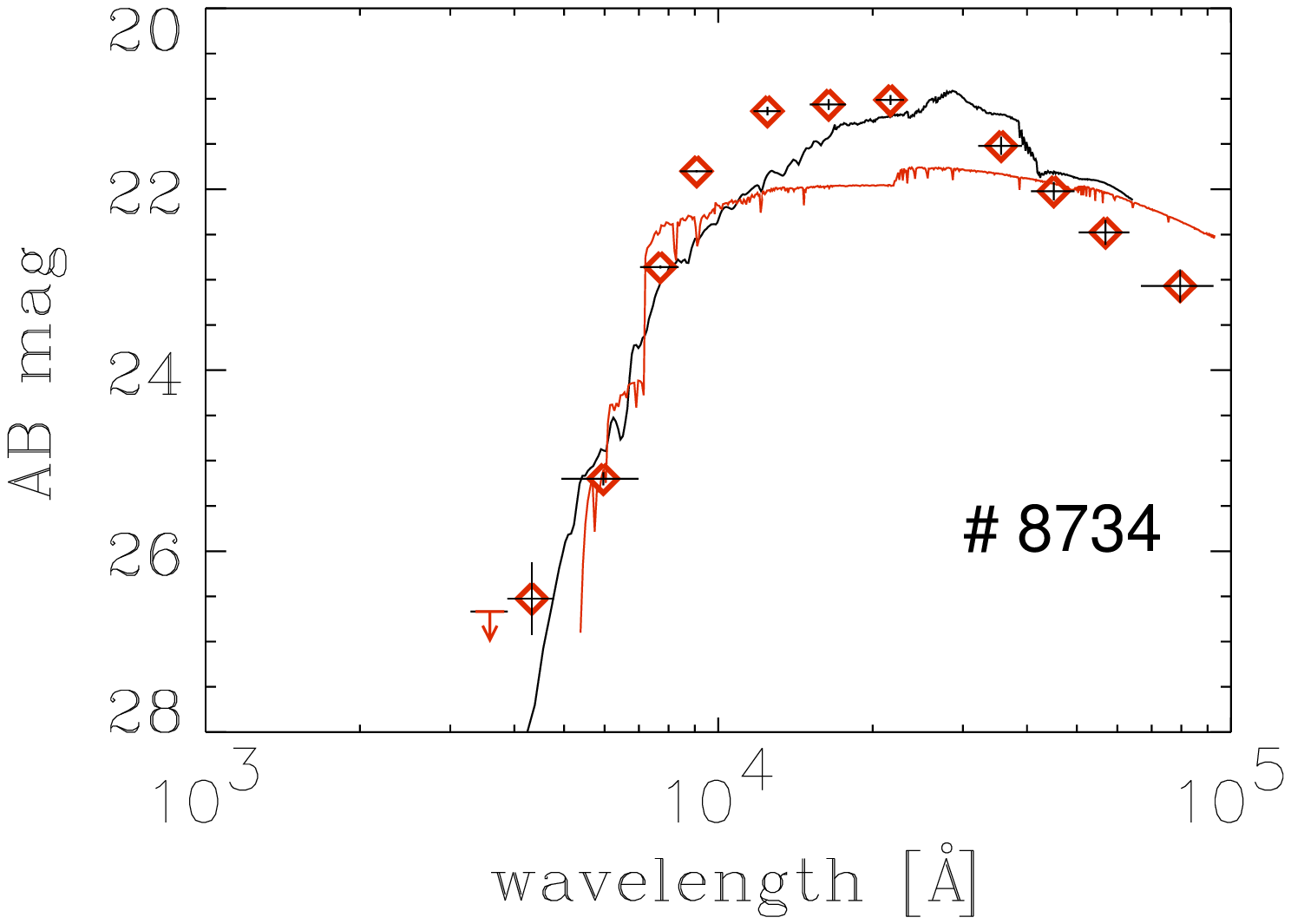} 
\includegraphics[width=\columnwidth]{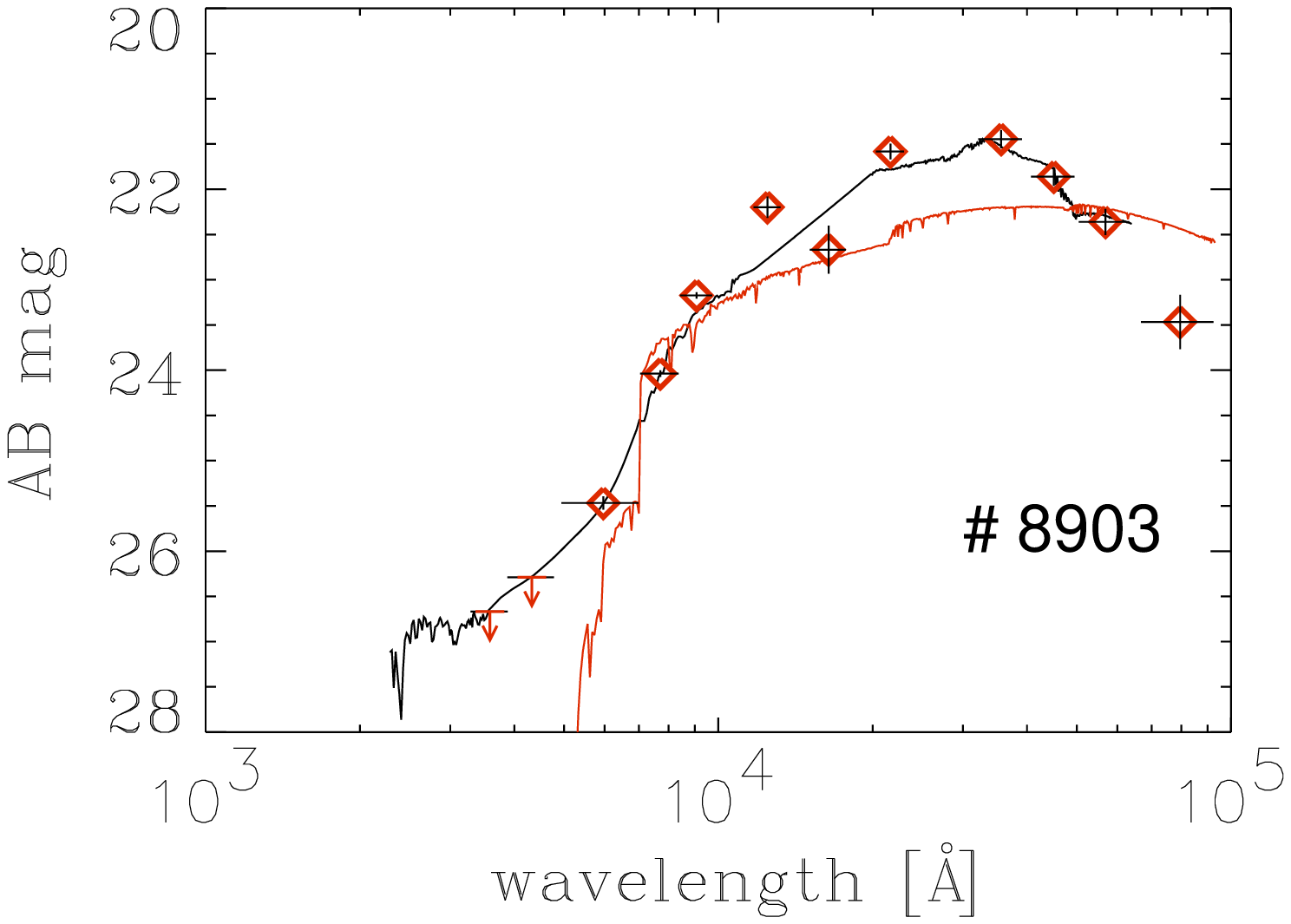} 
\caption{SEDs of the outliers in the $z_{spec}$-$z_{phot}$ space (see objects flagged in Fig.~\ref{fig:fig2}). Our best fits (black curves) produce  photometric
redshifts of $\sim$1.48, 0.68 and 0.98, respectively. We compare these results with the best fits (red curves) obtained by fixing the redshifts to the spectroscopic values
taken from the literature ($z=3.826$, $z=4.886$, and $z=4.762$; NED database). Obviously, the black curves fit the observed data better than the red ones. See also Tab.\ref{table:tab1}} 
\label{fig:fig3}
\end{figure}

\begin{figure}[htbp]
\begin{center}
\includegraphics[width=\columnwidth]{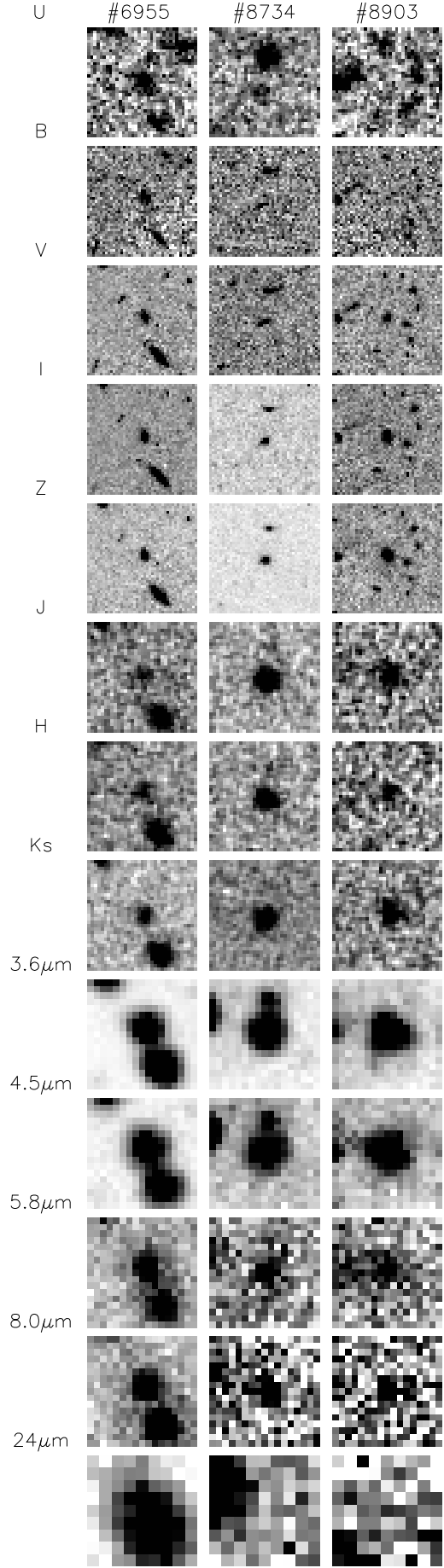} 
\caption{Multi-wavelength identification of the three outlier objects in the $z_{spec}$-$z_{phot}$ diagram. The cut-outs have a size of $10'' \times 10''$. North is up, East is at left.} 
\label{fig:fig4}
\end{center}
\end{figure}

 Since the current work is focused on galaxies at $z\geq 3.5$ we have
  carefully examined three outliers (\#6955,~\#8734 and~\#8903) in the
  photometric-spectroscopic redshift comparison with reported $z_{spec}\geq
  3.5$. These are noted with red diamonds in Fig.~\ref{fig:fig2}, and
  information about them is presented in Tab.~\ref{table:tab1}.    
 In Fig. \ref{fig:fig3} we show the photometric SEDs for these three
  objects and in Fig.~\ref{fig:fig4} we show cutout images from
  $U$-band through 24~$\mu$m.
 For the first two objects (\#6955 and \#8734)
  \citet{2001AJ....122..598D} report Keck spectra showing isolated
  emission lines with no continuum detection
  (their quality class 4; see Tab. 2 and 3 of that paper), which they identify
  as weak Ly-$\alpha$ emission. However
  object~\#6955 is detected at a significance $>3~\sigma$ limit in  
the $U$-band, ruling out the proposed $z = 3.826$.  More recent Keck/DEIMOS
  spectra from 2005 and 2006 (M.\ Dickinson \& D.\ Stern, private
  communication) detected an emission line at 9273\AA\ which is likely to be
  [OII]3727\AA\ at $z = 1.488$, consistent with our photometric redshift
  estimate for this object. Object \#8734 shows a red point source in the HST/ACS images, along with faint, bluer, extended emission
to the southeast. This is very likely to be a superposition of a  
cool galactic star with a faint galaxy. \citet{2004AJ....127.3137C} report
  that this is a star based on DEIMOS spectroscopy. The faint, extended
  component is certainly a galaxy, although it is clearly detected in the ACS
  $B$-band image, and is therefore unlikely to be at $z=4.887$ as reported by \citet{2001AJ....122..598D}. Overall, the photometry at all wavelengths redward
of the $V$-band is dominated by the point source component, and the  
photometric SED for this object Fig.~\ref{fig:fig3} clearly resembles that of a cool, late-type star (probably an M-dwarf).
Finally, for object \#8903, \citet{2001AJ....122..598D} report $z = 4.762$.  
However, ACS $V-i$ and
$i-z$ colors for this galaxy are not consistent with it being a $V$-- 
dropout Lyman break
galaxy, and fall within the locus of ordinary foreground galaxies.   
The ACS color images show a faint, red, early-type galaxy, and our photometric
  redshift estimate is $z_{phot} = 0.89$, very close to that proposed by \citet{2004AJ....127..180C}.  Although \citet{2004AJ....127.3137C} do not
report details about the spectral features that they observed, it is  
possible that faint [OII] emission at $z=0.88$ was mistaken for faint Ly$\alpha$ at $z = 4.76$.

\begin{table}
\begin{center}
\caption{Outliers of our $z_{phot}-z_{spec}$ comparison (Fig.~\ref{fig:fig2}). Their best fit SEDs are shown in Fig.~\ref{fig:fig3}. The coordinates and the spectroscopic redshifts are from NED database. The last column is the distance to the NED position (arcsec).}
\label{table:tab1}
 \begin{tabular}{cccccc}
id&RA&Dec&$z_{phot}$&$z_{spec}$&dist($''$)\\
\hline
\#6955&12:36:40.510&+62:13:34.90&1.48  &3.826& 0.27\\
\hline
\#8734&12:36:37.630&+62:14:53.70&0.68  &4.886& 0.16\\
\hline
\#8903&12:37:21.048&+62:15:01.73&0.98  &4.762& 0.41\\
\hline
\end{tabular}
\end{center}
\end{table}

\subsection{Galaxy mass estimate}\label{masses}

We used the code {\itshape Hyperzmass} to estimate the mass for each of our
sources. The {\itshape Hyperzmass} code works in a similar way to \hypz \citep[see][]{2007A&A...474..443P}. In this task the redshifts were fixed at the derived photometric values, or spectroscopic ones, when available.
The observed data points were fitted using both the synthetic spectral models
of \citet[][hereafter BC03]{2003MNRAS.344.1000B} and \citet[][hereafter
  MA05]{2005mmgf.conf..290M}. 
Both template sets are composed by models of stellar populations with Star Formation Rates (SFRs) exponentially decreasing with time ($\psi(t)=\tau^{-1}exp^{-t/\tau}$). In our analysis we considered models with different values of the time-scale $\tau$ (in unit of Gyr) : 0.1, 0.3, 1, 2, 3, 5, 10, 15, 30 and $\infty$. The last model corresponds to a stellar population with a constant SFR.

For the BC03 models we used the \citet{2003PASP..115..763C} Initial Mass
Function (IMF). Instead, the MA05 models adopt the
\citet{2001MNRAS.322..231K} IMF, which is relatively similar to the Chabrier IMF.
To compare the different results we computed a systematic correction to convert 
the masses obtained with Kroupa IMF to Chabrier IMF ($\log M_{Chabrier}=\log M_{Kroupa} - 0.04$).

In the SED fitting analysis the extinction parameter, $A_{\rm V}$,  was allowed
to span the range of values $0~\leq~A_V~\leq~6$, with a step of 0.1. 
This large reddening range is required given that synthetic stellar population models are not intrinsically reddened as the observed ones. 
We used $A_V=6$ as an upper limit for the reddening, since there is no evidence of objects with $A_V$ higher than $\sim 5$. 
 The galaxy age
was left as a free parameter, spanning the range from $5\times^6$~yr to $11.7\times10^9$~yr (and always
requiring ages less than the age of the Universe at each redshift). 
To minimize the number of free parameters we limited the fits to solar metallicity.

\section{Massive galaxies at $z~\ge 3.5$}\label{zge3.5}
As previously discussed,  evidence for populations of massive
galaxies at $z>3$ has been found by several studies. However, results are not yet
conclusive since the number of objects is low, and frequently biased
against objects that are faint or undetected in optical and near-IR
bands. Moreover for most of these objects no spectroscopic redshift information is
available. Also, the possible contamination by lower redshift interlopers is a source of uncertainty.
To shed some light on the nature and the number density of the high redshift population of massive galaxies we selected a final sample of $z_{phot}\ge 3.5$ 
candidates, by applying the following selection criteria (summarized in Table \ref{table:tab2}). Among the initial $4\,142$  
IRAC-4.5 $\mu$m selected galaxies we found 190 
galaxies with photometric redshifts of $z_{phot}\ge 3.5$.

 All such objects were carefully inspected for possible blending issues.
	One of the main problems in using \spi\ data in crowded fields is, in fact, the
  blending between nearby sources. Neighboring objects that can be
  well separated in optical and near-IR images, might happen to be blended in the IRAC
  bands, due to the limited spatial resolution of the \spi\ telescope. Hence,
  in order to build a bona-fide sample of high-$z$ galaxies we had to take into account that a
  fraction of the IRAC-selected objects could be affected by blending. We
  decided to visually inspect all the $z\geq3.5$ objects, and we excluded from the sample any
  sources that  the \sex\ software was not able to
  deblend. To identify them we visually inspected examined the \sex\ ``aperture image'' in
  each filter.
   This is a `check-image' given as output by the software, where the circular 
  apertures are shown for all the objects detected above the threshold level.
  We classified an object as blended if the IRAC position did not correspond to
  a single source in the higher resolution optical and/or near-IR images
  (e.g. objects that are well separated in the optical or near-IR images may
  be blended in the IRAC images due to their poorer angular resolution). About the 25\% 
  of the pre-selected $z_{phot}\ge 3.5$ objects were flagged as blended and therefore
  discarded.   

By matching our catalogue with the publicly available X-ray catalogue of the
Chandra Deep Field-North Survey\footnotemark[3], we found 14 objects that were
detected in the hard X-ray band, and one with soft X-ray flux. 
We excluded these 15 sources from the final sample since X--ray emission indicates the presence of AGNs in those galaxies, 
that can alter the observed flux and induce an overestimate of the stellar masses.

 Finally we introduced a further criterion, based on IRAC colours. 
We considered as genuine $z\geq 3.5$ candidates only those sources showing
an SEDs peak at the 8.0~$\mu$m observed frame (1-2~$\mu$m rest-frame for $z\geq
3.5$ galaxies) in the AB scale.
The reason is that the near-infrared emission of all but extremely young
galaxies show a distinct peak at 1.6~$\mu$m in the rest-frame, that is
redshifted in the 8.0~$\mu$m IRAC band for $z\geq 3.5$ sources.
This peak corresponds to the combination
of the Planck spectral peak of cool stars and the effects of a minimum in the
$H^{-}$ opacity in the stellar atmospheres. 
It has been used as photometric redshift indicator in several previous studies \citep[e.g.][]{2002AJ....124.3050S, 2007A&A...467..565B}.
This criterion was applied because at this step we expect that our
  high-$z$ sample will be contaminated by a substantial fraction of lower-redshift
  galaxies,  because of the intrinsic uncertainty of the photometric redshift
  estimate.  The degeneracy between reddening
  and redshift could introduce lower-redshift contaminants in the final
  high-$z$ sample (i.e. dust reddened galaxies at lower redshift can be 
  mistaken for higher-$z$ old stellar populations).  
  Given that galaxies at $z\geq 3.5$ represent only a 
  minority ($\sim$ 4\%) of the galaxy sample selected at $24\mu$m,  while most of 
  the objects are at $z < 2$, from a statistical point of view we could 
  expect a more substantial contamination by lower-$z$ sources entering our $z>3.5$ redshift selection,
  largely overbalancing the number of genuine $z>3.5$ galaxies lost as due to errors in their photometric redshifts.
  In addition, as already pointed out, the 8~$\mu$m data-point was excluded from the
  SED fitting analysis because for objects at $z<1.8$ it could sample the PAH
  emission lines for dusty galaxies, and the stellar population models that we
  used do not include dust emission. In this way, by excluding the 8~$\mu$m
  band from the fitting procedure, we may fail to distinguish galaxies at
  $z\sim2-3$ where the 1.6~$\mu$m peak lies in the 5.8~$\mu$m band from those at $z>3.5$.
  Requiring the SED to rise from 5.8 to 8~$\mu$m can help to exclude these
  possible lower-redshift contaminants.

The parameterization of the colour criterion we used is the following:

\begin{eqnarray}
&(m_{8.0} \leq  m_{5.8}+0.1)  \cap (m_{8.0} \leq  m_{4.5}+0.1) \cap &\label{crit}\\
& \cap (m_{8.0} \leq  m_{3.6}+0.1) &\nonumber 
\end{eqnarray}
the 0.1~mag terms are introduced to account for the median uncertainty in the IRAC photometry of $\sim 0.1$.

\begin{figure}[htbp]
\includegraphics[width=\columnwidth]{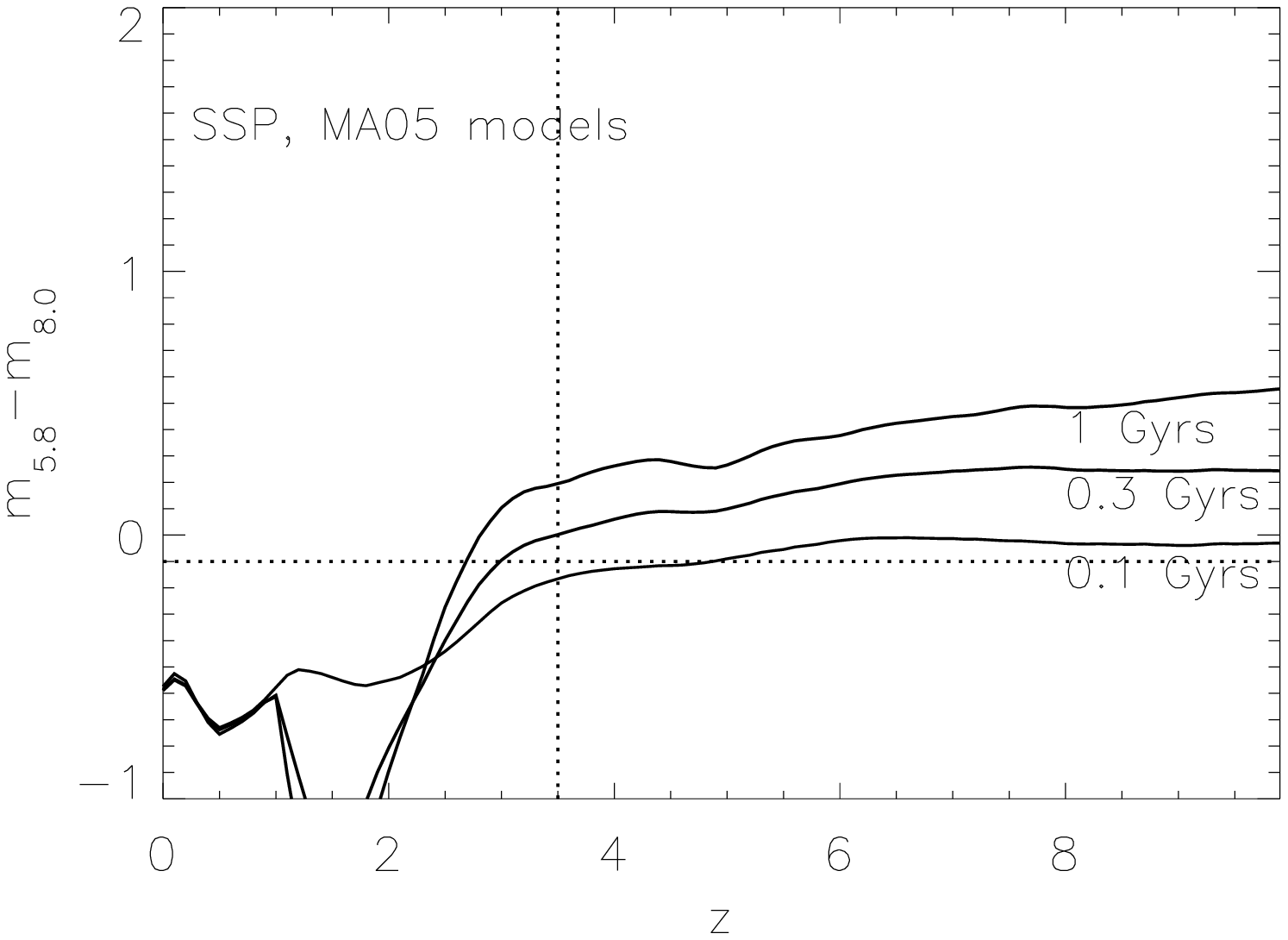}
\includegraphics[width=\columnwidth]{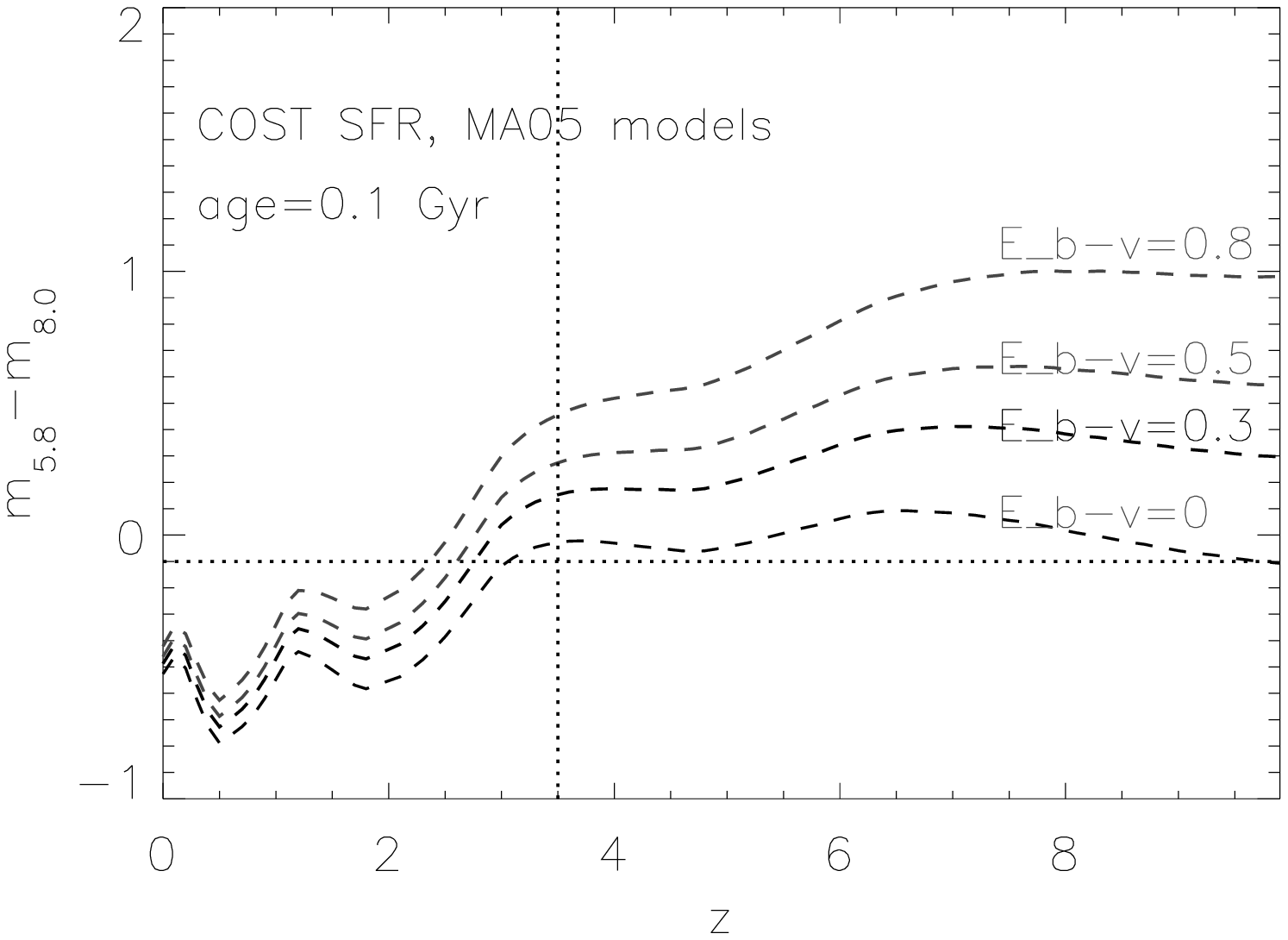}
\caption{The two panels show the $m_{5.8}-m_{8.0}$ colour in function of redshift for the Maraston stellar population models \citep{2006ApJ...652...85M}. Top panel: Single-burst Simple Stellar Populations (SSPs) with three different ages (0.1, 0.5 and 1~Gyr). Bottom panel: 0.1~Gyr stellar populations with constant star formation rate (SFR) and four different extinction values ($E_{B-V}$~=0.0, 0.3, 0.5, 0.8)}.
\label{fig:fig5}
\end{figure}

In the two panels of Fig. \ref{fig:fig5} we show the expected $m_{5.8}-m_{8.0}$ colour as a
function of redshift for different stellar population models. The top panel shows Simple Stellar Populations (SSPs) with three different ages from 
0.1 to 1 Gyr, and the bottom panel shows a constant star formation rate model with fixed age of 0.1~Gyr and 
reddening  from $E_{B-V}$~=0 to 0.8. The templates shown are from the Maraston stellar population models \citep{2005mmgf.conf..290M}.
 These models, however, are not supposed to be used for $\lambda >
  2.5~\mu$m where they display a strong, unphysical, discontinuity. This discontinuity affects the IRAC colours at $0< z < 2.7$.
Hence we tested the reliability of this colour
  criterion also with BC03 templates, that are not affected by this problem.
We verified that the expected IR colours for BC03 stellar population
  models at $z<2.7$ are equally bluer than the colour cuts defined by the equation \ref{crit}.

From the upper panel of Fig.~\ref{fig:fig5} it is clear that the
  1.6~$\mu$m peak criterion may lead to exclude some blue and unreddened (E$_{B-V}=0$) galaxy at $z>3.5$ with age
  $<$0.1~Gyr. This is because galaxies dominated by young stellar populations
  do not show the 1.6~$\mu$m peak in their SEDs~\citep[see][]{2002AJ....124.3050S}. In fact, due to this criterion we excluded from the
  final sample a spectroscopic
  confirmed LBG~\citep[object \#5455, from][with $z_{spec}=5.34$ and $z_{phot}=5.18$]{1998AJ....116.2617S},
  for which the best fit result predicted age $\sim 7$~Myr, and mass $\sim
  10^{10}$~M$_{\odot}$. Nonetheless we decided to apply 
  this criterion as a conservative way to obtain
  a reliable sample of the most massive galaxies at $z\geq 3.5$. The objects missed in this way are very few (see next section), 
  and are also the least
  massive ones. Their inclusion would not strongly affect our
  estimate of the comoving stellar mass density (for more details see Sect.~\ref{dropout}, \ref{incompleteness} and \ref{tot_mass_dens}).

\begin{table}[htbp]
\begin{center}
\caption{Sample selection criteria. In the left and right columns the subsequent steps to select the final sample and the corresponding number of objects are shown, respectively.}
\label{table:tab2}
\begin{tabular}{|c|c|}
\hline
Selection Criteria & Number of Objects\\
\hline
$m_{4.5}~\le 23$& 4142\\ 
\hline
$z_{phot}~\ge3.5$ & 140\\ 
\hline
No hard X-ray emission & 125\\
\hline
IRAC/channel 4 peak & 53\\
\hline 
\end{tabular}
\end{center}
\end{table}

 This criterion is very efficient in excluding SSPs galaxies with redshifts
 lower than $\sim2.5$. On the other hand, 
we should point out that it cannot completely remove the contamination 
  by interlopers at $2.5<z<3.5$, as SSPs models with ages 
  $>1$~Gyr and $3<z<3.5$, and dusty star forming galaxies at $2.5< z < 3.5$ can still satisfy this criterion
  (see the bottom panel of the same figure). 

Finally it should be noted that the adopted
colour criterion would be met  also by the so-called `IRAC power-law sources'. 
These objects are generally selected by requiring a rising SED over the four 
IRAC bands, and they are believed to be generally $z<2.5$ AGNs with a hot dust component swamping the 
stellar emission \citep{2006ApJ...640..167A,2007ApJ...660..167D}. 
Since \citet{2007ApJ...660..167D} selected a sample of (62) IRAC
  power-law galaxies in the Chandra Deep Field Nord (CDFN), whose 27 are in the GOODS-N field, we investigate the overlap with their sample.
We found 10 sources in our 4.5~$\mu$m IRAC-selected ($m_{4.5}\leq23$) sample (of 4142 objects) in common with the `IRAC power-law' sample.
All these sources but one have the infrared part of the SED peaked at
8.0~$\mu$m, but most of them had been excluded from our $z\geq3.5$ final sample
because of their lower photometric redshifts. Only one object of the \citet{2007ApJ...660..167D} sample fulfill all of our selection criteria.
This is object \#14793 in our sample (see Tab.~\ref{table:tab4} and Fig.\ref{fig:fig14}), corresponding to the object
CDFN:[DRP2007]19080 in Donley et al. This galaxy has no X-ray emission
($<4.41\times10^{-16}$erg s$^{-1}$ cm$^{-2}$) and a rather high $24~\mu$m
flux (111~$\pm3.5 \mu$Jy).  
The high redshift solution is still consistent with these properties. 

As result we obtained a final sample of 53 galaxies ($\sim 43\%$ of the pre-selected $z\geq 3.5$ sample), that we consider to be valid candidates at $z~\geq 3.5$. 
About the 12\% of the 125 pre-selected high-$z$ candidates have been discarded
due to the low S/N ratio in the 8.0~$\mu$m-band, but we cannot exclude that
some of them could be at high redshift. The other 45\% is likely mainly
composed either by younger systems (age~$<0.1$~Gyr) at the same redshift, or by
lower redshift $z<3.5$ contaminants. We will show in the following that the former are only a minority (see Sect.~\ref{dropout}). 
In Table~\ref{table:tab6}, the multi-band photometry of the final sample is shown.

As a further test on the quality of this sample we checked how many candidates
  satisfy the empirical blending criterion tested on extensive simulations by the GOODS
  team and used by \citet{2007ApJ...670..156D} and
  Dickinson et al. (2008 in prep.).
  This is based on 
  the measurements of the angular separation ($\Delta\theta$) between IRAC 
  positions and the $K$-band counterparts (we used also optical coordinates for $K$-undetected sources).   
  According this criterion, if a galaxy has $\Delta\theta >0\farcs5$, it is
  likely contaminated by blending in the IRAC bands. 
  However we should point out that this criterion proved to be successful on
  objects detected at the 5-10~$\sigma$ in both of the $K$ and IRAC
  bands. For sources with fainter optical and near-IR 
  detections the shift of the centroid position throughout the different bands 
  could be larger due to higher noise. Hence a larger $\Delta\theta$ could derive from 
  coordinate fluctuations rather than from blending contamination.
  Moreover due to the extremely red colours of our candidate high-$z$
  galaxies, this criterion is not extensively applicable to our full sample
  being half of it both optical and near-IR undetected.  
  Out of the sources with at least a reliable optical or near-IR 
  detection ($>$3-5~$\sigma$) we verified that the majority (18/25) did pass this
  criterion.
  On the other hand we found 7 galaxies (\#1098, \#2172, \#2796, \#13857, \#15268, \#15541,\#15761) with 
  $\Delta\theta >0\farcs5$. However for these objects the lack of visual evidence of blending led us to conclude
  that the quite large $\Delta\theta$ separation, is a S/N issue. Moreover
  for two of these galaxies the confirmed spectroscopic redshifts (\#13857,
  \#15541, see Sect.~\ref{scuba}) prove that the measured IRAC photometry
  leads to reliable SED fitting results.
 
\subsection{Comparison between the BC03, MA05 models}\label{sec:templates}
The SED fitting analysis with BC03 and MA05 templates gives somewhat discordant
 results for both galaxy masses and ages estimate. This is an issue recently
 debated by several authors \citep[e.g.][]{2006ApJ...652...85M,
 2007ApJ...657L...5K, 2007astro.ph..3052B},  the major source of this discrepancy being the different treatment of
 the Thermally Pulsing Asymptotic Giant Branch (TP-AGB) phase of stellar
 evolution in the population models. 

In the MA05 models stars in the TP-AGB phase contribute a dominant source of
bolometric and near-infrared light for stellar populations in the age range
0.2 to 2~Gyr, much larger than is found in the BC03 models. For those models the TP-AGB phase has been calibrated with local stellar populations. Another difference is in the treatment of convective overshooting. In their work \cite{2006ApJ...652...85M} tested both the BC03 and MA05 models on a sample of high-$z$ galaxies, with reliable spectroscopic redshifts ($1.4\lesssim~z~\lesssim2.5$) in the Hubble Ultra Deep Field (HUDF). They found that the results of MA05 models imply younger ages by a factors up to 6 and lower stellar masses, by $60\%$ on average, with respect to BC03 (and in general to the others models of population synthesis). The overestimate of BC03 galaxy stellar masses, produced by the lack of the TP-AGB contribution to the integrated luminosity, become considerable for evolved stellar populations with ages between 0.5 and 2~Gyr.

\begin{figure}[htbp]
\includegraphics[width=\columnwidth]{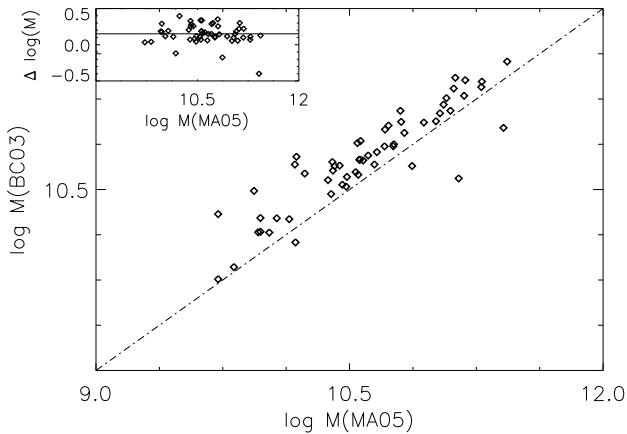}
\caption{Comparison between estimated masses by means of different template
  libraries for our final sample of 53 objects: log M(BC03) vs log M(MA05). It
  is visible that BC03 models tend to overestimate galaxy masses with respect
  to the MA05 ones. The inserted panel shows the distribution of the log
  M(BC03)-log M(MA05) relation  as a function of the galaxy mass.}
\label{fig:fig6}
\end{figure}

In Fig. \ref{fig:fig6} we compare  the stellar  masses estimated for the $z\geq 3.5$ galaxies with the two different template libraries (BC03 vs MA05). It appears
that the BC03 models tend to overestimate galaxy masses of our sample with respect to the MA05 ones, by 0.15~dex on average
(see the smaller panel on the top of Fig.~\ref{fig:fig6}), in good agreement with the previous studies \citep[e.g.][]{2006ApJ...652...85M, 2007ApJ...655...51W, 2008A&A...482...21C, 2004ApJS..154....1W}.
In the rest of this paper we will adopt the MA05 stellar templates to compute the galaxy stellar masses and the comoving stellar mass density. 
However, in some cases we also use the BC03 models results for comparisons with the literature.

\subsection{Comparison with Lyman break selection techniques}\label{dropout}
One of the most popular method used to identify high-$z$ galaxies is the `dropout' technique, a colour selection based on the red-shifted 912~\AA\ Lyman-break caused by neutral hydrogen in the galaxy SEDs. This technique was introduced by \citet{1990ApJ...357L...9G} and \citet{1992AJ....104..941S} to select Lyman Break Galaxies (LBG) at $z\simeq3$ (u-band dropout). It has been largely used also to identify galaxies in the redshift range $z\sim4-6$, as b-, v- or i-dropout \citep{1999ApJ...519....1S, 1998hdf..symp..219D, 2003ApJ...592..728S, 2004ApJ...600L..93G}. However this colour criterion is strongly biased against red massive galaxies at high redshift, that are faint or totally undetected in optical and near-IR bands. 

 To obtain a quantitative estimate of the fraction of red objects missed by
  combining optical magnitude limited selection and dropout techniques, we
  matched our final sample of IRAC-selected high redshift galaxies (53
  objects) with a sample of $B$- and $V$-dropout LBGs in the GOODS-N field from \citet{2004ApJ...600L.103G}. We found that only 7 
of our massive high-$z$ candidates are
  included in the $B$-dropout sample, and none in the $V$-dropout sample. This
  means that the remaining 
IR bright (m$_{4.5}<23$) $z\geq 3.5$ candidates have been missed by those selection criteria, due to their faint emission at optical and near-IR wavelengths (most of them having $z_{850}>27$).  

While the vast majority of our candidates are missed by the dropout selection
technique, there could also be a significant population of massive
galaxies with $m_{4.5}<23$  that are identified through the dropout technique  
and that have been missed by our
selection criteria.
We tested for the presence of such galaxies using again
the sample of \citet{2004ApJ...600L.103G}.

\citet{2004ApJ...600L.103G} listed 684 LBGs candidates ($B$-dropout +
$V$-dropout). Only 7 of those are also part of our sample of $z>3.5$ candidates. 
Only 32 of the Giavalisco et al. galaxies have $m_{4.5}<23$
(15 $B$-dropout and 17 $V$-dropout). Of these, 20 were excluded from
our sample because of their photometric redshifts $z<3.5$ (8 $B$-dropout and 12 $V$-dropout). 
One more object ($V$-dropout) was excluded because of the presence of X-ray emission
and 4 objects ($V$-dropouts) were excluded because of the lack of an IRAC-$8.0~\mu$m peak.
One of these 4 $V$-dropout galaxies
has a confirmed spectroscopic redshift of $z=5.2$ (see also Sect.~\ref{zge3.5}). 
It is reasonable to suppose that all the 4 $V$-dropouts galaxies that we excluded 
are actually genuine $z>3.5$ galaxies with young ages, lacking a distinct peak 
in the IRAC-$8.0~\mu$m band.
Even taking them into account would increase our sample of $z>3.5$ galaxy candidates by only 7\%
and would reduce to 81\% the fraction of $z>3.5$ massive galaxies that are lost by the UV dropout selection
techniques.

 It is also interesting to make a comparison with the Lyman-break galaxies
 samples selected by \citet{2007MNRAS.374..910E} and
 \citet{2007ApJ...659...84S} in the GOODS-South field. The former work reports
 6 galaxies at $z\sim 6$ with masses in the range $1-2.4\times 10^{10}$
 M$_{\odot}$, and 4.5~$\mu$m magnitude m$_{4.5}>$23. Since the authors
 used Salpeter IMF and BC03 models to derive galaxy SEDs, we have to consider
 that galaxy masses for these 6 objects are overestimated by a factor of
 $\sim2.4$ \citep[2.4=$1.4\times 1.7$, being 1.4 and 1.7 the conversion factors from
 BC03 to MA05 and from Salpeter to Chabrier IMF,
 respectively][see also Sect.~\ref{sec:templates} and Sect.~\ref{tot_mass_dens}]{2007A&A...474..443P,2008A&A...482...21C}. When rescaled,
 masses span the range $0.4-1\times 10^{10}$ M$_{\odot}$, with only one galaxy
 with M$\sim10^{10}$~M$_{\odot}$. Lyman
Break Galaxies similar to the ones found by \citet{2007MNRAS.374..910E} 
would not be included in our sample, since these are fainter than our
magnitude cut m$_{4.5}=23$. This issue is in agreement with the lower
masses found for these LBGs with respect to our high-$z$ sample.    

The latter work also reports 6 galaxies, spectroscopically confirmed at
$z\sim4.4-5.6$. We used the same conversion factor to rescale galaxy masses
from BC03 and Salpeter IMF and we found that only three galaxies of the
\citet{2007ApJ...659...84S} sample have masses
$>10^{10}$~M$_{\odot}$. Moreover only two out of these three sources have m$_{4.5}<23$.    

Hence only 4 galaxies out of the 12 LBGs at $4.4\leq z\leq 6$ from the two works, fall within our mass bracket (M$\sim10^{10}-10^{11}$~M$_{\odot}$), and only
two would meet our magnitude selection ($m_{4.5}<23$). 
 The two IRAC brightest and most massive sources from
\citet{2007ApJ...659...84S} have ages $\sim 150$~Myr, if rescaled by a
factor of 6 to account for the overestimate produced by BC03 models
\citep{2006ApJ...652...85M}, and $E_{B-V}\sim0.0$. 
We can suppose that they are similar to the youngest unreddened population
excluded from our final sample by the 8~$\mu$m peak criterion (see
Sect.~\ref{zge3.5} and Fig.~\ref{fig:fig5}, top panel), such as the 4 V-band
dropout galaxies of the sample of \citet{2004ApJ...600L..93G}.
The other two galaxies with masses $\sim 10^{10}$~M$_{\odot}$ and fainter than 
our IRAC magnitude selection would be even younger (ages $<100$~Mry). 

In conclusion, it seems that we are not missing a significant population
of sources with our selection criteria. A few blue star forming galaxies 
with young ages are indeed excluded but these objects have generally 
low stellar masses that would not meet our stellar mass completeness limits
(M$_{\star}\geq 5\times 10^{10}$~M$_{\odot}$; see Sect.~\ref{incompleteness} and \ref{tot_mass_dens}).

\subsection{24~$\mu$m emission}\label{24mic}

We used observations at  $24~\mu$m with \spi\ +MIPS to give some 
constraints on the ``activity'' of our high redshift galaxy candidates. 
Detection of $24~\mu$m flux for $z>3.5$ galaxy candidates could be explained 
in terms of radiation coming from star formation activity lately 
re-processed and re-emitted by the dust (emission lines from PAH
molecules), or with the presence of an highly obscured AGN in
relatively quiescent galaxies. We divided our sample of high redshift candidates
in two sub-samples: MIPS-detected (MIPS-d) and MIPS-undetected (MIPS-u) samples
and discuss them separately in the following.

\subsubsection{The MIPS-u sample}

21 objects in our sample are undetected in MIPS down to the limit of $20~\mu$Jy. The best fit Spectral Energy Distributions (SEDs) of the 
MIPS-u sub-sample is presented on the left  panels of Fig.~\ref{fig:fig13}. The right panels of the same figure show the $\chi^2_{\nu}$ distribution as a function of
redshift for each of the sources. We also report the most 
probable photometric redshift solution ($z_{phot}$). The best fitting parameters and the estimated galaxy masses are presented in Table \ref{table:tab3}. 
Most of the MIPS-u redshifts are spread
around a median value of $z\sim 4$, in a redshift range $z$=3.5--5.
We find that the best fitting SEDs give $\chi^2_{\nu}< 2$ for 20 out of 21 objects and no degeneracy on the redshift solution for the majority of the sources.
On the contrary, we find also lower redshift solutions within the 90\% 
confidence intervals of $\chi^2_{\nu}$ distribution for 5 objects (\# 4008, \# 9561, \# 12327 and \#14130), 
caused mainly by the redshift-reddening degeneracy (see Tab.~\ref{table:tab3} and Fig.~\ref{fig:fig13}).

\begin{table*}[htbp]
\begin{center}
\renewcommand{\footnoterule}{}
\begin{tabular}{llllllllllll}
	\hline\hline
\#ID & $z_{phot}$ & $\chi^2_{\nu}$ & $P_{\chi^2_{\nu}}$ & $A_{V}$ & $z_{inf-90\%}$ & $z_{sup-90\%}$ & $z_{2-phot}$ &  $P_{\chi^2_{\nu},2}$ & log M(MA05)\\
	\hline\hline

\#522  &  3.928  &  0.410  &  94.28	& 0.80  & 3.060  &  5.230  & 10.060  &  43.52  &  9.960 \\
\#1098  &  4.369  &  1.465  &  14.55   & 0.20  &  4.250 &   4.460  &  4.922  &  0.04  & 10.143 \\
\#2574  &  3.620  &  0.307  &  97.98   & 0.40  &  2.570 &   4.110  &  6.098  &  0.47  & 10.581 \\
\#3081  &  3.543  &  0.466  &  91.26   & 0.70  &  3.340 &   3.760  &  0.456  &  0.00  &  9.816 \\
\#3302  &  4.292  &  0.306  &  98.01   & 0.30  &  3.620 &   4.600  & 10.060  &  0.00  & 10.565 \\
\#4008  &  9.024  &  0.201  &  99.63   & 0.40  &  2.150 &  10.060  &  3.753  &  99.57  &10.069$^{\mathrm{a}} $\\
\#5073  &  4.481  &  0.738  &  68.90   & 0.20  &  4.250 &   4.670  &  3.417  &  20.62  & 11.073 \\
\#5460  &  4.236  &  0.854  &  57.65   & 0.00  &  3.480 &   4.390  &  3.683  &  50.87  & 10.402 \\
\#6876  &  3.620  &  0.683  &  74.09   & 0.20  &  2.220 &   4.040  &  3.172  &  72.67  & 10.485 \\
\#7042  &  3.774  &  0.646  &  77.54   & 0.50  &  3.340 &   3.900  &  3.529  &  75.10  &  9.723 \\
\#7286  &  3.564  &  5.582  &   0.00   & 0.10  &  2.850 &   3.620  &  3.193  &  0.00  & 10.610 \\
\#7309  &  3.613  &  1.267  &  24.30   & 0.00  &  3.130 &   3.830  &  3.417  &  22.90  &  9.723 \\
\#7785  &  3.970  &  1.964  &   3.28   & 0.30  &  3.900 &   4.040  &  0.540  &  0.00  & 10.391 \\
\#8403  &  4.264  &  2.018  &   2.76   & 0.00  &  4.110 &   4.390  &  3.347  &  0.57  & 10.710 \\
\#9537  &  4.103  &  0.150  &  99.89   & 0.40  &  3.200 &   4.390  &  6.651  &  0.00  & 10.801 \\
\#9561  &  4.026  &  0.838  &  59.14   & 0.00  &  2.080 &   4.320  &  3.396  &  51.28  & 10.412 \\
\#11682  &  4.320  &  0.367  &  96.11   & 0.00  & 4.040  &  4.530  &  2.941  &  64.89  & 10.186 \\
\#12327  &  4.971  &  0.202  &  99.41   & 0.70  & 2.920  &  7.750  &  3.256  &  96.17  & 11.185 \\
\#14130  &  4.432  &  0.315  &  97.06   & 0.10  & 1.590  &  5.090  &  3.067  &  96.30  & 10.235 \\
\#14260  &  4.152  &  1.243  &  25.71   & 0.50  & 4.040  &  4.250  &  0.652  &   0.01  &  9.935 \\
\#15268  &  4.775  &  2.534  &   0.47   & 0.80  & 4.390  &  5.090  &  7.260  &   0.07  & 10.373 \\

	\hline\hline
\end{tabular}
\caption{Best fitting parameters for the MIPS-u sample. In the order: identification number, z (best fit solution), $\chi^2_{\nu}$, probability relative to z, reddening, upper and lower bound for the $90\%$ confidence interval, $z_{2-phot}$(second best solution), probability relative to $z_{2-phot}$. We should point out that, for galaxies with $z_{phot}\sim
  8-10$, the best fit redshifts result uncertain, and the equally probable
  solution range ($z_{inf}-z_{sup}$), within the 90\% of confidence, is very large. Also for some of these objects -marked by the
  footnote$^{a}$- the \textit{hyperzmass} procedure was not able to find any
  physical solution by fitting the data
  points with MA05 models the extreme redshifts.}
\begin{list}{}{}
\item[$^{\mathrm{a}}$] \scriptsize{This value for the galaxy mass was found for the secondary solution of photometric redshift, $z=3.753$, because the \textit{hyperzmass} software did not find a plausible solution at $z= 9.024$.}
\end{list}
\label{table:tab3}
\end{center}
\end{table*}

\begin{figure}[h]
\includegraphics[width=\columnwidth]{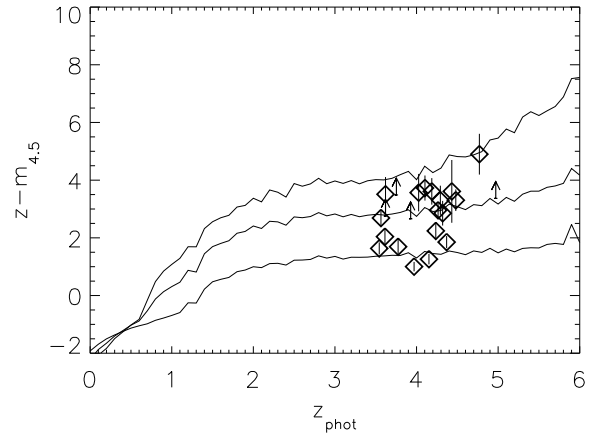}
\caption{Comparison between the observed $z-m_{4.5}$ colours of the MIPS-u galaxies (black open diamonds) and the ones predicted at the same redshift for SSPs models (MA05) with different ages (black curves in the figure, with age of  0.1, 0.3 and 0.5~Gyr, respectively, from the lower part upward). Most MIPS-u objects have $z-m_{4.5}$ colours consistent 
with SSPs with ages in the range of 0.3--0.5~Gyr. On the contrary, a smaller part of the sub-sample (7 objects) shows bluer colors, more compatible with younger SSPs (0.1--0.3~Gyr), probably exhibiting star formation activity.}
\label{fig:fig7}
\end{figure}

 To further constraints on the nature of the stellar populations of
  MIPS-u galaxies we
  compared their $z-m_{4.5}$ colours with the expected ones for MA05 SSP
  models with different ages ( 0.1, 0.3 and 0.5~Gyr).
The results are shown in Fig.~\ref{fig:fig7}. Although the MIPS-u objects
  span a large range of $z-m_{4.5}$ colours, most of them show colours
  consistent with older starlight (0.3--0.5~Gyr). On the contrary, a smaller
part of the sub-sample (7 objects) shows bluer colours, more compatible with
younger SSPs (0.1--0.3~Gyr), probably implying ongoing star formation activity.

The SED fitting analysis and the lack of mid-IR $24~\mu$m emission 
could thus suggest that many of the MIPS-u galaxies might be `quiescent', non dusty candidates 
at $z\geq3.5$. However, we should not forget that the MIPS observations in the field
are not deep enough to unequivocally constrain the lack of star formation in these galaxies.
The $24~\mu$m emission is generally used as SFR 
indicator in high-redshift galaxies \citep[$z \sim$ 2 to 3;][]{2005ApJ...626..680D, 2007ApJ...668...45P, 2007ASPC..380...35Y,2007arXiv0711.1902R}. Following
\citet{2001ApJ...556..562C} we can estimate that a source at $z\sim4$ with a $24~\mu$m flux
of $20~\mu$Jy will have a total IR luminosity of $\sim1.5\times10^{13}$~L$_{\odot}$, corresponding to a SFR of $\sim$ 1500~M$_{\odot} yr^{-1}$ for a Chabrier IMF \citep{1998ARA&A..36..189K}. 
Therefore the $24~\mu$m non-detection imposes only a very loose upper limit to the 
SFR. However, the $24~\mu$m non-detection is a necessary condition for quiescent systems (given the MIPS flux limit). In the 
following analysis we will use the MIPS-u galaxy sample to derive an upper 
limit to the space density and stellar mass density of `quiescent' candidates at $z\sim4$ (Sect. \ref{dens_mips-u}).

\subsubsection{The MIPS-d sample}
Out of the 53 galaxies of our high-$z$ sample, more than half
(32) are detected at $24~\mu$m. The best fitting SED results for the
MIPS-d objects and their relative $\chi^2_{\nu}$ distributions within the
$90\%$ confidence intervals are shown in Fig. \ref{fig:fig14}. In Table
\ref{table:tab4} we report the best fitting parameters for the MIPS-d
sample. The quality of the fit for the MIPS-d SEDs is very good,
with $\chi^2_{\nu}$ values  below 1.5 for almost all objects (except for \#5511, which has $\chi^2_{\nu} = 2.46$). 
However most galaxies in the sub-sample present degenerate redshift solutions, caused by the different models used and the amount of reddening present (\#~702,
\#~2560, \#~2796, \#~3393, \#~3850, \#~5157, \#~5367, \#~6099, \#~9301, \#~9306, \#~10176, \#~11304, \#~13129, \#~14105). 

It is difficult to understand
the nature of these objects,  given that often high and lower photometric
redshift solutions result equally probable. In these cases the $24~\mu$m emission could be interpreted 
in different ways, depending on the redshift of the objects and on the reddening parameter. Assuming that $z\geq3.5$ is correct, and taking into account the lack of
hard-X ray emission, the MIPS flux could ensue from highly obscured AGNs hosted in  relatively quiescent galaxies. However, still consistently with the
high redshift solution, the 24$\mu$m emission could be also due to star formation at extreme high rates.
In the next section we will examine this possibility using sub-mm/mm photometry from the SCUBA `supermap'\citep{2003MNRAS.344..385B,2005MNRAS.358..149P}, and the
MAMBO \citep{2008arXiv0806.3106G} and AzTEC \citep{2008arXiv0806.3791P} imaging surveys of GOODS-N.

On the other hand if the lower redshift solution ($z\sim 2-3$) is considered, the $24~\mu$m emission could be explained in terms of PAH emission from dust heated by star formation activity, and those objects would be accounted as dusty star-burst galaxies at $z\sim2-3$. 
\begin{table*}[htbp]
\begin{center}
\begin{tabular}{llllllllllll}
	\hline\hline
ID&$z_{phot}$&$\chi^2_{\nu}$&$P_{\chi^2_{\nu}}$&$A_{V}$&$z_{inf-90\%}$&$z_{sup-90\%}$&$z_{2-phot}$&$P_{\chi^2_{\nu},2}$ & log M(MA05)\\
	\hline\hline
\# 588 &  4.285 &  0.785 &  64.36  & 0.60 & 3.690 &   4.740 &  9.388 &	0.05 &  9.973 \\
\# 702 &  8.814 &  0.067 &  99.99  & 0.40 & 1.450 &  10.060 &  3.340 &  99.99 & 10.563 \\
\#1720 &  3.522 &  0.526 &  87.33  & 0.60 & 2.850 &   4.250 &  3.921 &  84.30 & 10.399 \\
\#2172 &  3.711 &  0.536 &  86.56  & 0.70 & 3.550 &   4.180 &  4.117 &  70.42 & 11.097 \\
\#2560 &  3.830 &  0.044 & 100.00  & 0.40 & 2.780 &   8.520 &  7.274 & 100.00 & 10.762 \\
\#2796 & 10.060 &  0.279 &  98.59  & 0.80 & 3.360 &  10.060 &  3.536 &  95.36 & 10.870$^{\mathrm{a}} $ \\
\#2894 &  3.893 &  0.628 &  79.14  & 0.70 & 2.990 &   4.460 & 10.060 &	0.37 & 10.441 \\
\#3393 &  5.503 &  0.477 &  90.59  & 0.80 & 3.830 &   5.860 &  5.167 &  84.73 & 10.647 \\
\#3850 &  4.166 &  0.063 & 100.00  & 0.70 & 3.410 &   9.290 &  8.009 &  99.90 & 10.940\\
\#5157 &  4.740 &  0.344 &  96.90  & 0.00 & 4.110 &   7.400 &  6.504 &  92.59 & 11.433\\
\#5367 &  4.355 &  0.494 &  89.51  & 0.80 & 3.900 &   4.880 & 10.060 &	0.03 & 10.806\\
\#5511 &  4.152 &  2.521 &   0.50  & 0.50 & 4.040 &   4.250 &  1.310 &	0.00 & 10.551\\
\#6099 &  5.251 &  0.367 &  96.10  & 0.10 & 3.200 &   8.450 &  4.999 &  96.02 & 11.178\\
\#6245 &  5.062 &  0.108 &  99.98  & 0.00 & 4.460 &   5.370 &  2.934 &  81.31 & 10.662\\
\#6463 &  9.437 &  0.193 &  99.68  & 0.10 & 3.690 &  10.060 &  4.215 &  99.64 &  9.973$^{\mathrm{a}} $\\
\#8520 &  3.683 &  0.573 &  83.71  & 0.00 & 3.340 &   3.830 &  3.137 &  44.99 & 10.730\\
\#9007 &  3.669 &  1.117 &  34.42  & 0.60 & 2.780 &   4.110 &  2.969 &  34.41 & 10.824\\
\#9301 &  8.226 &  0.059 & 100.00  & 0.20 & 2.640 &  10.060 &  8.037 & 100.00 & 11.035\\
\#9306 &  8.660 &  0.127 &  99.95  & 0.80 & 2.360 &  10.060 &  8.219 &  99.92 & 11.011\\
\#9782 &  3.928 &  1.472 &  14.28  & 0.20 & 3.480 &   4.320 &  8.723 &	0.02 & 11.116\\
\#10176 &  5.818 &  0.002 & 100.00  & 0.20 &3.550  &  9.570 &  5.517 & 100.00 & 11.280\\
\#11304 &  4.145 &  0.378 &  95.67  & 0.80 &3.550  &  8.100 &  5.461 &  89.78 & 10.759\\
\#12699 &  3.543 &  1.760 &   6.21  & 0.30 &3.200  &  3.690 &  3.354 &   5.62 & 10.558\\
\#13129 &  7.960 &  0.030 & 100.00  & 0.20 &3.410  & 10.060 &  8.296 & 100.00 & 11.056\\
\#13857 &  3.746 &  0.863 &  56.74  & 0.10 &3.270  &  4.040 &  3.403 &  43.23 & 10.707\\
\#14505 &  3.564 &  0.014 & 100.00  & 0.80 &2.920  & 10.060 &  7.757 & 100.00 & 10.025\\
\#14668 &  4.600 &  0.209 &  99.56  & 0.10 &1.870  &  5.230 &  4.789 &  99.53 & 10.177\\
\#14722 &  4.145 &  1.816 &   5.23  & 0.40 &4.110  &  4.180 &  0.617 &   0.36 & 10.180\\
\#14793 &  4.474 &  1.225 &  27.41  & 0.00 &4.040  &  5.160 &  9.185 &   0.01 & 11.411\\
\#15541 &  3.949 &  1.216 &  27.45  & 0.40 &3.690  &  4.110 &  0.680 &   0.00 & 11.146\\
\#15761 &  4.292 &  0.177 &  99.78  & 0.10 &3.900  &  4.530 &  6.567 &   0.00 & 11.126\\
\#15771 &  3.564 &  0.360 &  96.35  & 0.20 &3.060  &  3.830 &  6.854 &   0.00 & 11.283\\

	\hline\hline
\end{tabular}
\caption{Best fitting parameters for the MIPS-d sample. In the order:
  identification number, z (best fit solution), $chi^{2_\nu}$, probability
  relative to z, reddening, upper and lower bound for the $90\%$ confidence
  interval, $z_{2-phot}$(second best solution), probability relative to
  $z_{2-phot}$.  For the objects with extremely high redshift best fit
  solutions, the same caveat we gave in Tab.~\ref{table:tab3} should be taken
  into account}.
\begin{list}{}{}
\item[$^{\mathrm{a}}$] \scriptsize{These values for the galaxy masses were found for the secondary solutions of photometric redshift, because the \textit{hyperzmass} software did not find a acceptable solution at $z\sim 9-10$.}
\end{list}\label{table:tab4}
\end{center}
\end{table*}

\subsection{Sub-mm detections : a large population of massive starburst galaxies at $z~\sim4$?}\label{scuba}
Sub-mm/mm selected galaxies (hereafter SMGs) are the brightest star forming galaxies known, being much more more luminous than the local ultra-luminous IR galaxies
-ULIRGs- with $L_{IR}\gg10^{12}$~L$_{\odot}$. These galaxies are massive young objects seen during their formation epochs, with very high SFR
\citep{1999astro.ph..3157L, 2002MNRAS.331..817S}, about one order of magnitude larger than that of typical systems with similar masses \citep{2007ApJ...670..156D}. 
They are also fairly rare objects, probably due to the short duration of their bright phase \citep[$<< 100~Myr$; ][]{2005MNRAS.359.1165G}. 
These objects could represent the common early phase in the formation of massive elliptical galaxies, and hence they might be
crucial link to understand the massive galaxy formation process.

The numerous sub-mm datasets available in  the GOODS-N field can be used to improve the constraints on the nature of our $z\geq 3.5$ candidates.

We made direct comparison with three sub-mm maps of the field: 
\begin{itemize}
 \item the Sub-millimeter Common User Bolometer Array \citep[SCUBA, ][]{1999MNRAS.303..659H} `supermap' of 35 SMGs selected with S/N$\geq 4$ at $850~\mu$m \citep{2003MNRAS.344..385B, 2005MNRAS.358..149P, 2006MNRAS.370.1185P}; 
\item the Max Planck Millimeter Bolometer Array \citep[MAMBO, IRAM 30-m telescope, ][]{2000astro.ph.10553B} $1200~\mu$m map from \citet{2008arXiv0806.3106G}, composed by 27 sources detected with $3.5\leq$~S/N~$ <4$ and 20 with S/N~$\geq4$;
\item the AzTEC \citep[][ ,15 m James Clerk Maxwell Telescope -JCMT]{2008MNRAS.386..807W} 1100~$\mu$m map of 28 sources detected with S/N $\leq 3.75$ \citep{2008arXiv0806.3791P}.
 \end{itemize}

We used a simple approach, looking for overlap of our sample with the position
of the submm/mm selected galaxies in the above works. We have used radial 
separation limits of 7'', 6'' and 9'', to search for matches with galaxies 
in the SCUBA, MAMBO and AzTEC, respectively.

We found eight of our high-$z$ candidates in the error box of the SCUBA `supermap' sources (\#5157, \#6099, \#6463, \#8520, \#10176, \#11682, \#13857, \#15541), but only seven (except \#11682) correspond to the IRAC counterparts identified by \citet{2006MNRAS.370.1185P}.
For two of these galaxies, GN20 and GN20.2, (objects \#15541, \#13857 in
our sample) spectroscopic redshifts around $z\sim4$ were recently determined
by \citet[][hereafter D08]{Daddietal2008}. Those objects are among the most distant SMGs spectroscopically confirmed known at the present. Our photometric redshifts
for these galaxies are consistent within the error ($\sim 0.2$) with the spectroscopic ones (being 3.75 and 3.95 for GN20 and GN20.2 respectively). It is interesting
to remark that D08 discuss the evidence of a galaxy proto-cluster at $z\sim4$ in the GN20 and GN20.2 area, due to the finding of an over-density of B-dropout
galaxies. Four objects in total from our $z>3.5$ sample (\#13857, \#14722, \#15268, \#15541) are also within a 25'' radius centered on the coordinates
of GN20.
This represents an overdensity of a factor of 18 with respect to the expected
number density of IRAC-selected galaxies in the field. Massive galaxies seem
to be tracing this proto-cluster structure at $z=4.05$.

We found five sources of our high-$z$ sample matching the MAMBO galaxies within the correlation distance. Out of them only \#15541 (GN20) is also SCUBA-detected. 
In addition, 8 objects from the AzTEC survey correspond to galaxies in our $z>3.5$ sample, within the adopted search radius. Three of them are also SCUBA detected, and three are in common with the MAMBO survey from G08. 

Candidate $z>3.5$ galaxies in our sample that are consistent with being counterpart of  SCUBA, MAMBO and/or  AzTEC sources  are listed in Tab.~\ref{table:tab5}.

\begin{table*}[htbp]
\begin{center}
{\small  
\renewcommand{\footnoterule}{}
\begin{tabular}{lllllllllll}
\hline \hline
ID&  SCUBA& MAMBO/AzTEC&  $z_{phot}$& $z_{min 90\%}$ & $z_{max 90\%}$ & $z_{phot}$ & $z_{radio-IR}$& S$_{850}$ & S$_{1200/1100}$ & d\\
(this paper)& name& name&(this paper)& & & (P06)&(D08)&[mJy]&	[mJy]&[arcsec]\\
\hline \hline
\#588     &$...$ &$...$/AzGN12 &4.28 & 3.69 & 4.74 & $...$ & $...$& $...$&$3.07$&5.2 \\
\#3850    &$...$ &$...$/AzGN20N & 4.17 & 3.41 & 9.29 & $...$ & $...$&$...$&$2.79$& 2.3\\
\#5367    &$...$ &$...$/AzGN27N & 4.35 & 3.90 & 4.88 & $...$ & $...$&$...$&$2.31$& 6.5\\
\#14505  &$...$ &GN1200.9/AzGN28 & 3.56 & 2.92 & 10.06 & $...$ & $...$&$...$&$2.7/2.31$& 4.1/4.2 \\
\#522      &$...$ &GN1200.5/AzGN06 & 3.93 & 3.06 & 5.23 & $...$ & $...$&$...$&$3.8/4.13$&2.9/1.1\\
\#7785     &$...$ &GN1200.46/$...$ & 3.97 & 3.90 & 4.04 & $...$ & $...$& $...$ & 2.1$^{\mathrm{a}} $&4.5\\
\#11304    &$...$ &GN1200.13/$...$ & 4.14 & 3.55 & 8.10 & $...$ & $...$& $...$&$2.2$& 0.8\\
\#6099     &GN03  &$...$           & 5.25 & 2.95 & 10.0 & 2.0  & 1.9& $1.8$& $...$&2.5\\				
\#10176    &GN09 & $...$/AzGN31  &  5.818 & 3.55 & 10.0 & 2.4 & 13.42 & 8.9& 2.13& 3.9/5.3\\			
\#5157        &GN11 &$...$  & 4.74   & 4.11 & 7.40 & 2.3 & 2.45    & 7.0 &$...$&2.4\\	\#8520  &GN12 & $...$/AzGN08  & 3.68  & 3.34 &  3.83 & 3.1 & 3.20& 8.6 & 3.83 &5.16/8.7\\				
\#6463     &GN18 &$...$&9.437/4.215     &  3.7 &  10.0 & 2.2 & 2.50& 3.2 & $...$&3.2\\	\#15541 &GN20 & GN1200.1/AzGN01 & 3.949 & 3.69 & 4.11 & 2.95 &{\bf 4.055}&20.3& 9.3/10.69 &1.3/1.9/1.4\\
\#13857  &GN20.2a &$...$&3.746 & 3.27 & 4.04  & 2.83 & {\bf 4.051} & 9.9&$...$&6.2 \\
\hline \hline
\end{tabular}
}
\caption{SCUBA, MAMBO and AzTEC counterparts found in our $z\geq 3.5$ of MIPS-d detected sources. In the columns we report: the ID number of the objects (this paper); the names from the sub-mm surveys; photometric redshift estimated in this paper and their 90\% confidence intervals; photometric redshifts by \citet[][P06]{2006MNRAS.370.1185P}; radio-IR redshifts by \citet[][D08]{Daddietal2008} - with the two spectroscopic values in boldface- ; the SCUBA, MAMBO and AzTEC fluxes. The last column is the distances between the proposed IRAC-4.5~$\mu$m counterparts and the sub-mm positions.}
\begin{list}{}{}
\item[$^{\mathrm{a}}$] \scriptsize{Not deboosted flux}
\end{list}

\label{table:tab5}
\end{center}
\end{table*}

It should be noted that the bulk of the sub-mm counterparts, except the
objects \#522 (GN1200.5/AzGN06) and \#7785 (GN1200.46), are 24~$\mu$m-detected
(MIPS-d sample). The latter is also a less secure MAMBO detection,
being among the sources with a S/N below the 4.5~$\sigma$ limit, and
is not part of the `robust deboosted catalogue'. 

\begin{figure}[htb]
\includegraphics[width=\columnwidth]{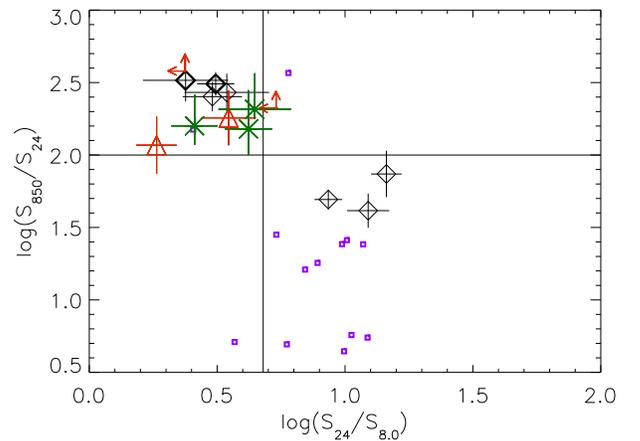}
\caption{S$_{850}$/S$_{24}$ vs S$_{24}$/S$_{8.0}$ colours for the SCUBA(black diamonds), MAMBO(red triangles) and AzTEC(green crosses) detections of our high-$z$
sample. The comparison with the GN20 and GN20.2 \citep{Daddietal2008} seems to support the high-$z$ hypothesis for most of these sources (left part of the diagram;
see Sect.\ref{scuba}). The solid black line represent the colour limit suggested
by \citep{Daddietal2008} for sub-mm galaxies at $z\gtrsim4$ (log(S$_{24}$/S$_{8.0}) <0.7$). The small violet squares are the SCUBA `supermap' sources that do not overlap with our high-$z$ sample, most of them spectroscopically confirmed at $z\sim1-2.5$.}
\label{fig:fig8}
\end{figure}

 We can check the reliability of the
  high-$z$ solution for the sub-mm detections by comparing their
  S$_{850}$/S$_{24}$ vs S$_{24}$/S$_{8.0}$ colours with the ones found by D08
  for the two confirmed SMGs galaxies at redshift $\sim 4$ (GN20 and GN20.2).
We show this comparison in Fig.~\ref{fig:fig8}, where GN20 and GN20.2a are
  represented as black bold diamonds.   
 As suggested by D08, starburst galaxies at $z\sim 4$ would have
  relatively blue Spitzer MIPS-IRAC colours, similarly to GN20 and GN20.2a (log(S$_{24}$/S$_{8.0}$) $<0.7$,
  log(S$_{24}$/S$_{4.5}$) $<1.11$). Moreover, because of the negative sub-mm
  k-correction the S$_{850}$/S$_{24}$ ratio should increase with
  redshift. Hence we expect that starburst at $z\sim 4$ would occupy the
  upper-left part of the diagram in Fig.~\ref{fig:fig8}, with
  log(S$_{850}$/S$_{24}$)$>2$ (such as GN20 and
  GN20.2a).

Out of the SCUBA sources (black diamonds), two (\#10176/GN09 and \#8520/GN12)
  have mid-IR colours comparable with GN20 and GN20.2a, and
  log(S$_{850}$/S$_{24}$)$>2$). This finding can be seen as a further confirmation
  of the $z>3.5$ redshift estimated for these objects. It is also in agreement, within
  the 99\% confidence interval, with the radio-IR redshifts estimated by D08 (see their Tab. 3). 
In the same figure also the `SCUBA-blank' MAMBO (red triangles) and AzTEC (green crosses) detections are shown. To represent them in this diagram we scaled the
S$_{1200}$ and S$_{1100}$ fluxes to S$_{850}$, multiplying by a factor of 2, i.e. the typical S$_{850}$/S$_{1200}$ ratio found for GN20 and GN20.2 (D08). This
is also consistent with the S$_{850}$/S$_{1100}$ flux ratio derived by \citet[][$2.08\pm 0.18$]{2008arXiv0806.3791P}. 
For the MIPS-u \#522\ and \#7785\ we show the limits in the two-colours diagram, computed with S$_{24}$=20~$\mu$Jy (the 5~$\sigma$ $24~\mu$m-MIPS flux lower limit). Also the sub-mm-IR colours of these sources are consistent, within the error bars, with the expected colours for high-$z$ SMGs. 
For comparison we show also the positions occupied in Fig~\ref{fig:fig8} by the other sub-mm galaxies from the \citep{2006MNRAS.370.1185P} sample with secure IRAC counterparts, that do not overlap with our sample. Almost all these sources (violet small squares) are spectroscopically confirmed at $z\sim1-2.5$ \citep[see][]{2006MNRAS.370.1185P}. 
The evidence that most of them occupy the lower-right part of Fig~\ref{fig:fig8} strongly support the above statements. Other two galaxies from the SCUBA `supermap' have colours similar to GN20 and GN20.2, i.e. GN10, and GN22. The first one is claimed by several authors \citep[D08,][]{2008ApJ...673L.127D, 2007ApJ...670L..89W, 2008arXiv0805.3503W} to be a $z>3.7$ starburst, although \citep{2006MNRAS.370.1185P} report a $z=2.2$ photometric redshift. The second one is a spectroscopically confirmed galaxy at  $z=2.509$ \citep{2005ApJ...622..772C, 2006MNRAS.370.1185P}. It represents the only case in the SCUBA sample of a lower-redshift sub-mm/mm emitter with both S$_{850}$/S$_{24}$ ratio and IRAC colours similar to spectroscopic confirmed $z\geq3.5$ SMGs. Nevertheless from a statistical point of view the diagram shown in Fig~\ref{fig:fig8} seems to be a valid diagnostic to prove high-$z$ starburst for sub-mm/mm sources.

Hence we can conclude that, for $\sim$ 78\% of the galaxies in our sample that are likely sub-mm/mm detected, these results
 support the hypothesis of extreme SFR activity at high-$z$. 
Instead the sub-mm (SCUBA) sources lying in the right part of
 Fig.~\ref{fig:fig8}(objects \#5157, \#6099 and \#6463 -$\sim22$\% of the sub-mm detections) 
 are more probably at lower redshifts, as proposed by \citet{2006MNRAS.370.1185P}.

\begin{figure}[htb]
\includegraphics[width=\columnwidth]{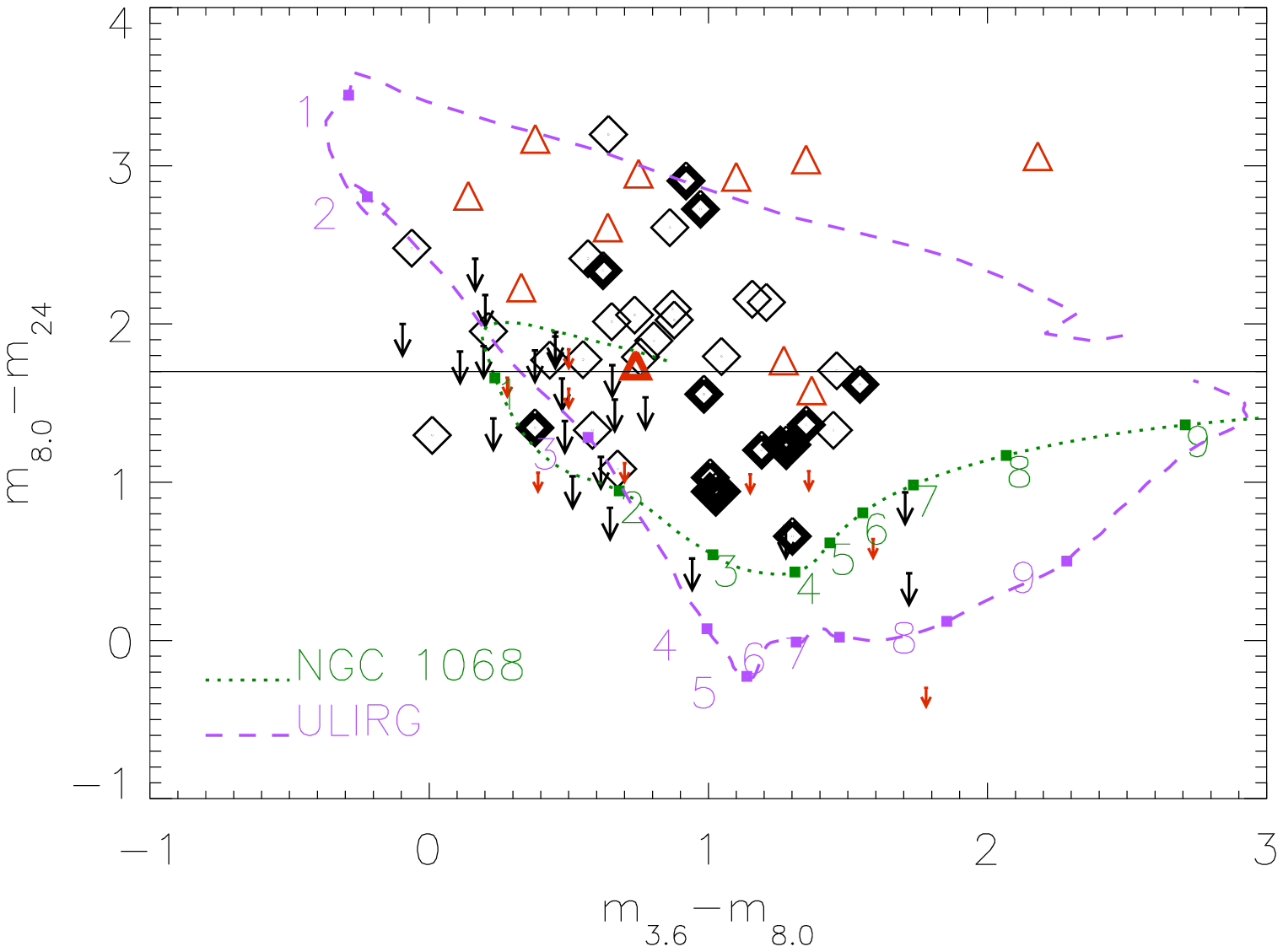}
\caption{Infrared colours $m_{8.0}-m_{24}$ vs $m_{3.6}-m_{8.0}$ of our
  high-$z$ sample of galaxies (black diamonds) compared with the
  objects found by \citet{2007A&A...470...21R} (red open triangles). The bold red triangle represents HUDF-JD2
  \citep{2005ApJ...635..832M}. Bold black open diamonds are the SCUBA detected
  galaxies of our sample. The two objects with confirmed $z\sim 4$, GN20 and
  GN20.2, are highlighted (filled black diamonds). The MIPS-undetected objects
  are shown as upper limits in the $m_{8.0}-m_{24}$ colour: black and red
  arrows for our sample and \citet{2007A&A...470...21R} sample, respectively.
Two evolutionary colour tracks of an ULIRG \citep[violet dot-dashed
  curve;][]{2001ApJ...556..562C}, and of a galaxy hosting an obscured AGN
  (dashed green curve; Seyfert 2 galaxy NGC 1068), with increasing redshift
  values, are also shown. The black horizontal line represents the upper limit suggested by D08 and also shown in Fig.~\ref{fig:fig8} for colours of
  starburst galaxies at $z\geq 3.5$.}
\label{fig:fig9}
\end{figure} 

\subsection{Mid-IR vs IR colours}
The properties and the overall spectral energy distributions make the
MIPS-d galaxies comparable to one of the most debated objects in the
literature, HUDF-JD2. This object, identified as a $z\sim 3.4$ galaxy by
\citet{2004ApJ...616...63Y} and \citet{2004ApJ...615..603C}, was recently
suggested by \citet{2005ApJ...635..832M} as a massive post star-burst galaxy
at $z\simeq 6.5$ hosting an obscured AGN. However subsequent studies \citep{2007MNRAS.376.1054D,
  2007A&A...470...21R, 2007ApJ...665..257C} claimed that HUDF-JD2 is more likely
a lower redshift ($z\sim 1.5$-2.2) star forming galaxy with very strong dust reddening. 
Our galaxies are also similar to candidate $z>4$ galaxies found by
\citet{2007A&A...470...21R} and \citet{2008ApJ...676..781W}. 
In Fig.~\ref{fig:fig9}, the infrared colours
  ($m_{8.0}-m_{24}~vs~m_{3.6}-m_{8.0}$) of our high-$z$ sample (open black
  diamonds) are compared with the objects found by \citet[][open red
  triangles]{2007A&A...470...21R}. The bold red triangle represents HUDF-JD2
  \citep{2005ApJ...635..832M}. Bold black open diamonds are the sub-mm
  detected galaxies of our high-$z$ sample. The two objects with confirmed
  $z\sim 4$, GN20 and GN20.2, are represented by filled black diamonds. The MIPS-undetected objects
  are shown as upper limits in the $m_{8.0}-m_{24}$ colour  \citep[black and red
  arrows for our sample and the sample of][respectively]{2007A&A...470...21R}. 
The evolutionary colour tracks of an ULIRG \citep{2001ApJ...556..562C} and of a galaxy hosting an obscured AGN (the Seyfert 2 galaxy NGC 1068) with increasing
redshift values are also shown. It is clear, though, that such locally calibrated templates  are not very useful to interpret
the colors of distant galaxies. This can be argued just considering that the observed $m_{8.0}-m_{24}$ colours of GN20 and GN20.2 are $\sim 1$~mag redder with
respect to the ULIRG track at $z\sim4$. As discussed in Daddi et al. (2008), this is most likely due to a lower  stellar mass to light ratio 
in the distant galaxies.

The black horizontal line represents the upper limit suggested by D08 and also shown in
Fig.~\ref{fig:fig8} for colours of starburst galaxies at $z\geq
3.5$ (log(S$_{24}$/S$_{8.0}) <0.7$, in AB
magnitude: $m_{8.0}-m_{24}<1.7$; see also Sect.~\ref{scuba}). If the
high-$z$ solution is correct, objects bluer than this limit should be strongly
dominated by star formation, such as GN20 and GN20.2. The issue that most of the
the sub-mm detected galaxies in our sample are down to the black line in the
figure also suggest that they have properties similar to the two
 starbursts spectroscopically confirmed at $z\sim4$. The MIPS-d sources which have
 similar colours but are lacking in a sub-mm counterpart could be also high-$z$ starbursts. 
 Alternatively, the
lack of sub-mm detection should be justified either by the shallow sensitivity
of the currently available sub-mm instruments (SCUBA, MAMBO, AzTEC), or by
particularly ``warm'' SEDs for this objects.

On the other hand, the $z\geq 3.5$ candidates with $m_{8.0}-m_{24}>1.7$ in
Fig.~\ref{fig:fig9} could represent a different population. Their redder
colours, more similar to the bulk of the \citet{2007A&A...470...21R}
sample, could be due to heavily obscured AGN, as also suggested by
\citet{2007A&A...470...21R} for most of their sources, or they might be $z<3.5$
contaminants. The position of object HUDF-JD2, overlapping with the solid black line in the middle part of the diagram, does not allow us to give any further constraint on its nature. 

\section{The comoving stellar mass density at $3.5 < z < 5$}\label{complet}
 Recent studies \citep{2005ApJ...619L.131D,2006A&A...459..745F,2007ApJ...659...84S,2007MNRAS.377.1024V,2007MNRAS.374..910E,2006ApJ...651...24Y} have extended measurements of the total stellar mass density up to $z\sim4-5$, using deep multicolour data from optical and near-IR selected samples. 
One of the  goals of this paper is to determine how our IRAC-4.5 selected sample contributes to the comoving stellar mass density at $z \geq 3.5$.  In fact, as already discussed, 
observations of the high redshift Universe are generally biased against red galaxies with large mass-to-light ratios, 
potentially missing galaxies that could potentially make a strong contribution to the total mass density \citep[see][]{2006A&A...459..745F}. 
Here, we emphasize that we can only determine a lower limit to the comoving
stellar mass density in the redshift interval $3.5\leq z\leq5$, for several reasons.
First, the combination of our IRAC 4.5~$\mu$m magnitude limit and the 1.6~$\mu$m peak SED criterion 
leads to missing objects with lower masses at these redshifts, such as the fainter, bluer Lyman break galaxies, 
as we have discussed in Sect.~\ref{zge3.5}. More generally, a magnitude-limited sample (selected at
any wavelength) is not the same as a mass--limited sample, as we will discuss further in the next section.

\subsection{Incompleteness effects and mass selection criteria}\label{incompleteness}
In estimating the comoving stellar mass density we must take into account that our sample could suffer from incompleteness. 
As several other authors \citep{2004A&A...424...23F, 2004ApJ...608..742D, 2005ApJ...624L..81L} have noted,
IR-selected samples are still not equivalent to mass-selected samples.
In fact, at any redshift, the galaxies detected above the sample magnitude limit can have a fairly large range of 
possible M$_{\star}/$L ratios, depending on their spectral properties, ages, dust extinction, metallicities, etc. 
The effect is that magnitude-limited samples, at higher redshifts, are progressively biased against objects with lower masses and high M$_{\star}/$L ratios, such as galaxies that are not currently forming stars, or which are highly extincted.
 
\begin{figure}[htbp]
\includegraphics[width=\columnwidth]{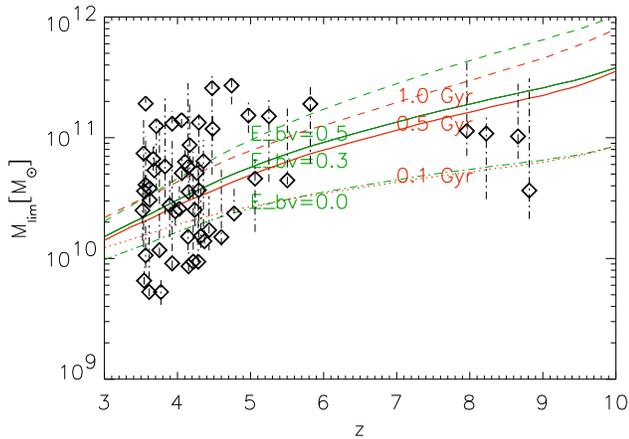}
\caption{The redshift dependence of the stellar mass limit for our IRAC
  magnitude-limited sample ($m_{4.5}\leq23$), derived from MA05 stellar population 
  synthesis models.  The red curves represent SSP models with different ages (0.1, 0.5 and 1~Gyr). 
  The green curves are models with constant SFR, an age of 0.5~Gyr, and dust extinction, with three 
  different values of the colour excess E$_{B-V}$ (0.0, 0.3 and 0.5). The points indicate the objects
  of our $z\geq 3.5$ sample, with error bars indicating the 68\% confidence ranges on their stellar masses.
}
\label{fig:fig10}
\end{figure}

We have attempted to evaluate the consequences and minimize the effects of 
incompleteness by determining the threshold mass limit as a function of
redshift for a variety of galaxy models spanning a range of ages and degrees
of extinction.  For this purpose we used the MA05 template library with the 
Chabrier IMF, the same as was used in the preceding sections to derive the 
galaxy stellar masses.  We consider both the simplest case of SSP
models with different ages (0.1, 0.5 and 1~Gyr), and a set of 
models with a constant star formation rate, an age of 0.5~Gyr, and
different amounts of extinction \citep[E$_{B-V}=0.0, 0.3, 0.5$, applied using
the Calzetti attenuation law;][]{2000ApJ...533..682C}.  The observed $4.5~\mu$m 
flux as a function of redshift was derived from these models, and from this
we translate our 4.5~$\mu$m magnitude limit ($m_{4.5} \leq 23$) into a limiting
stellar mass at each redshift.  The results are shown in Fig. \ref{fig:fig10}.
For passively evolving galaxies (i.e., the SSPs) with ages  less than 0.5~Gyr, 
and for star-forming galaxies with E$_{B-V} < 0.3$, our sample should be fairly
complete at $M > 5\times10^{10}$~M$_{\odot}$ out to $z < 5$.  This is another advantage 
of our $4.5\mu$m sample selection from the very deep GOODS IRAC data, which is complete
to significantly lower mass thresholds at these very high redshifts compared to 
$K$--selected samples\citep[see][]{2006A&A...459..745F}. 

Assuming that the redshift estimate for the objects in our high-$z$ sample are correct, we derive their contribution to the comoving number density and to the total stellar mass density, both for the total sample (MIPS-u+MIPS-d) and for the MIPS-u sample only, in the redshift interval $3.5\leq z\leq5$ as described in the following sections.
The comoving volume in this redshift range over the total GOODS-N solid angle ($\sim165$ arcmin$^2$, coincident with the ACS z-band area) is $V_{com}\sim 0.725\times 10^6$~Mpc$^3$. 

\subsection{Contribution to the stellar mass density from our total sample in the redshift bin $z=3.5-5$}\label{tot_mass_dens}
Based on the preceding analysis, we assume a limiting stellar mass M$\sim5\times10^{10}$~M$_{\odot}$ as the completeness threshold
for our magnitude-limited sample.  For quiescent galaxies with intermediate ages ($\sim0.5$~Gyr), we should be complete above
this mass threshold out to $z \sim 5$.  However, for $z > 4$, we may be incomplete near this mass limit for the oldest possible 
passive stellar populations, or for heavily dust-obscured galaxies.

Given this potential incompleteness, and the fact that we are undoubtedly missing a large percentage of bluer, lower--mass LBGs 
in the same redshift range (as discussed in section \ref{dropout}), we must consider our estimate for the stellar mass density 
at $3.5\leq z\leq 5$ to be a lower limit.  
Fig.~\ref{fig:fig10} shows the derived stellar masses versus redshift for the objects in our 
IRAC-selected sample.  We count objects and sum their stellar masses for galaxies above our
mass threshold (M$\sim5\times10^{10}$~M$_{\odot}$).  The resulting comoving number density and 
the stellar mass density for this sample are 
$2.6\times10^{-5}$~Mpc$^{-3}$ and $(2.9\pm 1.5)\times10^{6}$~M$_{\odot}$~Mpc$^{-3}$, respectively.

\begin{figure}[htbp]
\includegraphics[width=\columnwidth]{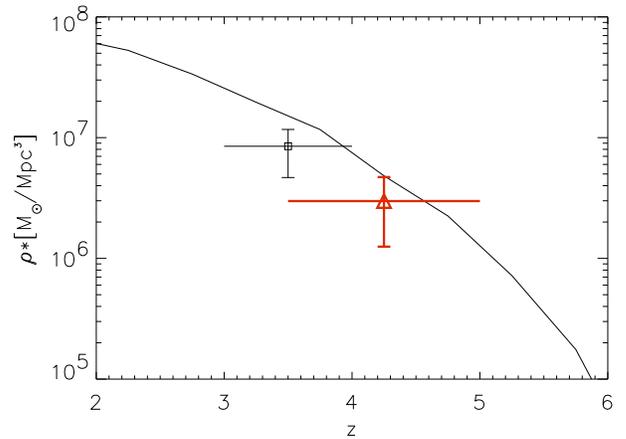}
\caption{Comparison between our estimate of the lower limit of the stellar mass density ($\rho_{\star}$) 
at $3.5 < z < 5$ (red triangle) and the stellar mass density estimated by \citet[][F06]{2006A&A...459..745F} at $3 < z < 4$,
for galaxies in the mass range M$_{\star} \sim 3\times10^{10}- 3\times10^{11}$~M$_{\odot}$. The curve shows the evolution of stellar mass density with redshift from the Millennium Simulation models, computed for galaxies above our threshold limit of $\ge 5 \times 10^{10}$~M$_{\odot}$.}
\label{fig:fig11}
\end{figure}

In Fig.~\ref{fig:fig11} we show our lower limit to the comoving stellar mass density ($\rho_*$) 
at $3.5 < z < 5$, and compare it to another estimate \citep[][F06]{2006A&A...459..745F} 
at $3 < z < 4$ for galaxies in a similar mass range (M$_{\star}=3\times 10^{10}-3\times 10^{11}$~M$_{\odot}$).
We take the F06 value from the compilation published by \citet{2008arXiv0801.1594W},
and have corrected it for the $M/L$ differences between the Salpeter and Chabrier 
IMFs (log M(Chabrier)=log M(Salpeter)-0.23), as well as by a factor of 1.4 to account for 
the difference between the BC03 used by F06 and the MA05 models used in our study
(Sect. \ref{sec:templates}).  We also show a curve derived from the Millennium Simulation models for the redshift evolution of the stellar mass density, computed for galaxy masses above the threshold limit of $\ge 5 \times 10^{10}$~M$_{\odot}$. Assuming that the photometric redshifts for all the selected high-$z$ candidates are correct, our results are in good agreement with the Millennium Simulation predictions.

\subsection{Stellar mass density for the MIPS-u candidates in the redshift bin $z=3.5-5$}\label{dens_mips-u}
As described above, our sample should be largely complete for `quiescent', non-star-forming galaxies with 
ages $\leq 0.5$~Gyr and masses $M > 5\times 10^{10}$~M$_{\odot}$.  However, at these large redshifts, 
the flux limit at 24~$\mu$m, even for very deep data from GOODS, corresponds to quite large star formation
rates $> 1000$~M$_\odot$/yr.  Therefore, we cannot affirm that the MIPS-undetected (MIPS-u) objects
in are sample are truly `quiescent', and thus we can only derive an upper limit for the number and stellar mass 
densities of quiescent galaxies.  For the MIPS-u subsample with stellar masses $\geq 5 \times 10^{10}$~M$_{\odot}$ at $z=3.5-5$, we find a comoving number density of $0.97\times10^{-5}$~Mpc$^{-3}$, and a stellar mass density of $(1.15\pm 0.7)\times10^{6}$~M$_{\odot}$Mpc$^{-3}$ (MA05 templates and Chabrier IMF). 

\begin{figure}[htbp]
 \includegraphics[width=\columnwidth]{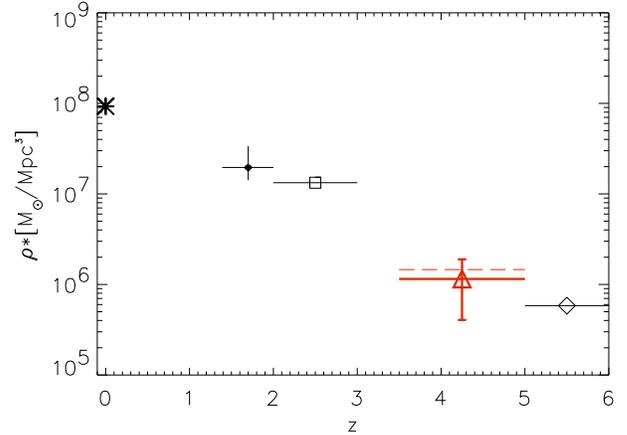}
\caption{Comparison between the upper limit to stellar mass density ($\rho_*$)
  found in this paper for `quiescent' galaxies in the redshift bin $z=3.5-5$
 and results from the literature at other redshifts. All points have been scaled
  to Chabrier IMF and MA05 masses (see Sect.~\ref{tot_mass_dens}). Our estimate is represented as a red bold triangle. The horizontal red dashed line represent the stellar mass density we derive by considering all MIPS-u galaxies to be SSPs with age of 0.5~Gyr.
The local mass density in passive galaxies is shown by the black asterisk \citep{2004ApJ...600..681B}. The black open square is for Distant Red Galaxies (DRGs) at $z = 2-3$ in the Hubble Deep Field South \citep{2005ApJ...624L..81L}. The black filled rhombus is from \citet{2005ApJ...626..680D} for passive-BzK galaxies in the Hubble Ultra Deep Field (HUDF) at $z=1.39-2$, and the open rhombus 
is for Balmer-break MIPS undetected galaxies at $z = 5$--7 in GOODS-South from \citet{2008ApJ...676..781W}.}
\label{fig:fig12}
\end{figure}

\begin{figure*}[t]
\includegraphics[width=0.8\textwidth]{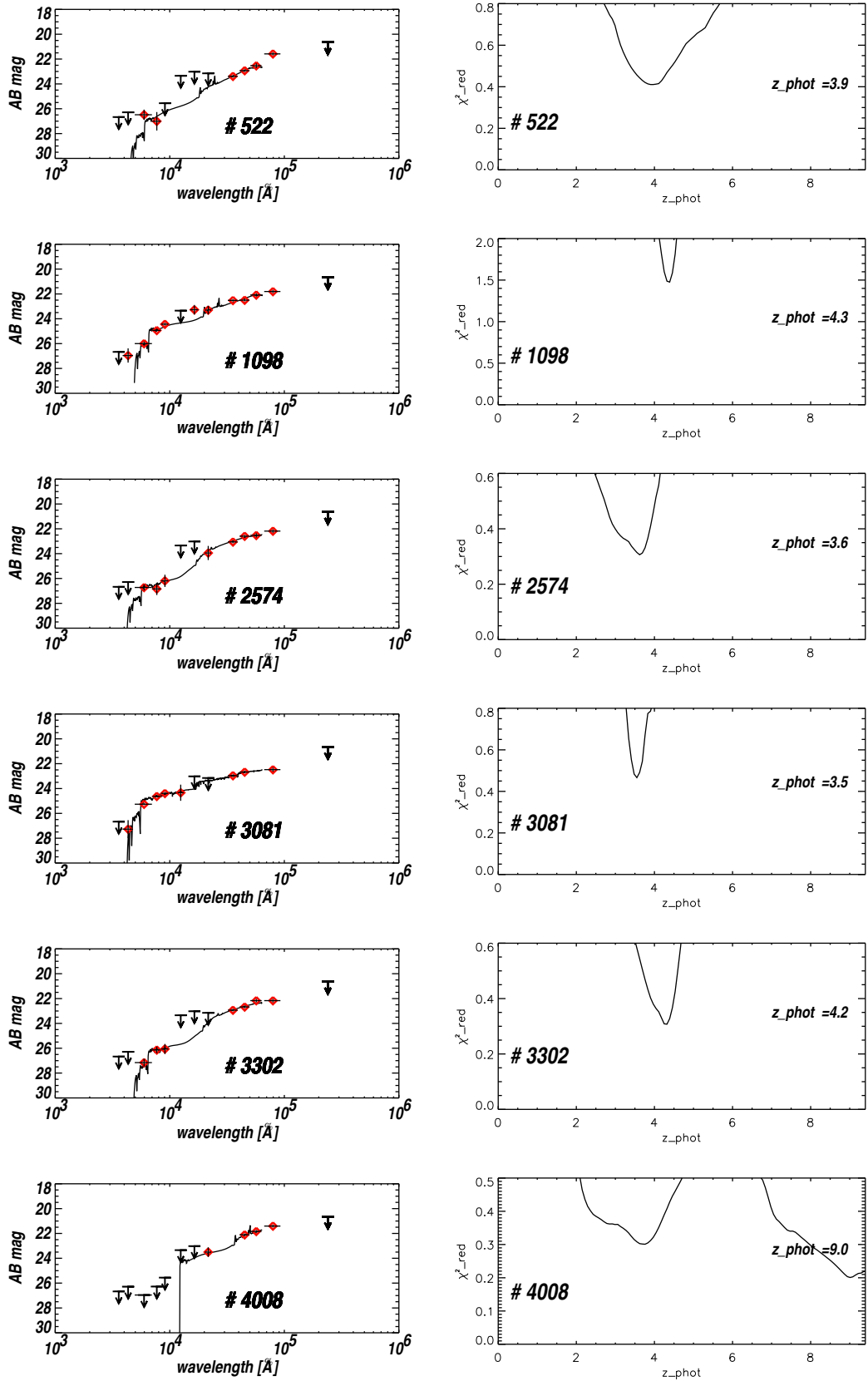}
\caption{Spectral Energy Distribution best-fit models for the MIPS-u sample in
  units of AB magnitude. The red diamonds represent the observed magnitudes
  from the photometric data. Non-detections are represented by upper limits. See also Tab.~\ref{table:tab3}.}
\label{fig:fig13}
\end{figure*}

In Fig.~\ref{fig:fig12} we compare our results for the MIPS-u sample with estimates
of the stellar mass density for quiescent galaxies at other redshifts from the literature.
To facilitate this comparison, here we have scaled all values to the Chabrier IMF 
and MA05 models by means of the $M/L$ conversion relations given above (see
Sect.\ref{tot_mass_dens}).
 We have also computed the comoving stellar mass density of the
  MIPS-u galaxies considering the unlikely case that they were all truly ``passive'',
  consistent with single burst stellar populations (SSPs) with no dust. 
  The masses were derived using the ratio between the observed IRAC-$4.5~\mu$m 
  flux for each galaxy and the expected flux for an MA05 SSP model at the 
  same redshift, assuming an age of $\sim 0.5$~Gyr and mass normalized to 1~M$_{\odot}$. 
  In this case, the number density of MIPS-u candidates above the mass threshold 
  would be larger ($1.52\times10^{-5}$~Mpc$^{-3}$). The corresponding stellar mass density
  would be $(1.46\pm 0.7)\times10^6$~M$_{\odot}$Mpc$^{-3}$. This value is also represented in Fig.~\ref{fig:fig12} by the red dashed line. From this exercise
  we concluded that even if all the MIPS-u galaxies were considered as old SSPs,  the number density of `quiescent' galaxies at $z=3.5-5$ would remain very small in comparison with that found at $z\leq 2-2.5$ by previous studies.

The stellar mass density we found for the MIPS-u galaxies in the redshift range $z=3.5-5$ is $\sim2$ orders of magnitude lower than the local value \citep[black asterisk; ][]{2004ApJ...600..681B}, and more than 10 times lower than the values found at $z\sim1.7$ by \citet[][; black filled rhombus]{2005ApJ...626..680D} for the passive-BzK galaxies in the Hubble Ultra Deep Field (HUDF) and at $z\sim2.5$ by \citet[][black open square]{2005ApJ...624L..81L} for Distant Red Galaxies (DRGs) in the Hubble Deep Field South. Moreover our result is also in good agreement with the mass density estimated by \citet{2008ApJ...676..781W} for $z \sim 5.5$ Balmer-break-selected galaxies without MIPS detections in the GOODS South Field at redshift $z\sim5.5$ (open rhombus in the figure). Considering that our estimate is an upper limit to the comoving stellar mass density of massive and quiescent galaxies, we conclude that there is evidence of strong evolution beyond $z\sim 2$ \citep[see also][]{2005ApJ...626..680D}. The decreasing number density of quiescent objects at high redshifts could mean that most massive galaxies at early epochs were experiencing strong star formation activity.

\section{Summary and conclusions}\label{concl}
In recent years, a substantial number of massive galaxies have been found at redshifts as large as $z\sim5$. However, although the range of stellar populations and observed colors among galaxies at all redshifts studied to date has proven to be quite heterogeneous, most surveys at $z > 3.5$ to date have primarily selected samples of star-forming blue galaxies, and have been biased against red massive objects that are faint or completely undetected in optical and near-IR bands.

In this work we have studied galaxies selected at IRAC-$4.5~\mu$m in the GOODS-N field, in a manner that is complementary with respect to optical or near-IR selection. This allows us to pick out massive, evolved galaxies at high redshift, for which the optical rest frame emission (which is most nearly related to the stellar mass) is sampled by the redshifted, rest-frame near-infrared portion of the spectrum.  On the other hand, this selection may also be sensitive to `active' obscured galaxies at high redshift, for which IR emission could be reprocessed radiation from absorbing material surrounding an AGN, or from dust heated by star formation, like that seen in dusty starburst galaxies such as local ULIRGs.

Recently, other studies based on near-IR selection have detected red, massive galaxy candidates at $z_{phot}\geq4$, but have not obtained complete samples. In most of these studies, the samples have been selected in the $K$-band, with the possibility of missing $K$-undetected objects visible in the longer-wavelength IRAC filters \citep[e.g.][]{2005ApJ...635..832M,2008ApJ...676..781W}. 
\citet{2007A&A...470...21R} used IRAC-$3.6~\mu$m selection ($m_{3.6}<23.26$) to identify a sample of high-$z$ galaxies missed by optical and $K$-band selection. However they assumed as additional conditions the non-detection of their candidates in optical bands, and a detection close to the sky threshold limit in the K band ($K>23$, AB system), which make their sample {\it a priori} incomplete.

Here, we attempt to select a sample of massive high redshift red galaxies that is as complete as possible, by means of IRAC selection, that could allow us to recover massive objects that may have been missed by previous studies.  The most important results of this paper are summarized in the following items:

\begin{description}
 \item $\bullet$~{By selecting at IRAC-$4.5~\mu$m ($m_{4.5}\leq 23$), we
 extracted a sample of $4\,142$ objects in the $\sim 165~$arcmin$^2$ of the
 GOODS-N field. We performed an SED fitting analysis to estimate photometric
 redshifts for those objects, and found fifty-three candidates at $z\geq 3.5$, 
 also requiring that peak of the $f_\nu$ stellar spectral energy distribution
 (at $1.6~\mu$m in the rest-frame) falls in the IRAC-$8.0~\mu$m band. We excluded 
 unobscured AGN from our final sample, identified by their hard and/or soft-X ray
 emission.  Almost $81\%$ of our candidates are completely missed by the $B$- and 
 $V$-band Lyman break dropout techniques designed to identify UV-bright, star-forming
 galaxies at similar redshifts. For each object, we evaluate the reliability and the 
 confidence limits of our photometric redshift estimates with a $\chi^2$ test. 
 Two objects have spectroscopically confirmed redshifts \citep[GN20 and GN20.2;][]{Daddietal2008} 
 in good agreement with our estimates.}

\item $\bullet$~{We divide our final sample in two sub-samples of (32) MIPS-detected and (21) MIPS-undetected 
 objects (the MIPS-d and MIPS-u samples, respectively).\\
{\bfseries MIPS-d sample:} In the subsample detected at 24$\mu$m, we found 18 galaxies with unambiguous high photometric redshift solutions and 14 with degenerate solutions. If the high redshift solution is correct, the $24~\mu$m emission could indicate the presence of a heavily obscured AGN (perhaps Compton thick due to the absence of X ray emission), or perhaps strong star formation activity in a dusty, hyperluminous starburst.  We cannot firmly exclude the possibility for these galaxies to be dusty starburst galaxies at lower redshifts, for which the $24~\mu$m flux could be explained with PAH emission at $z\sim2-3$.\\
{\bfseries MIPS-u sample:} 21 of our $z \geq 3.5$ candidates are undetected at 24$\mu$m. The lack of $24~\mu$m emission is a necessary but not sufficient condition to indicate that these are quiescent galaxy. The upper limit to the SFR set by MIPS detection limit ($f_{24}<20~\mu$Jy) corresponds to a star formation rate $< 1500$~M$_{\odot} yr^{-1}$, assuming bolometric corrections derived from spectral templates for local ultraluminous infrared galaxies. For this reason we can use this sub-sample only to give an upper limit to the total stellar mass of `quiescent' galaxies at $z\geq3.5$.}

\item $\bullet$~{14 galaxies from our $z \geq 3.5$ sample are detected at submillimeter and millimeter wavelengths with SCUBA \citep{2006MNRAS.370.1185P}, MAMBO \citep{2008arXiv0806.3106G} and AzTEC \citep{2008arXiv0806.3791P}. All but two of are also detected at 24$\mu$m.  Two objects, GN20 and GN20.2, have spectroscopically confirmed redshifts at $z\sim4$ \citet{Daddietal2008}. By comparing their colours with other SMGs galaxies in the S$_{850}$/S$_{24}$ vs S$_{24}$/S$_{8}$ diagram, we find that $\sim 78$\% of the sub-mm detected objects among our candidates have properties that are fully consistent with those of starburst galaxies at $z\geq3.5$.  The remaining $\sim 22\%$ are more likely at lower redshift.
}

\item $\bullet$~{We have computed the contribution to the number density and to the stellar mass density by the total (MIPS-d+MIPS-u) sample and by the subsample that is undetected at 24$\mu$m (MIPS-u) for galaxies with stellar masses above a limit of $5\times 10^{10}$~M$_{\odot}$, chosen to minimize incompleteness at $z < 5$.  For the redshift range $3.5 < z < 5$, we find a comoving number density $2.6\times 10^{-5}$~Mpc$^{-3}$ for average masses of $\sim 10^{11}$~M$_{\odot}$.
The corresponding stellar mass density is $(2.9\pm 1.5) \times 10^{6}$~M$_{\odot}$Mpc$^{-3}$. Our results are in good agreement with previous determinations in the literature, and with the prediction from the Millennium simulations.

For the MIPS-u sample, we determined a number density of $0.97\times 10^{-5}$~Mpc$^{-3}$ and a corresponding stellar mass density of $(1.15\pm 0.7)\times 10^{6}$~M$_{\odot}$Mpc$^{-3}$. We also considered the extreme case that all of the MIPS-undetected galaxies were purely quiescent, with ages $\sim0.5$~Gyr.  This results in larger upper limits for their number and stellar mass densities, $1.52\times 10^{-5}$~Mpc$^{-3}$ and $(1.46 \pm 0.7)\times 10^{6}$~M$_{\odot}$Mpc$^{-3}$ respectively.  Our MIPS-u sample of quiescent candidates accounts for $\sim 5-6\%$ of the mass density that was in place at $z\sim2$  and only for $\sim1\%$ of local stellar mass density \citep{2004ApJ...600..681B}.
This provides evidence for strong growth in the number of massive galaxies between redshift $z=4-5$ and $z=2$.}

\end{description}

\begin{acknowledgements}
We thank Kyoung-Soo Lee for her reductions of the KPNO-4m FLAMINGOS imaging
data of the GOODS-North field,and for information about the details of this
data set. CM and ED acknowledge support
from the French ANR grant numbers ANR-07-BLAN-0228 and ANR-08-JCJC-0008.
\end{acknowledgements}

\newpage
\begin{figure*}[t]
\includegraphics[width=0.8\textwidth]{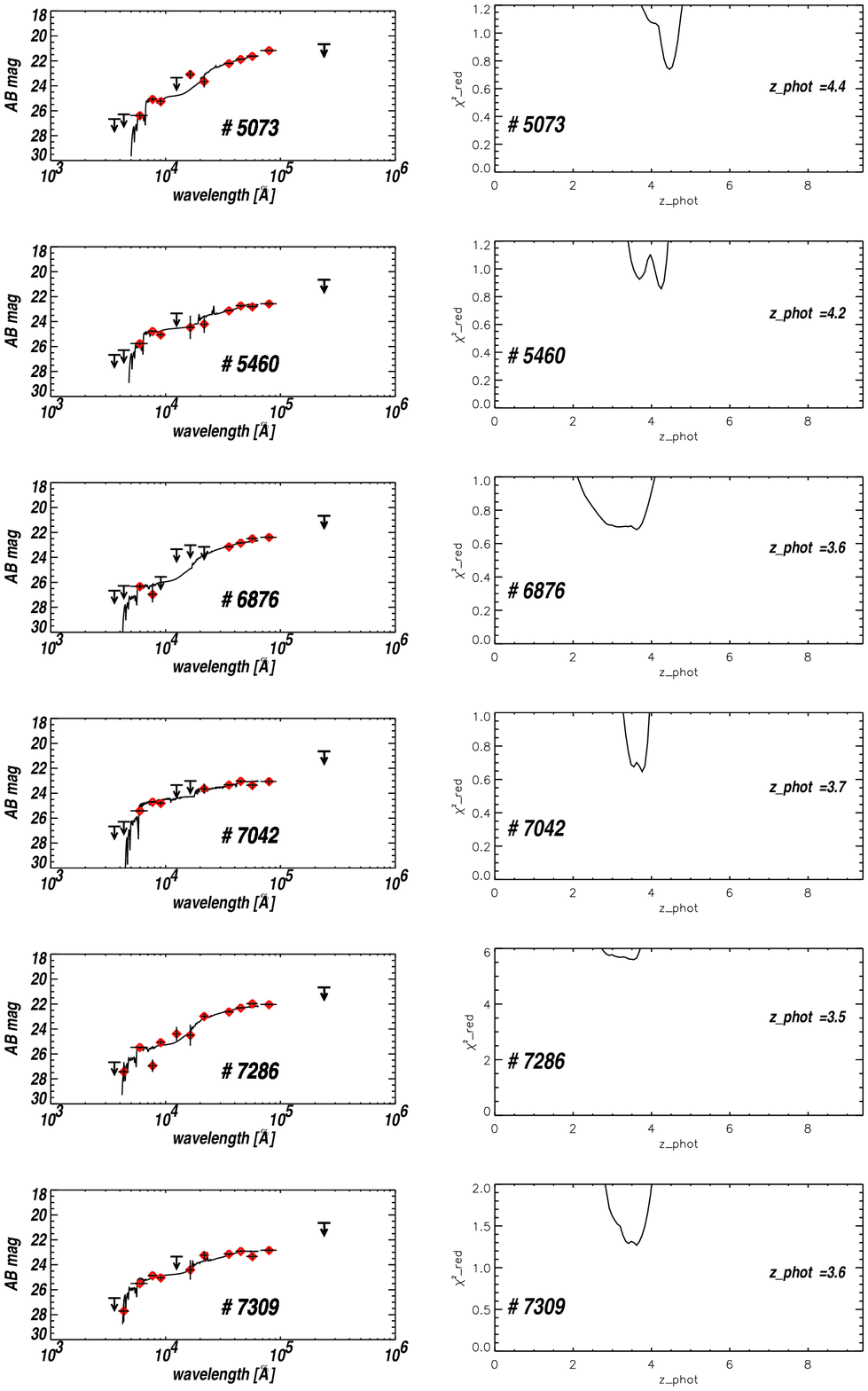}
\end{figure*}
\newpage

\begin{figure*}[t]
\includegraphics[width=0.8\textwidth]{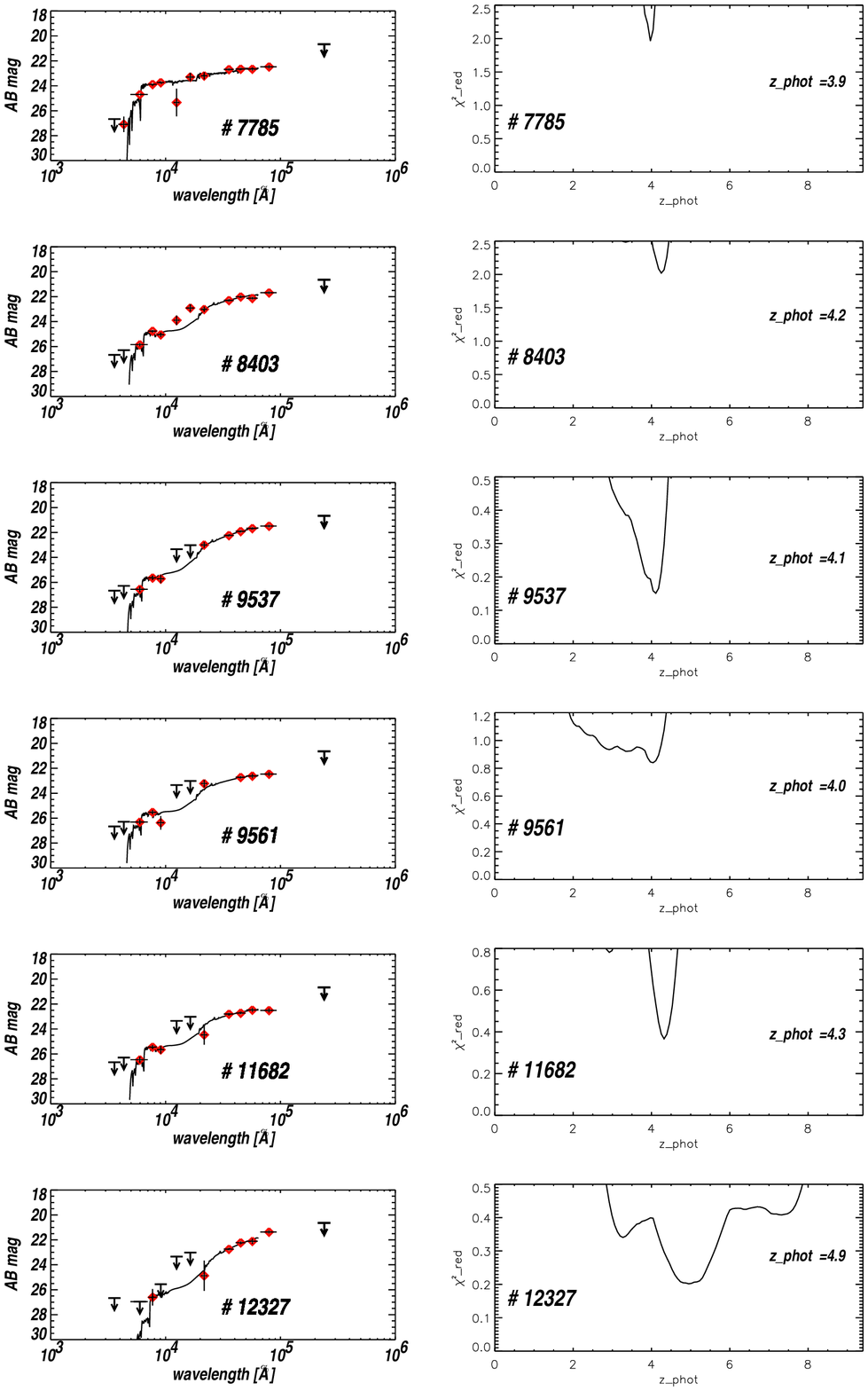}
\end{figure*}
\newpage

\begin{figure*}[t]
\includegraphics[width=0.8\textwidth]{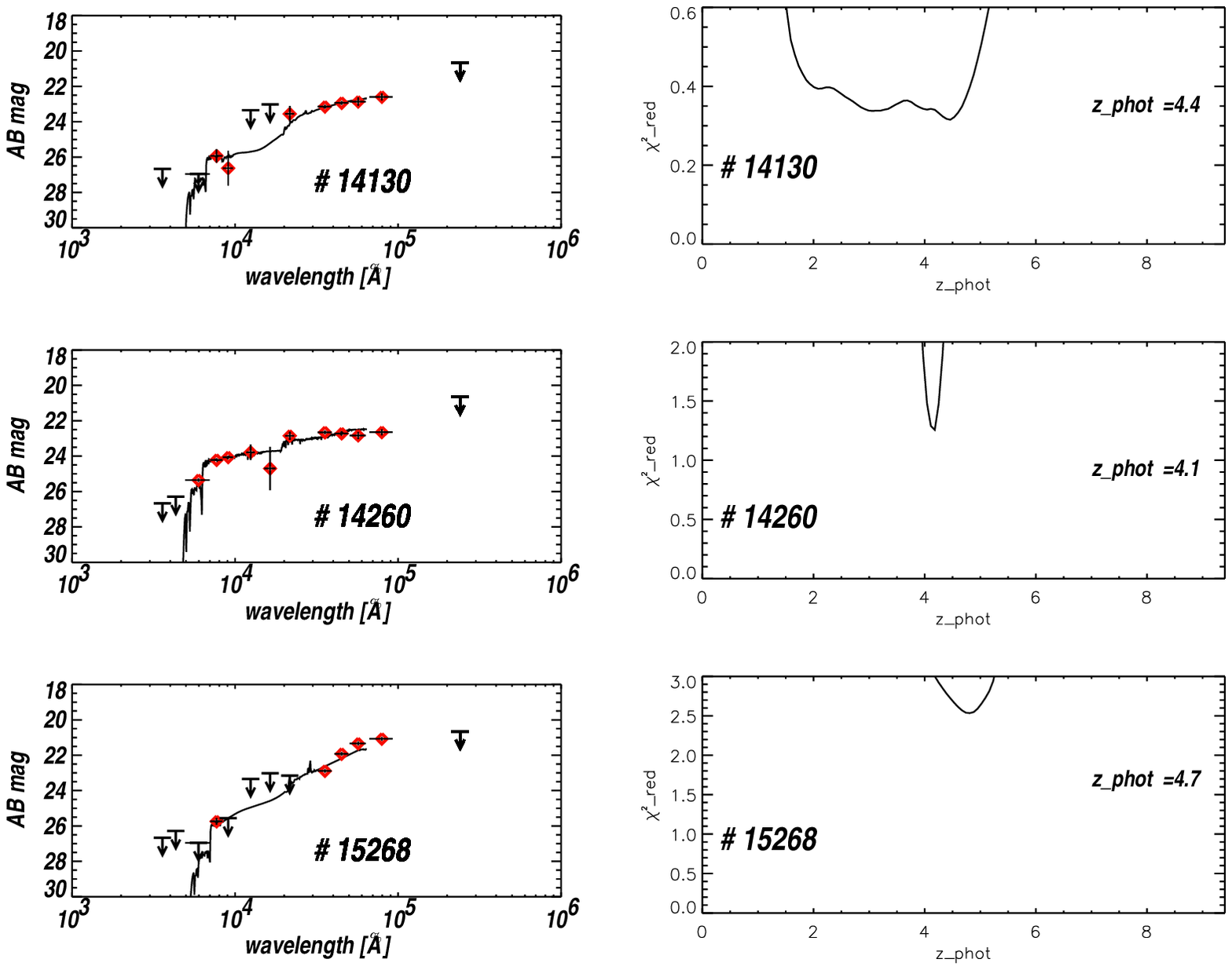}
\end{figure*}

\newpage

\begin{figure*}[t]
\includegraphics[width=0.8\textwidth]{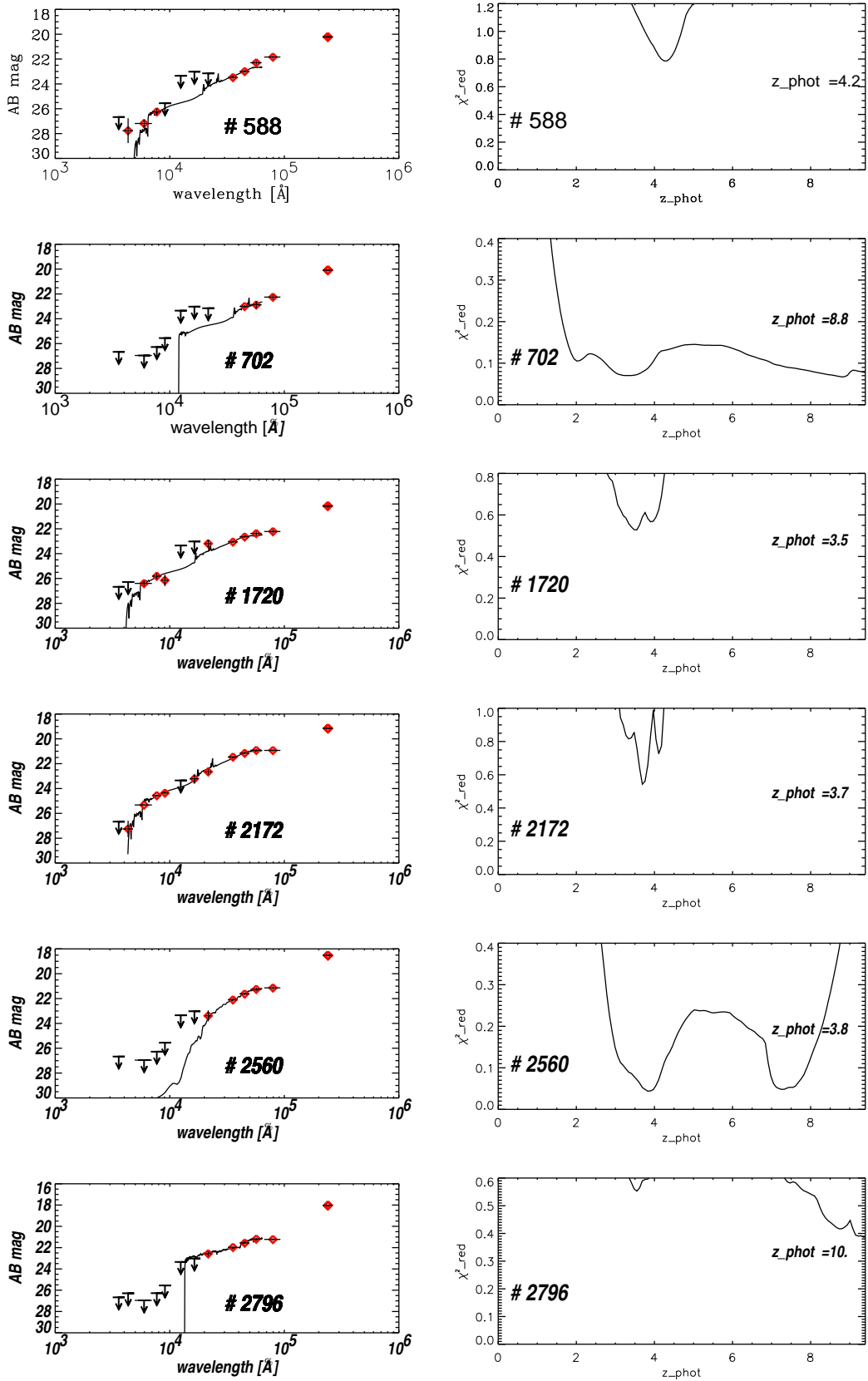}
\caption{Spectral Energy Distributions of MIPS-d galaxies of our final high-$z$ sample (on the left panels), and the correspondent $\chi^2_{\nu}$ distributions as a function of redshift (right panels) in $90\%$ confidence intervals.  For each candidate the most probable redshifts (first and second redshift solutions : $z_{phot}$ and $z_{phot2}$) and the $\chi^2_{\nu}$ values with the respective probability are also labeled. See also Tab.~\ref{table:tab4}.}
\label{fig:fig14}
\end{figure*} 

\newpage
\begin{figure*}[t]
\includegraphics[width=0.8\textwidth]{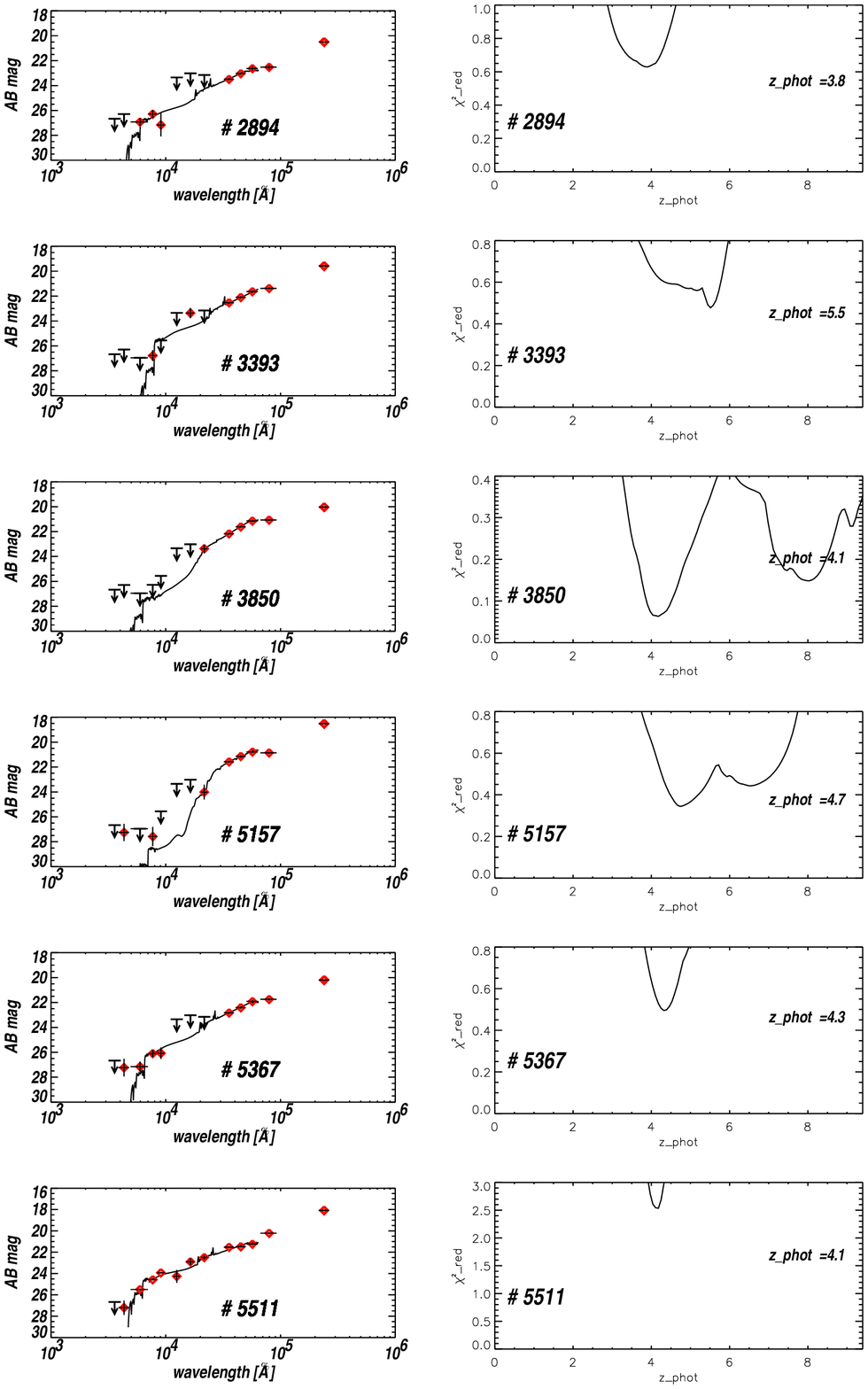}
\end{figure*} 

\newpage
\begin{figure*}[t]
\includegraphics[width=0.8\textwidth]{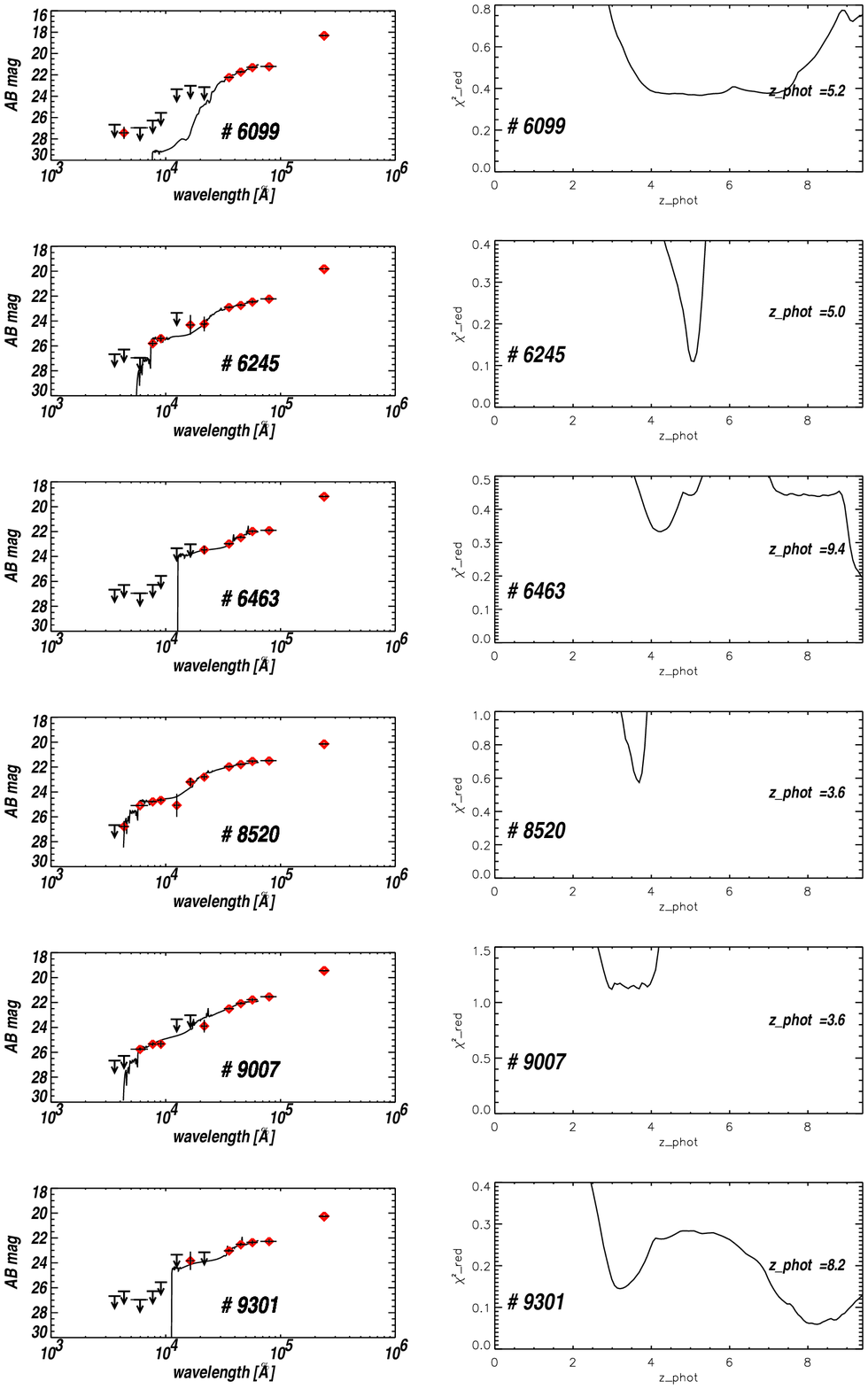}
\end{figure*} 

\newpage
\begin{figure*}[t]
\includegraphics[width=0.8\textwidth]{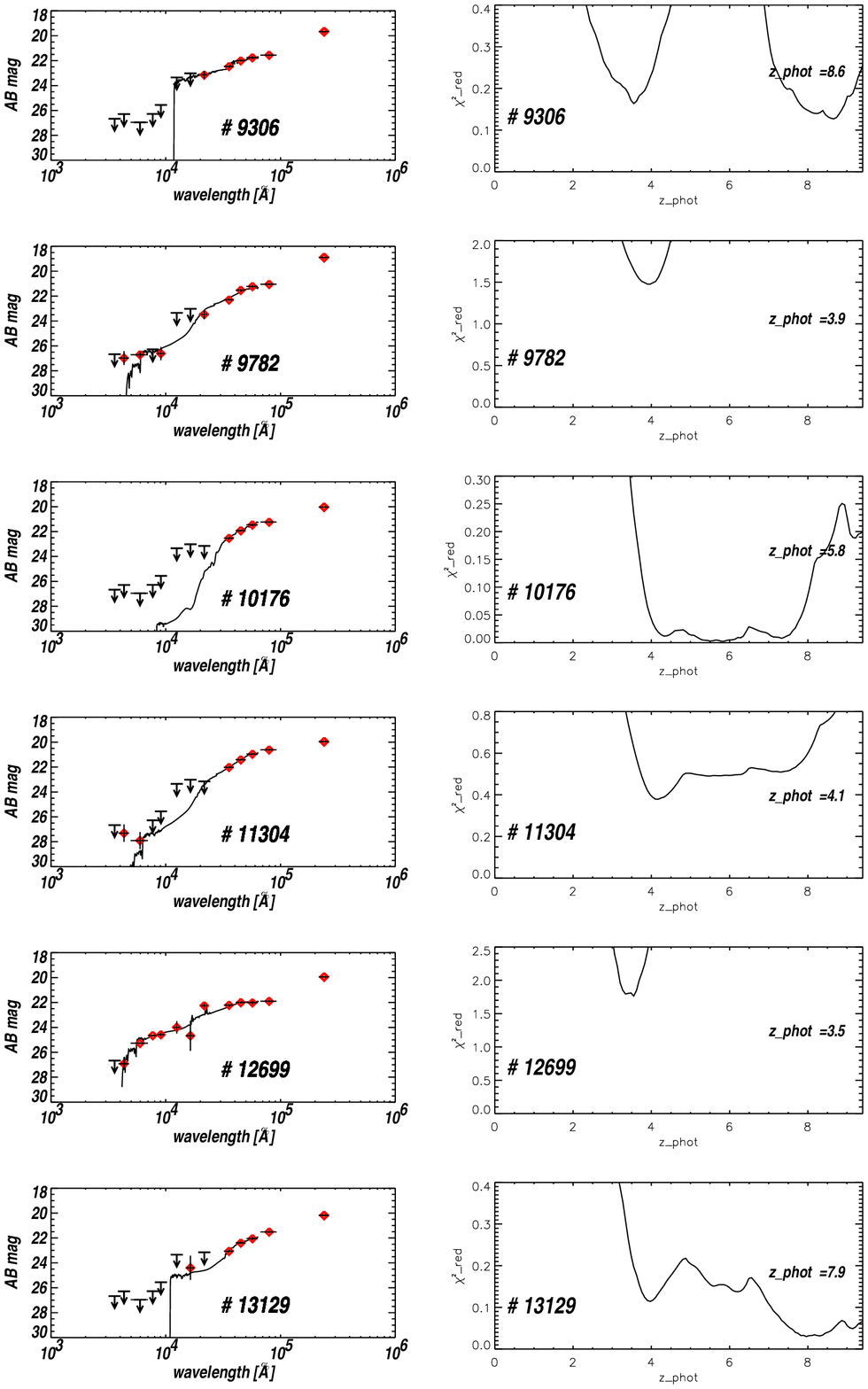}
\end{figure*} 

\newpage
\begin{figure*}[t]
\includegraphics[width=0.8\textwidth]{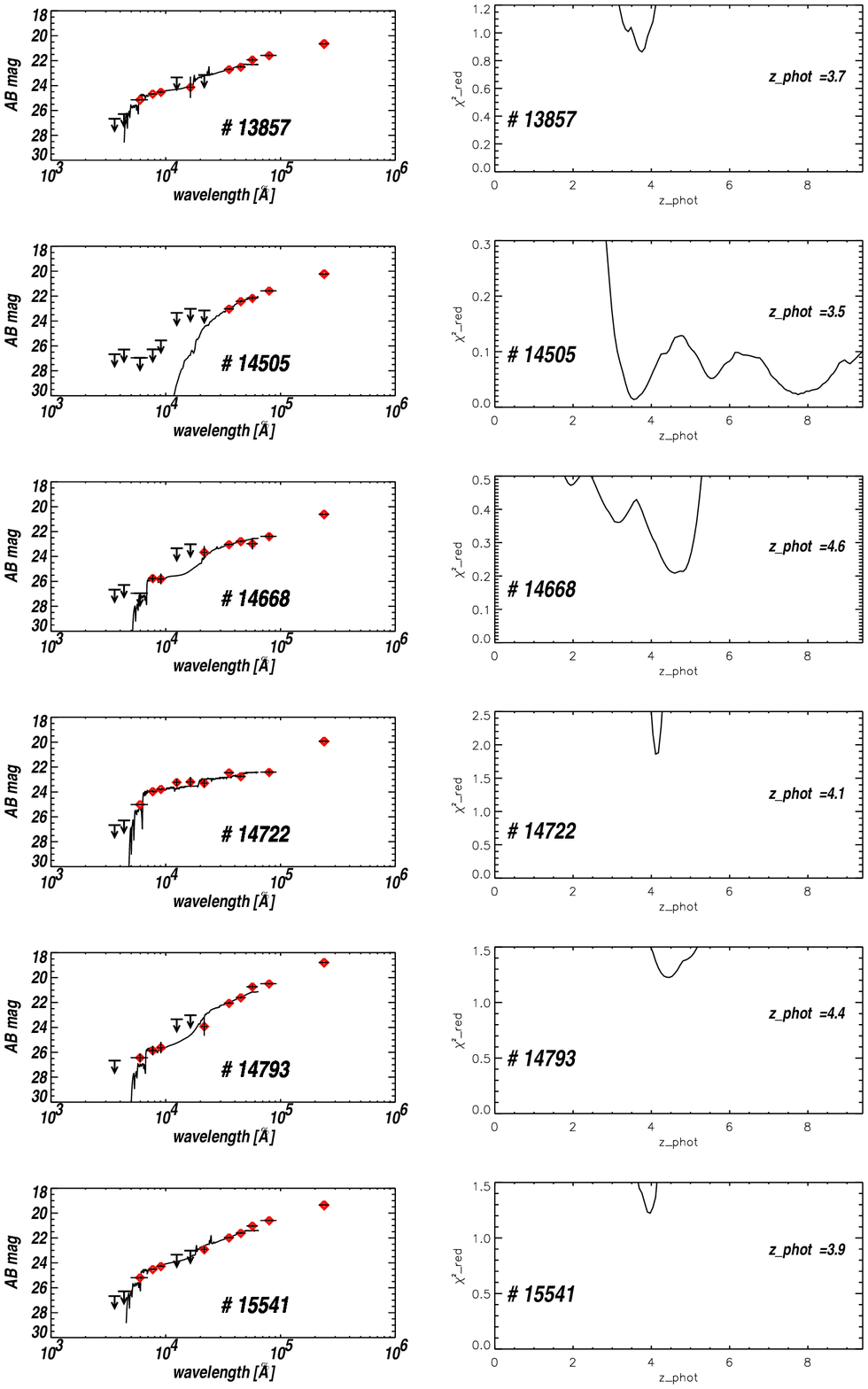}
\end{figure*} 

\newpage
\begin{figure*}[t]
\includegraphics[width=0.8\textwidth]{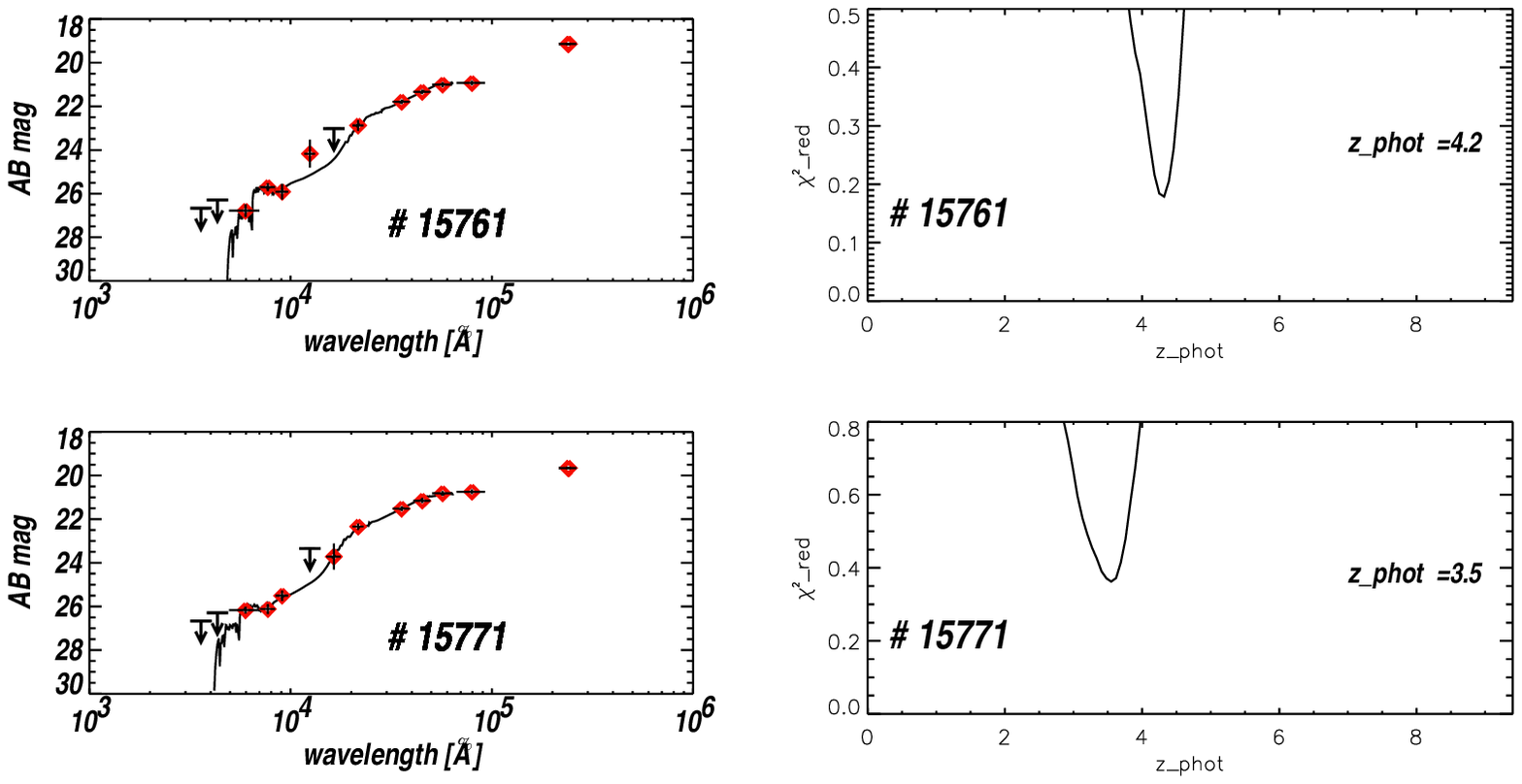}
\end{figure*} 

\newpage

\begin{landscape} 
\begin{figure*}
\begin{center}
\caption{Multi-wavelength images of the fifty-three galaxies of the final sample (MIPS-d+MIPS-u). \newline
The cut-outs have a size of $10'' \times 10''$. North is up, East is at left.}
\includegraphics[width=\textwidth]{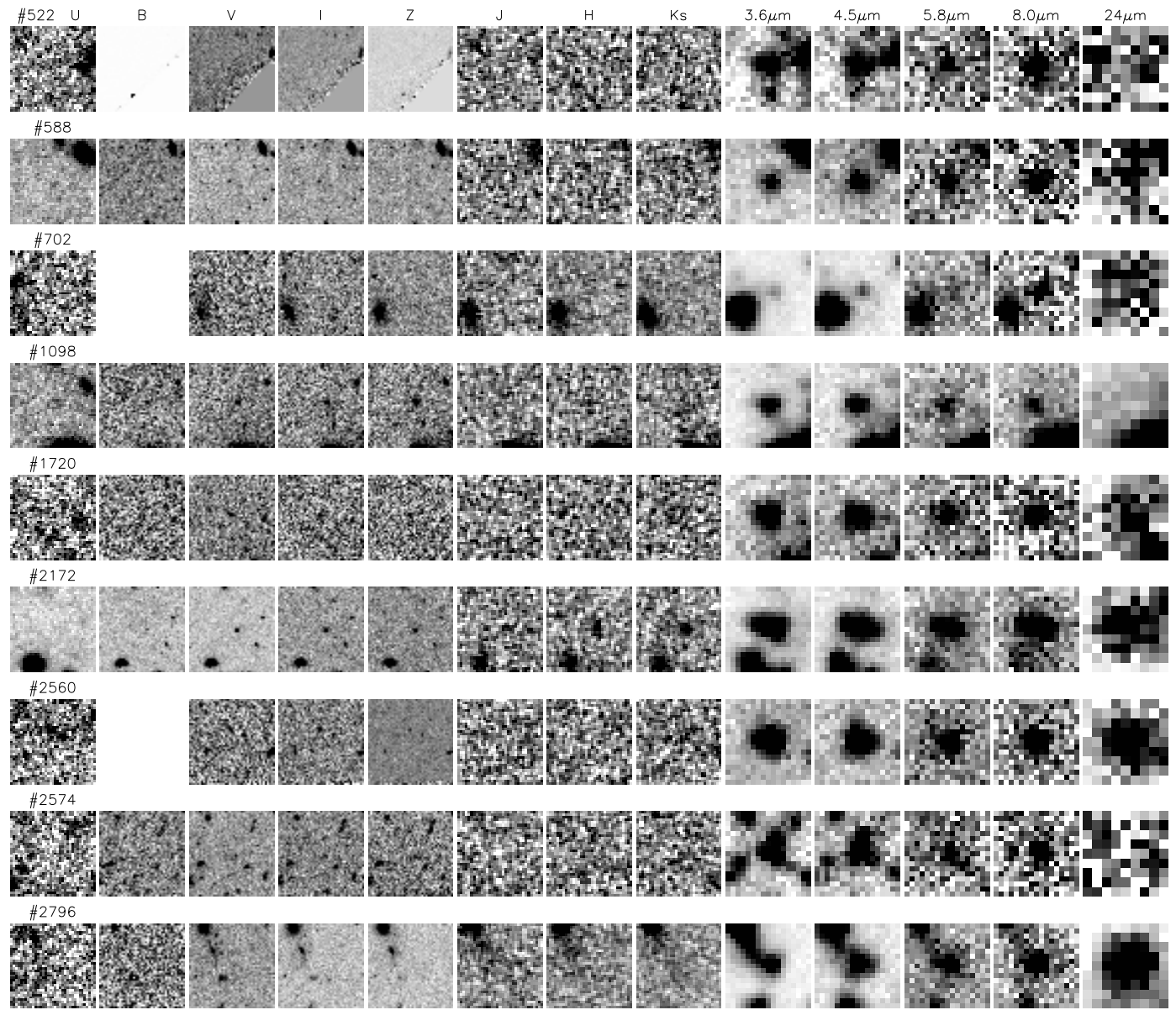}
\label{fig:fig15}
\end{center}
\end{figure*}

\begin{figure*}
\begin{center}
\includegraphics[width=\textwidth]{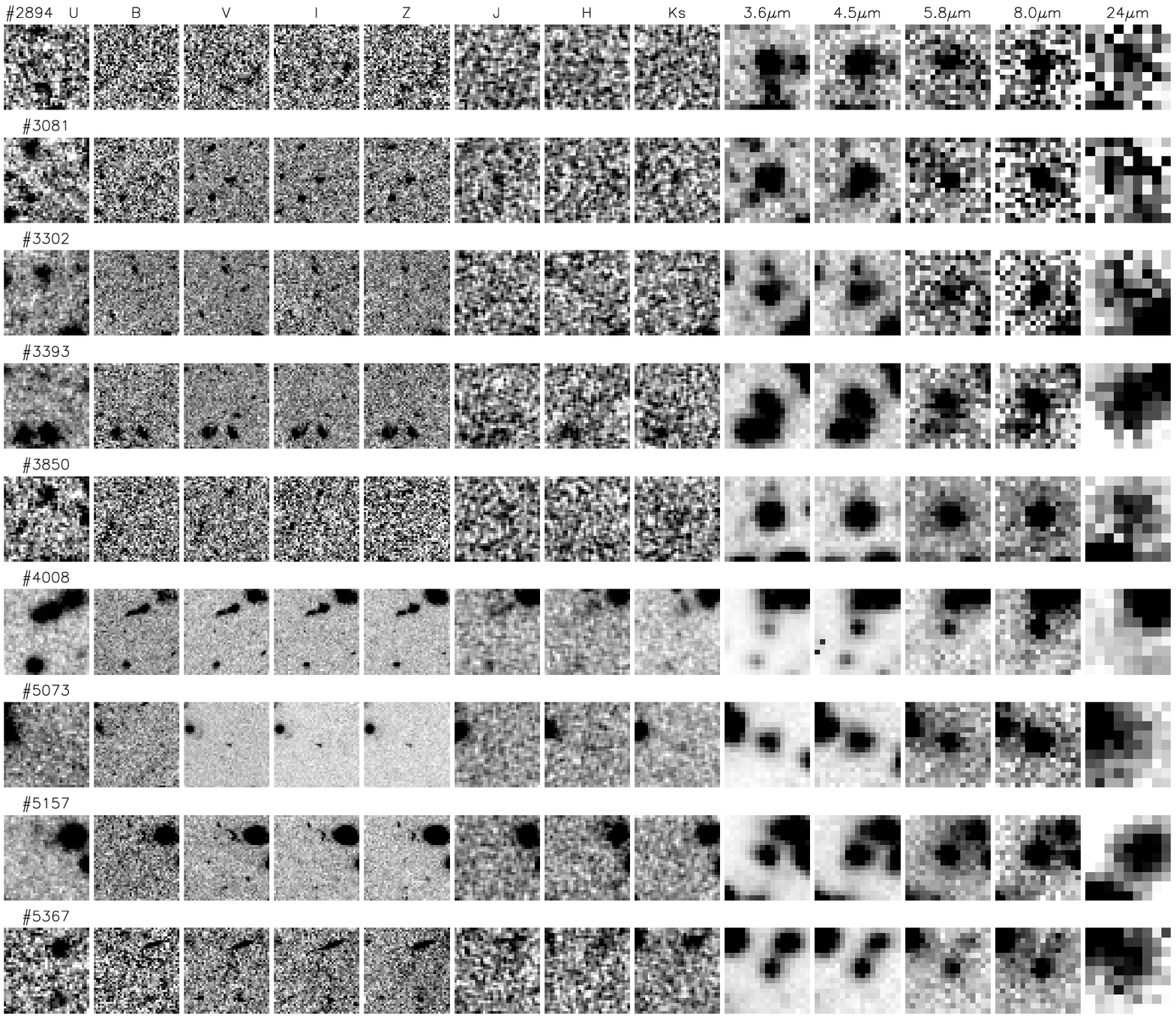}
\end{center}
\end{figure*} 

\begin{figure*}
\begin{center}
\includegraphics[width=\textwidth]{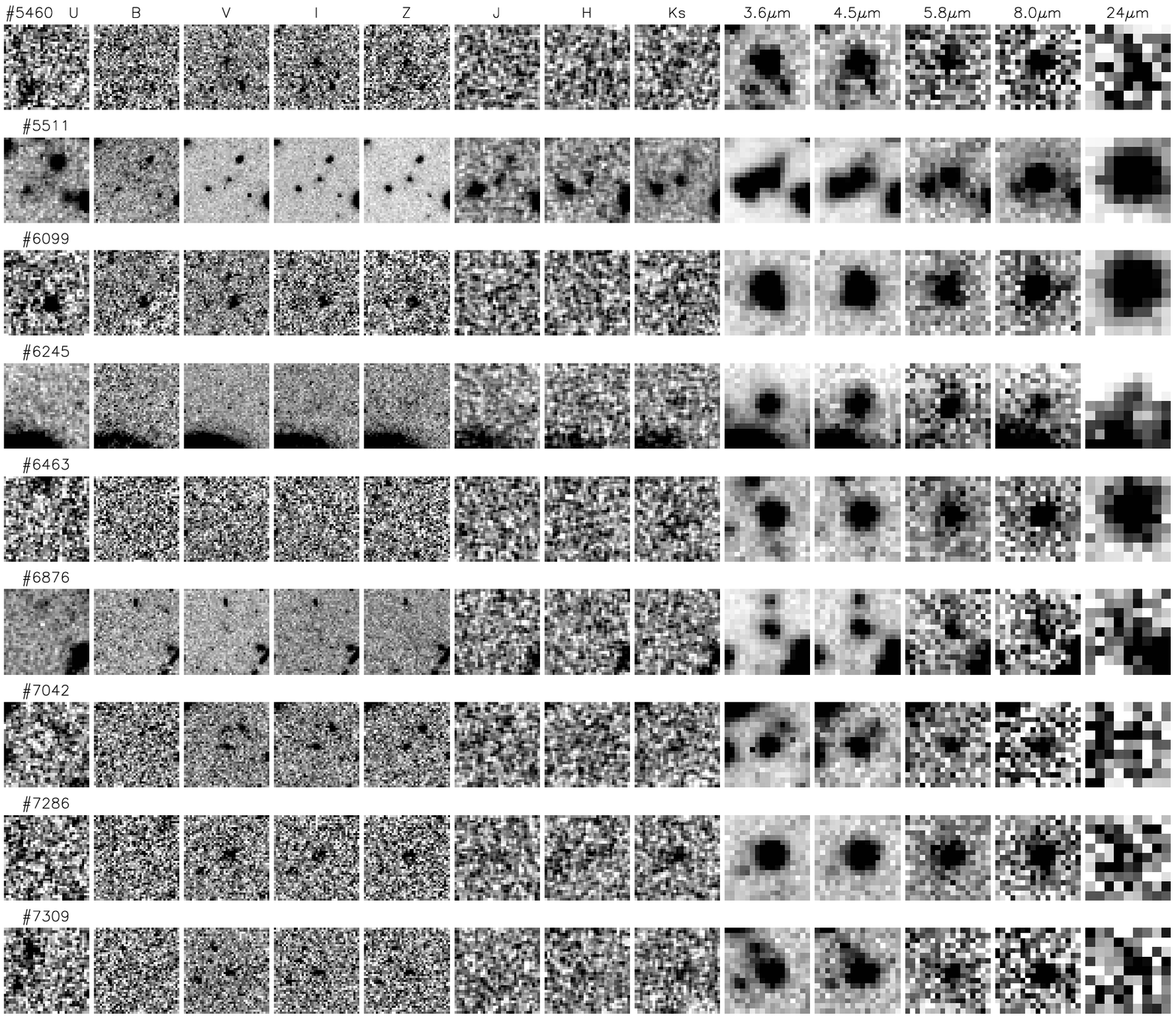}
\end{center}
\end{figure*} 

\begin{figure*}
\begin{center}
\includegraphics[width=\textwidth]{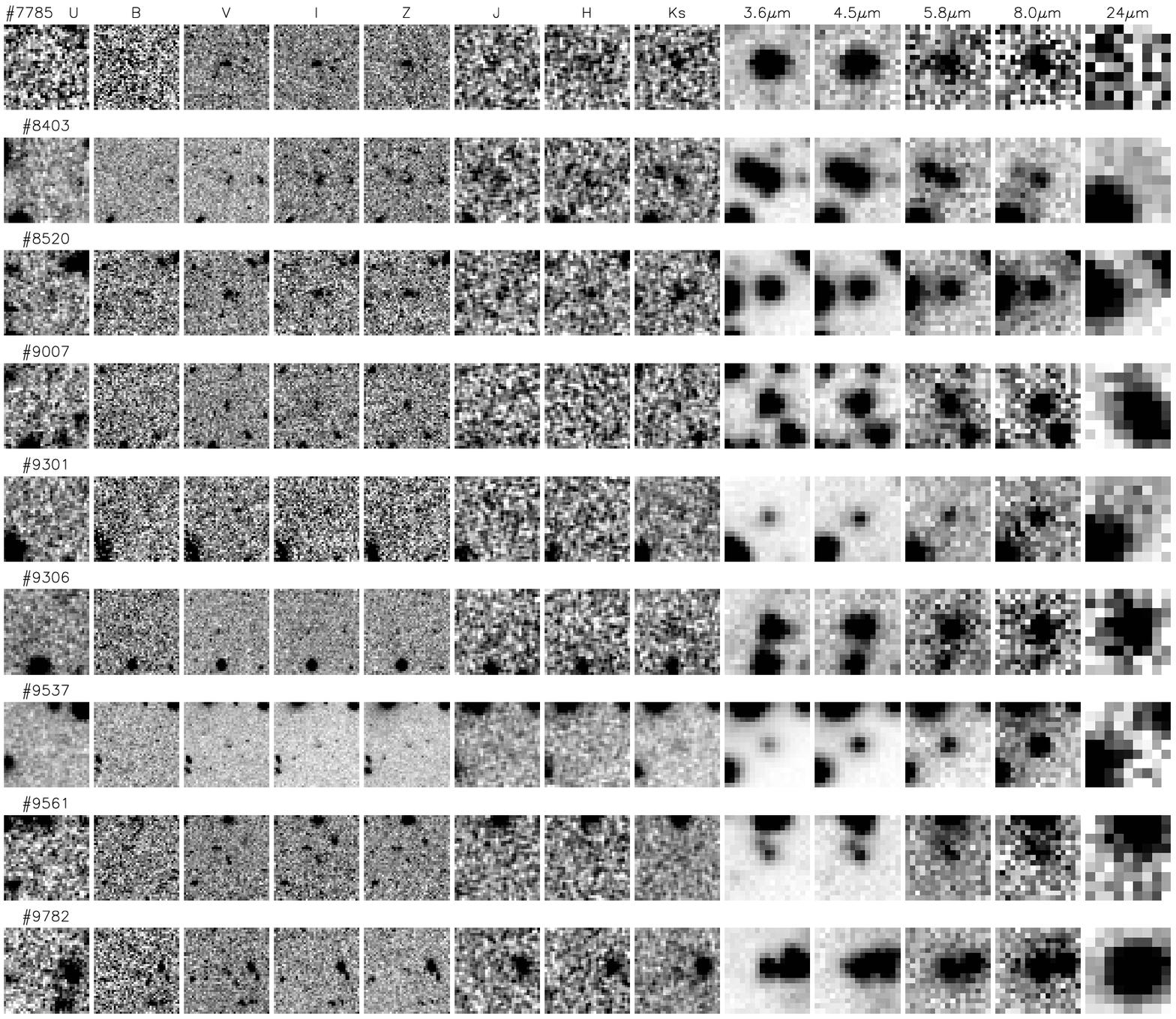}
\end{center}
\end{figure*} 

\begin{figure*}
\begin{center}
\includegraphics[width=\textwidth]{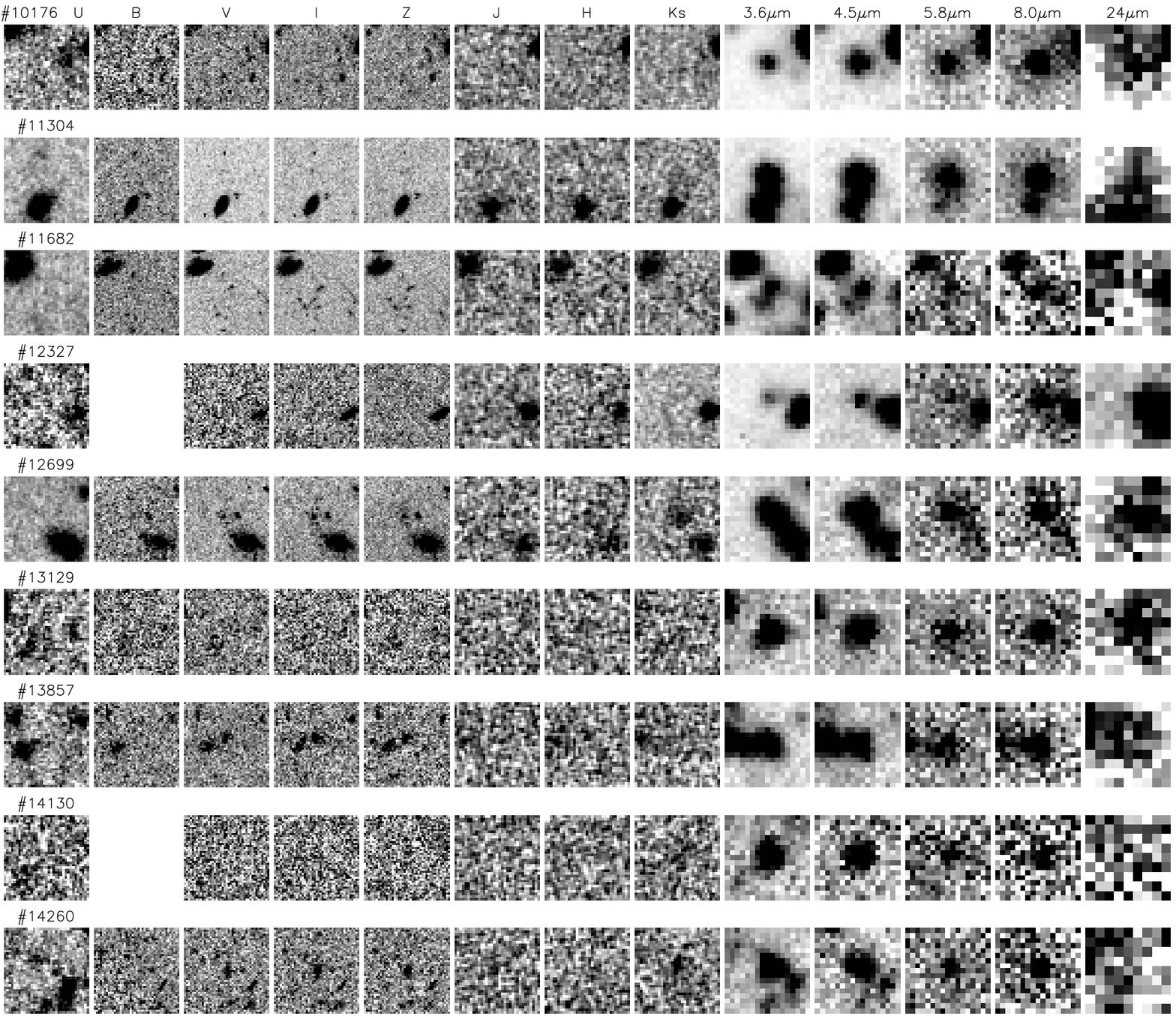}
\end{center}
\end{figure*} 

\begin{figure*}
\begin{center}
\includegraphics[width=\textwidth]{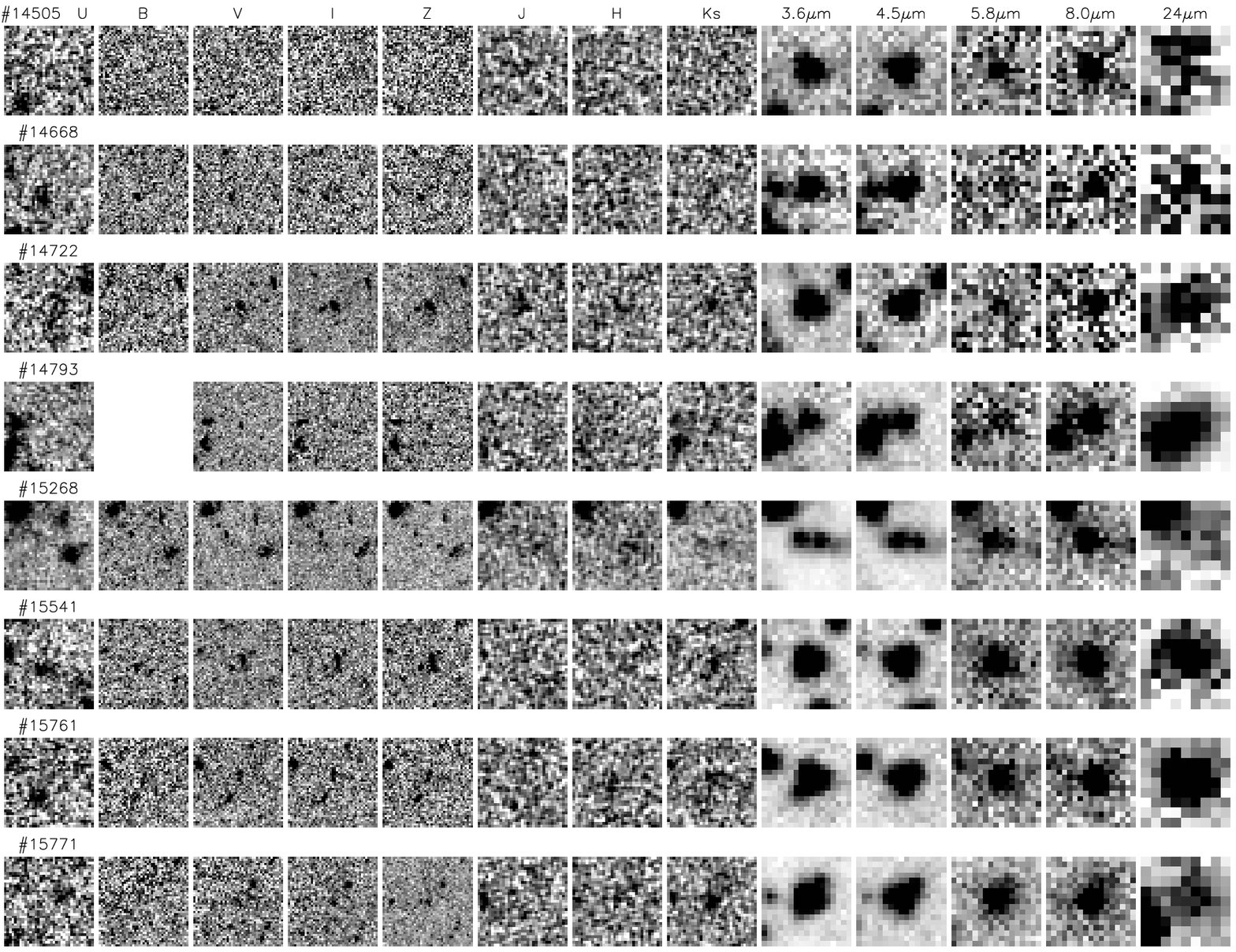}
\end{center}
\end{figure*} 
\end{landscape} 


\begin{landscape}
\begin{table}
\caption{\label{table:tab6}Multi-band photometry for the final sample of 44
  candidates at $z\ge 3.5$. The coordinates reported here are from the \spi\ data-set.}		 
{\tiny
\begin{tabular}{llllllllllllllll}
ID&  $U$& $B$& $V$& $i$& $z$& $J$& $H$& $Ks$& $m_{3.6}$& $m_{4.5}$& $m_{5.8}$& $m_{8.0}$& $m_{24}$& RA& Dec\\
\hline
\#   522  &  $>26.7$  &  $>26.3$  &  26.485  &  26.980  &  $>25.5$  & $>23.3$  &  $>23.0$  &  $>23.1$  &  23.288  &  22.895  &  22.517  &  21.583  &  $>20.6$   &  12:36:27.2 & 62:06:06.1 \\		      
\#   588  &  $>26.7$  &  27.774  &  27.187  &  26.210  &  $>25.5$  &  $>23.3$  &  $>23.0$  &  $>23.1$  &  23.379  &  22.960  &  22.274  &  21.836  &  20.218   	&  12:36:32.7 & 62:06:21.8 \\
\#   702  &  $>26.7$  &  $---$  &  $>26.9$  &  $>26.3$  &  $>25.5$  & $>23.3$  &  $>23.0$  &  $>23.1$  &  $---$  &  22.964  &  22.857  &  22.254  &  20.081  	&  12:36:15.7 & 62:06:43.4 \\		   
\# 1098  &  $>26.7$  &  26.968  &  25.998  &  24.907  &  24.317  & $>23.3$  &  23.242  &  23.127  &  22.422  &  22.465  &  22.065  &  21.807  &  $>20.6$      	&  12:36:36.8 & 62:07:14.0 \\		   
\#  1720  &  $>26.7$  &  $>26.3$  &  26.397  &  25.778  &  26.033  & $>23.3$  &  $>23.0$  &  23.021  &  22.954  &  22.611  &  22.354  &  22.218  &  20.161   	&  12:36:51.0 & 62:08:29.8 \\		   
\#  2172  &  $>26.7$  &  27.262  &  25.325  &  24.536  &  24.243  & $>23.3$  &  23.187  &  22.482  &  21.360  &  21.118  &  20.898  &  20.928  &  19.156   	&  12:36:13.3 & 62:09:02.0 \\		   
\#  2560  &  $>26.7$  &  $---$  &  $>26.9$  &  $>26.3$  &  $>25.5$  & $>23.3$  &  $>23.0$  &  23.222  &  21.999  &  21.587  &  21.227  &  21.137  &  18.527   	&  12:35:53.0 & 62:09:30.2 \\		   
\#  2574  &  $>26.7$  &  $>26.3$  &  26.722  &  26.810  &  26.070  &  $>23.3$  &  $>23.0$  &  23.778  &  22.958  &  22.559  &  22.510  &  22.183  &  $>20.6$   	&  12:36:04.7 & 62:09:25.5 \\	      
\#  2796  &  $>26.7$  &  $>26.3$  &  $>26.9$  &  $>26.3$  &  $>25.5$  &  $>23.3$  &  $>23.0$  &  22.404  &  21.869  &  21.528  &  21.167  &  21.227  &  18.029  &  12:36:03.1 & 62:09:47.5 \\		 
\#  2894  &  $>26.7$  &  $>26.3$  &  26.926  &  26.256  &  27.044  &  $>23.3$  &  $>23.0$  &  $>23.1$  &  23.399  &  22.997  &  22.608  &  22.521  &  20.496   	&  12:37:10.0 & 62:09:56.4 \\		 
\#  3081  &  $>26.7$  &  27.283  &  25.263  &  24.600  &  24.272  &  24.240  &  $>23.0$  &  $>23.1$  &  22.859  &  22.638  &  $---$  &  22.480  &  $>20.6$   	&  12:36:44.4 & 62:10:03.9 \\		 
\#  3302  &  $>26.7$  &  $>26.3$  &  27.155  &  26.106  &  25.940  &  $>23.3$  &  $>23.0$  &  $>23.1$  &  22.834  &  22.638  &  22.146  &  22.169  &  $>20.6$   &  12:35:59.9 & 62:10:08.4 \\		 
\#  3393  &  $>26.7$  &  $>26.3$  &  $>26.9$  &  26.751  &  $>25.5$  &  $>23.3$  &  23.329  &  $>23.1$  &  22.423  &  22.064  &  21.615  &  21.376  &  19.581   &  12:37:19.5 & 62:10:21.4 \\		 
\#  3850  &  $>26.7$  &  $>26.3$  &  $>26.9$  &  $>26.3$  &  $>25.5$  &  $>23.3$  &  $>23.0$  &  23.202  &  22.073  &  21.579  &  21.125  &  21.066  &  20.037  &  12:37:12.5 & 62:10:36.1 \\		 
\#  4008  &  $>26.7$  &  $>26.3$  &  $>26.9$  &  $>26.3$  &  $>25.5$  &  $>23.3$  &  $>23.0$  &  23.330  &  $---$  &  22.069  &  21.817  &  21.405  &  $>20.6$  &  12:36:53.7 & 62:11:12.9 \\		 
\#  5073  &  $>26.7$  &  $>26.3$  &  26.382  &  25.038  &  25.135  &  $>23.3$  &  23.058  &  23.489  &  22.107  &  21.831  &  21.602  &  21.165  &  $>20.6$   	&  12:36:36.1 & 62:11:54.7 \\		 
\#  5157  &  $>26.7$  &  27.265  &  $>26.9$  &  27.550  &  $>25.5$  &  $>23.3$  &  $>23.0$  &  23.846  &  21.479  &  21.102  &  20.764  &  20.856  &  18.519   	&  12:36:37.5 & 62:11:57.2 \\		 
\#  5367  &  $>26.7$  &  27.241  &  27.154  &  26.063  &  25.954  &  $>23.3$  &  $>23.0$  &  $>23.1$  &  22.729  &  22.373  &  21.895  &  21.745  &  20.189   	&  12:37:18.9 & 62:12:17.9 \\		 
\#  5460  &  $>26.7$  &  $>26.3$  &  25.754  &  24.742  &  24.932  &  $>23.3$  &  24.431  &  24.036  &  23.019  &  22.691  &  22.781  &  22.567  &  $>20.6$   	&  12:36:31.3 & 62:12:18.3 \\		 
\#  5511  &  $>26.7$  &  27.209  &  25.492  &  24.543  &  23.806  &  24.163  &  22.865  &  22.320  &  21.429  &  21.444  &  21.219  &  20.221  &  18.084   	&  12:36:37.1 & 62:12:31.1 \\		 
\#  6099  &  $>26.7$  &  27.438  &  $>26.9$  &  $>26.3$  &  $>25.5$  &  $>23.3$  &  $>23.0$  &  $>23.1$  &  22.129  &  21.678  &  21.269  &  21.209  &  18.305  &  12:36:08.6 & 62:12:51.2 \\		  
\#  6245  &  $>26.7$  &  $>26.3$  &  $>26.9$  &  25.766  &  25.270  &  $>23.3$  &  24.283  &  24.065  &  22.791  &  22.688  &  22.433  &  22.222  &  19.809   	&  12:37:20.6 & 62:13:00.2 \\		 
\#  6463  &  $>26.7$  &  $>26.3$  &  $>26.9$  &  $>26.3$  &  $>25.5$  &  $>23.3$  &  $>23.0$  &  23.287  &  22.875  &  22.426  &  21.948  &  21.903  &  19.177  &  12:37:02.5 & 62:13:02.6 \\		  
\#  6876  &  $>26.7$  &  $>26.3$  &  26.322  &  26.938  &  $>25.5$  &  $>23.3$  &  $>23.0$  &  $>23.1$  &  23.042  &  22.801  &  22.468  &  22.386  &  $>20.6$  &  12:36:30.4 & 62:13:33.0 \\		 
\#  7042  &  $>26.7$  &  $>26.3$  &  25.418  &  24.665  &  24.685  &  $>23.3$  &  $>23.0$  &  23.463  &  23.224  &  22.992  &  23.337  &  23.059  &  $>20.6$   	&  12:37:25.5 & 62:13:39.8 \\		 
\#  7286  &  $>26.7$  &  27.443  &  25.467  &  26.915  &  24.953  &  24.288  &  24.464  &  22.810  &  22.519  &  22.265  &  21.941  &  22.033  &  $>20.6$   	&  12:37:09.5 & 62:13:50.3 \\		 
\#  7309  &  $>26.7$  &  27.718  &  25.498  &  24.823  &  24.915  &  $>23.3$  &  24.377  &  23.068  &  23.031  &  22.875  &  23.306  &  22.830  &  $>20.6$   	&  12:36:41.6 & 62:13:50.1 \\		 
\#  7785  &  $>26.7$  &  27.099  &  24.698  &  23.850  &  23.618  &  25.240  &  23.265  &  23.026  &  22.584  &  22.616  &  22.637  &  22.473  &  $>20.6$   	&  12:36:55.9 & 62:14:12.5 \\		 
\#  8403  &  $>26.7$  &  $>26.3$  &  25.844  &  24.729  &  24.927  &  23.792  &  22.877  &  22.841  &  22.198  &  21.971  &  22.106  &  21.684  &  $>20.6$   	&  12:37:03.8 & 62:14:51.6 \\		   
\#  8520  &  $>26.7$  &  26.790  &  25.090  &  24.743  &  24.539  &  24.971  &  23.162  &  22.617  &  21.867  &  21.739  &  21.504  &  21.488  &  20.144   	&  12:36:45.1 & 62:14:48.8 \\		 
\#  9007  &  $>26.7$  &  $>26.3$  &  25.750  &  25.303  &  25.209  &  $>23.3$  &  $>23.0$  &  23.721  &  22.399  &  22.033  &  21.749  &  21.528  &  19.433   	&  12:37:34.7 & 62:15:15.8 \\		 
\#  9301  &  $>26.7$  &  $>26.3$  &  $>26.9$  &  $>26.3$  &  $>25.5$  &  $>23.3$  &  23.802  &  $>23.1$  &  22.924  &  22.484  &  22.332  &  22.270  &  20.255  &  12:38:03.6 & 62:15:30.5 \\		 
\#  9306  &  $>26.7$  &  $>26.3$  &  $>26.9$  &  $>26.3$  &  $>25.5$  &  $>23.3$  &  $>23.0$  &  22.973  &  22.360  &  21.967  &  21.735  &  21.555  &  19.661  &  12:37:33.8 & 62:15:32.1 \\		 
\#  9537  &  $>26.7$  &  $>26.3$  &  26.557  &  25.606  &  25.600  &  $>23.3$  &  $>23.0$  &  22.824  &  22.133  &  21.873  &  21.650  &  21.485  &  $>20.6$   	&  12:36:39.4 & 62:15:42.8 \\	 
\#  9561  &  $>26.7$  &  $>26.3$  &  26.310  &  25.492  &  26.253  &  $>23.3$  &  $>23.0$  &  23.056  &  $---$ &  22.686  &  22.597  &  22.463  &  $>20.6$   	&  12:37:52.7 & 62:15:49.3 \\		
\#  9782  &  $>26.7$  &  26.988  &  26.705  &  $>26.3$  &  26.493  &  $>23.3$  &  $>23.0$  &  23.298  &  22.203  &  21.487  &  21.199  &  21.046  &  18.889  	&  12:37:39.5 & 62:15:59.0 \\		
\# 10176  &  $>26.7$  &  $>26.3$  &  $>26.9$  &  $>26.3$  &  $>25.5$  &  $>23.3$  &  $>23.0$  &  $>23.1$  &  22.428  &  21.882  &  21.439  &  21.237  &  20.034 &  12:36:22.0 & 62:16:16.2 \\		 
\# 11304  &  $>26.7$  &  27.318  &  27.906  &  $>26.3$  &  $>25.5$  &  $>23.3$  &  $>23.0$  &  $>23.1$  &  21.918  &  21.373  &  20.939  &  20.617  &  19.959   &  12:36:31.9 & 62:17:14.8 \\		 
\# 11682  &  $>26.7$  &  $>26.3$  &  26.464  &  25.416  &  25.536  &  $>23.3$  &  $>23.0$  &  24.296  &  22.702  &  22.682  &  22.455  &  22.507  &  $>20.6$    &  12:37:39.2 & 62:17:36.8 \\		 
\# 12327  &  $>26.7$  &  $---$  &  $>26.9$  &  26.574  &  $>25.5$  &  $>23.3$  &  $>23.0$  &  24.705  &  22.645  &  22.189  &  22.091  &  21.367  &  $>20.6$    &  12:38:05.4 & 62:18:16.4 \\		  
\# 12699  &  $>26.7$  &  26.925  &  25.271  &  24.616  &  24.460  &  23.892  &  24.640  &  22.082  &  22.107  &  21.959  &  21.996  &  21.895  &  19.942   	&  12:37:24.0 & 62:18:33.7 \\		  
\# 13129  &  $>26.7$  &  $>26.3$  &  $>26.9$  &  $>26.3$  &  $>25.5$  &  $>23.3$  &  24.368  &  $>23.1$  &  22.963  &  22.364  &  22.029  &  21.515  &  20.186  &  12:37:28.1 & 62:19:20.5 \\		  
\# 13857  &  $>26.7$  &  $>26.3$  &  25.150  &  24.638  &  24.404  &  $>23.3$  &  24.107  &  $>23.1$  &  22.605  &  22.465  &  21.913  &  21.578  &  20.637   	&  12:37:08.8 & 62:22:02.2 \\ 
\# 14130  &  $>26.7$  &  $---$  &  $>26.9$  &  25.899  &  26.511  &  $>23.3$  &  $>23.0$  &  23.387  &  23.048  &  22.900  &  22.846  &  22.596  &  $>20.6$   	&  12:37:36.3 & 62:21:25.3 \\		  
\# 14260  &  $>26.7$  &  $>26.3$  &  25.352  &  24.183  &  23.941  &  23.682  &  24.668  &  22.678  &  22.552  &  22.680  &  22.814  &  22.647  &  $>20.6$   	&  12:37:13.0 & 62:21:11.4 \\		   
\# 14505  &  $>26.7$  &  $>26.3$  &  $>26.9$  &  $>26.3$  &  $>25.5$  &  $>23.3$  &  $>23.0$  &  $>23.1$  &  22.924  &  22.388  &  22.140  &  21.573  &  20.211 &  12:36:44.0 & 62:19:38.8 \\		   
\# 14668  &  $>26.7$  &  $>26.3$  &  $>26.9$  &  25.723  &  25.684  &  $>23.3$  &  $>23.0$  &  23.497  &  22.936  &  22.749  &  22.956  &  22.385  &  20.610   	&  12:37:12.3 & 62:23:03.4 \\		   
\# 14722  &  $>26.7$  &  $>26.3$  &  25.010  &  23.929  &  23.658  &  23.133  &  23.167  &  23.125  &  22.352  &  22.714  &  $---$   &  22.414  &  19.934   	&  12:37:11.5 & 62:21:55.8 \\		  
\# 14793  &  $>26.7$   &  $---$  &  26.436  &  25.824  &  25.501  &  $>23.3$  &  $>23.0$  &  23.757  &  21.952  &  21.566  &  20.716  &  20.492  &  18.787   	&  12:36:58.9 & 62:22:15.3 \\		   
\# 15268 &  $>26.7$  &  $>26.3$ &   $>26.9$ &    25.710  &  $>25.5$ &   $>23.3$&   $>23.0$  &  $>23.1$ &   22.790  &  21.887 &   21.316  &  21.071    & $>20.6$ &  12:37:09.6 & 62:22:02.5 \\
\# 15541  &  $>26.7$  &  $>26.3$  &  25.185  &  24.487  &  24.155  &  $>23.3$  &  $>23.0$  &  22.753  &  21.874  &  21.573  &  21.010  &  20.595  &  19.359   	&  12:37:11.9 & 62:22:12.4 \\		   
\# 15761  &  $>26.7$  &  $>26.3$  &  26.783  &  25.670  &  25.788  &  24.071  &  $>23.0$  &  22.707  &  21.688  &  21.299  &  20.985  &  20.931  &  19.140   	&  12:36:55.3 & 62:21:07.9 \\		   
\# 15771  &  $>26.7$  &  $>26.3$  &  26.170  &  26.087  &  25.386  &  $>23.3$  &  23.687  &  22.170  &  21.418  &  21.117  &  20.795  &  20.743  &  19.659   	&  12:37:01.5 & 62:20:25.3 \\		
\hline
\end{tabular}
}
\end{table}
\end{landscape}

\bibliography{ref_gn_ch2_081224}
\end{document}